\documentclass[a4paper,11pt]{report}
\usepackage{scriptie_package}

\begin{document}

\frontmatter

\begin{titlepage}

  \setcounter{page}{-1}

\setlength{\headheight}{14pt}

\begin{center}

\begin{figure}[H]
  \centering
  \begin{minipage}[b]{0.6\textwidth}
    \includegraphics[width=\textwidth]{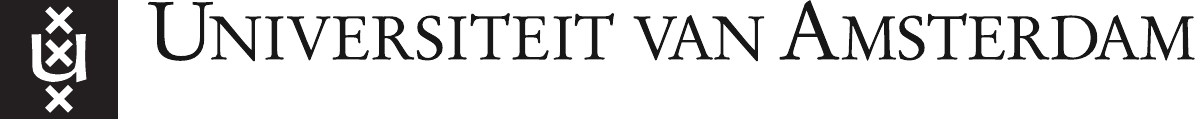}
  \end{minipage}
  \hfill
  \begin{minipage}[b]{0.3\textwidth}
    \includegraphics[width=\textwidth]{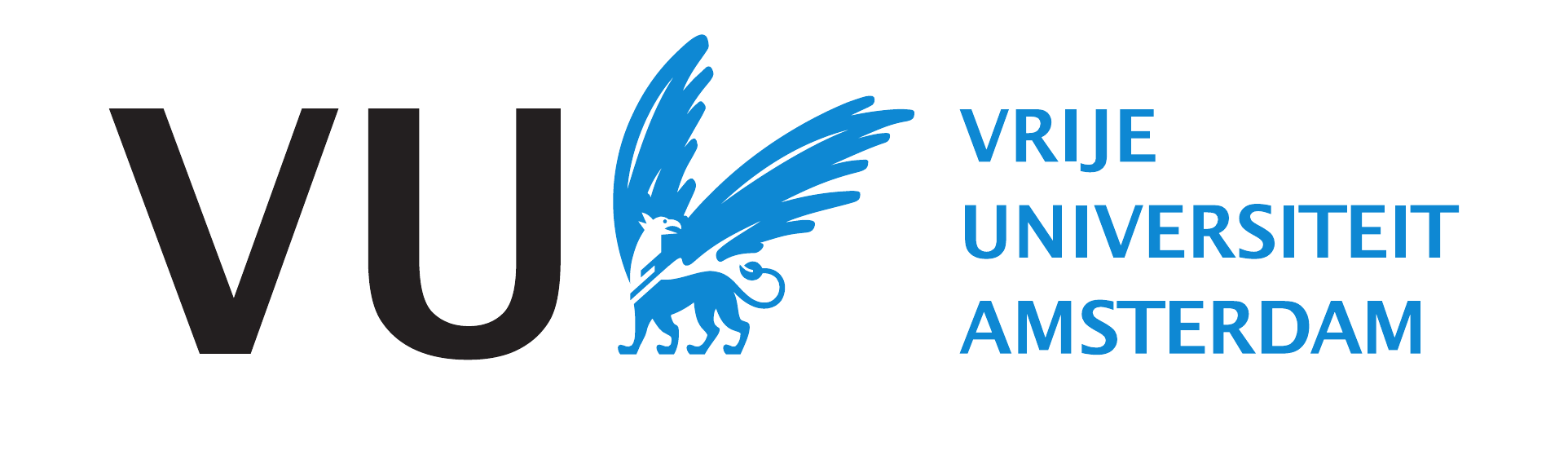}
  \end{minipage}
\end{figure}

\bigskip
\bigskip
\bigskip

{\makeatletter
\fontsize{25}{25}\selectfont{MSc Physics and Astronomy}
\makeatother}

{\makeatletter
{\medskip\fontsize{17}{17}\selectfont{Theoretical Physics}}
\makeatother}

\vspace{0.8cm}

{\makeatletter
\fontsize{24}{24}\selectfont{Master Thesis}
\makeatother}

\vspace{0.7cm}

\rule{\linewidth}{0.5pt}

\vspace{0.3cm}

{\makeatletter
\fontsize{33}{33}\selectfont\textbf{Positive thinking}
\makeatother}

{\makeatletter
\bigskip\fontsize{20}{20}\selectfont{A search for energy conditions}
\makeatother}
\vspace{0.15cm}
\rule{\linewidth}{0.5pt}

\begin{Large}
by
\end{Large}

\medskip

{\makeatletter
\fontsize{25}{25}\selectfont{Hidde Stoffels}
\makeatother}

\medskip

{\makeatletter
\fontsize{20}{20}\selectfont{14709120}
\makeatother}

\vspace{1cm}

\begin{tabular}{ll}
    Supervisor: & dr. Ben Freivogel \\
    Second reader: & dr. Diego Hofman \\
    Date:       & July 11, 2024 \\
    Period: & September 2023 - July 2024
\end{tabular}

\end{center}	
\end{titlepage}
\chapter*{Abstract}\addcontentsline{toc}{chapter}{Abstract}

Energy conditions are attempts to summarise the properties of realistic descriptions of matter via constraints on the energy-momentum tensor. This is, for example, useful when one wants to understand the types of spacetime geometry that can be realised in general relativity. However, it is currently unknown what (if any) energy condition a realistic quantum theory interacting gravitationally would obey.

In this thesis, we provide support for the conformally invariant averaged null energy condition (CANEC), a particular proposal towards a more generally valid energy condition. We motivate our interest in the CANEC by reviewing the role of the null energy condition (NEC) in classical gravity, summarising a general proof of the averaged NEC (ANEC) for quantum field theories in flat spacetime, and going over a holographic proof of the CANEC in a class of curved spacetimes. The context in which the CANEC applies is then extended with two proofs: a holographic proof in two types of spacetime to which the existing proof does not apply, and a field theoretical proof for conformal field theories in conformally flat spacetimes.

Finally, we consider light-ray operators in Minkowski spacetime. The connection between these operators and the CANEC is that the CANEC on a lightcone in Minkowski spacetime constrains a light-ray operator for an even number of spacetime dimensions $d$. Via an explicit computation, we show that many light-ray operators are positive-semidefinite in a class of non-minimally coupled but otherwise free real scalar field theories. One of the operators for which this holds is the CANEC-operator on the lightcone for even $d$.

\tableofcontents
\hypersetup{
    linkcolor=blue,
    }

\mainmatter

\chapter{Introduction}\label{chap_Intro}

General relativity describes gravity as a feature of the geometry of spacetime: absent other forces, matter and radiation move on geodesics in a curved spacetime. This curvature is in turn caused by matter and radiation, a relationship summarised in Einstein's field equations \cite{Einstein,HawkingEllis,Carroll_2019}:
\begin{equation}
    R_{ab} - \frac{1}{2}g_{ab}R = 8\pi G_N T_{ab}\ , \label{H1_EFE}
\end{equation}
The terms on the left-hand side (collectively known as the Einstein tensor) depend only on the metric tensor $g_{ab}$, which describes the geometry of spacetime; the energy-momentum tensor $T_{ab}$ on the right-hand side describes the matter and radiation in this spacetime in terms of quantities like the energy density and the pressure. 

When solving \eqref{H1_EFE}, a typical approach is to postulate an energy-momentum tensor (or calculate $T_{ab}$ for a given matter distribution) and solve for $g_{ab}$ to find the geometry associated with the postulated matter configuration. However, the inverse is also possible: fix a metric and use \eqref{H1_EFE} to find the $T_{ab}$ needed to support it. General relativity thus allows, in principle, for any spacetime geometry, including those with exotic features such as wormholes, closed timelike curves, and faster-than-light signalling. The fact that we have not (yet) observed these phenomena in nature raises the question: are these exotic phenomena physically realistic but for some reason hard to find, or are they merely mathematical artefacts? 

A related question can be asked with regard to black holes. It is well-known that the simplest black hole geometry (Schwarzschild spacetime) contains a point where the curvature diverges; similarly, explicit calculations have shown that the spherically symmetric collapse of a sufficiently large amount of pressureless matter leads to a divergent density \cite{Oppenheimer_1939}. Again, one may ask whether these singularities are physically relevant or artefacts of the high degree of symmetry in these scenarios.

Energy conditions provide a way to answer these questions (see \cite{Kontou_review} for a recent review). They are attempts to summarise the properties of realistic matter, usually expressed as an inequality that the energy-momentum tensor of a realistic theory satisfies. Examples of the properties encoded by energy conditions are the notion that realistic matter has a positive energy density or, through \eqref{H1_EFE}, that it gravitationally focuses light-rays. Such a constraint is often sufficient to exclude exotic spacetimes \cite{Olum_1998,AANEC}, which are supported by unrealistic matter (e.g. wormholes usually require some negative energy, but note \cite{wormholes_ANEC}).

The role of energy conditions in the study of singularities is most evident in the various singularity theorems \cite{Senovilla_1998}, for example the one formulated by Penrose \cite{Penrose_1965}. To state his theorem, Penrose introduced the concept of a trapped surface: a $d-2$ dimensional spacelike surface in a $d$ dimensional spacetime such that all light-rays emitted orthogonally from it initially converge. If the matter in the spacetime is then unable to defocus null geodesics (a property which can be expressed as an energy condition), Penrose showed that the spacetime must contain null geodesics that encounter an edge of spacetime, i.e. they are geodesically incomplete. In the context of a gravitationally collapsing matter distribution, this incompleteness is usually interpreted as indicating the presence of a singularity \cite{HawkingEllis}.

These are all results of classical general relativity. However, real matter is not classical; it is described by quantum field theories (QFTs), which allow an arbitrarily negative expectation value $\mean{T_{ab}}_\psi$ at any single point in spacetime \cite{EGJ_NietPuntsgewijs}. This has spurred the development of averaged energy conditions, which usually consider some complete causal (i.e. null or timelike) geodesic, contract the energy-momentum tensor with its tangent vector, and integrate the expectation value of the result along the geodesic. The most prominent of these averaged energy conditions is the averaged null energy condition (ANEC), which states that for any state $\ket{\psi}$
\begin{align}
    \int_\gamma\mean{T_{ab}}_\psi l^a l^b\,\d\lambda \geq 0\ , \label{scr_ANEC}
\end{align}
where $\gamma$ is a complete null geodesic with tangent vector $l^a$ and affine parameter $\lambda$. The ANEC can be proven to hold for general QFTs in Minkowski spacetime, using quantum information techniques (specifically the monotonicity of relative entropy) \cite{ANEC_Entropy} or an argument based on causality \cite{ANEC_causaliteit}. However, these arguments no longer hold in curved spacetimes, and the ANEC can be violated in e.g. Schwarzschild spacetime \cite{ANECviolation_Schwarzschild} or conformally flat spacetimes \cite{ANEC_violation_old,ANEC_violation_new}.

This violation opens up the possibility that exotic spacetimes could be viable once quantum effects are taken into account, and that these same quantum effects might prevent gravitational collapse from forming a singularity. To study this possibility, one may attempt to find an energy condition for QFTs that continues to hold in curved spacetimes. Several proposals have been made to generalise \eqref{scr_ANEC} to curved spacetimes, the most well-known being the self-consistent, achronal ANEC (AANEC) \cite{AANEC}. The AANEC posits that \eqref{scr_ANEC} holds in a curved spacetime if $\gamma$ is a complete achronal null geodesic (i.e. $\lambda$ is unbounded, and $\gamma$ does not connect timelike separated points), and the spacetime is semi-classically self-consistent, obeying \cite{Rosenfeld_1963}
\begin{equation}
    R_{ab} - \frac{1}{2}g_{ab}R = 8\pi G_N\mean{T_{ab}}_\psi\ . \label{H1_SemiclassEFE}
\end{equation}
Another approach is to first formulate an energy condition for QFTs in fixed spacetimes, in which the quantum fields do not affect the geometry. An indication for the existence of such energy conditions is provided by quantum energy inequalities \cite{Fewster_2012}, which are constraints similar to \eqref{scr_ANEC} but with a negative (possibly state-dependent) lower bound, and can be proven to hold for specific QFTs on spacetimes with a fixed geometry. One can also attempt to make the ANEC more local, for example by introducing a weighting function (see e.g. \cite{QEI_2D,DSNEC_scalar,DSNEC_fermion} for examples in flat spacetime); an important instance of such an energy condition in curved spacetimes is the conformally invariant ANEC (CANEC) \cite{CANEC_odd,CANEC_even}:
\begin{align}
    \int_{\lambda_-}^{\lambda_+}\eta^d\mean{T_{ab}}_\psi l^al^b\d\lambda \geq B\ . \label{intro_CANEC}
\end{align}
For the CANEC, we consider a small bundle of null geodesics in a $d$ dimensional spacetime and integrate along one of them, with affine parameter $\lambda$ and tangent vector $l^a$. The weighting function $\eta$ is the Jacobi field, which measures the distance between neighbouring geodesics in the bundle and vanishes at $\lambda = \lambda_\pm$; the bound $B$ vanishes for odd $d$, but is non-trivial for even $d$ due to the conformal anomaly.

\section{Summary of results and overview}

The goal of this thesis is twofold. On the one hand, we provide support for the CANEC as a generalisation of the classical null energy condition to quantum fields in curved spacetimes by extending the context in which it applies. On the other hand, we will demonstrate that certain light-ray operators \cite{LO_original,LO_Mathys} only have non-negative expectation values for a non-minimally coupled real scalar field in flat spacetime. These two topics are connected by the observation that for even $d$, there is a light-ray operator which corresponds to the CANEC-operator on the Minkowski spacetime lightcone.

In light of these aims we review several results from the literature that motivate our interest in the null energy, in its average value, and finally in the CANEC itself. Furthermore, we do several novel computations, of which we would like to highlight three results.

\paragraph{Extended holographic proof.} The original proof of the CANEC \cite{CANEC_odd,CANEC_even} considered holographic CFTs (conformally invariant QFTs which, by the AdS/CFT correspondence, also describe semi-classical gravity) in a specific class of spacetimes. We demonstrate that this class can be somewhat extended in two ways. For $d=3$, we use Gaussian normal coordinates to show that the steps in the proof of \cite{CANEC_odd} are valid in a general spacetime, although they only lead to a non-trivial energy condition in spacetimes that satisfy a geometrical constraint. This constraint is violated by maximally symmetric and Ricci flat spacetimes, but we nevertheless find that holographic CFTs on (a subset of) these spacetimes obey an energy condition which either is the CANEC or closely related to it. For this proof, we follow the method of \cite{holographic_ANEC}, and our result in maximally symmetric spacetimes essentially supplements the results of \cite{holographic_ANEC} and \cite{Rosso_2020} with a proof in AdS spacetime.

\paragraph{Field theoretical proof in conformally flat spacetimes.} Since holographic CFTs are rather special, we also consider more general CFTs. Following up on \cite{Rosso_2020}, we show that for general CFTs, the Weyl transformation of the Minkowski spacetime ANEC can be interpreted as the CANEC in a conformally flat spacetime. To ensure that this Weyl transformation makes sense, we demand that the conformal factor is positive and finite everywhere except possibly at the edges of the flat and conformally flat spacetimes; it must also be positive and finite along a complete null geodesic in Minkowski spacetime. The CFT is constrained by the demand that the fields are smooth in both the flat and conformally flat spacetimes, which implies that it should be possible to impose Dirichlet boundary conditions on the CFT.

\paragraph{Positive light-ray operators.} The aforementioned proof of the CANEC for general CFTs relies on the fact that the Minkowski spacetime ANEC can be interpreted as the CANEC on a null plane. Inspired by the possibility of a CANEC on more general null congruences in Minkowski spacetime, we turn to a set of related objects: light-ray operators \eqref{LO_def} \cite{LO_original,LO_Mathys}. We study them for a non-minimally coupled real scalar field in Minkowski spacetime and find that the light-ray operators which satisfy \eqref{LO_NonMinConstraint} only have non-negative expectation values. Although this provides a large number of constraints on the theory, we argue that it does not generally imply that the CANEC holds on arbitrary null congruences in Minkowski spacetime.

\paragraph{} This thesis is organised as follows. Chapter \ref{chap_ClassEC} first motivates our interest in the null energy by discussing several classical energy conditions and proving Penrose's singularity theorem from the null energy condition. Then, chapter \ref{chap_ANEC} demonstrates the failure of classical energy conditions in QFT, sketches the proof of ANEC for QFTs in Minkowski spacetime from \cite{ANEC_Entropy}, and briefly describes the ANEC's failure in fixed curved spacetimes. This motivates a search for generalisations of the ANEC to curved spacetimes, two of which are discussed and, to some degree, proven in chapter \ref{chap_curveANEC}: the CANEC, following \cite{CANEC_odd}, and the AANEC, following \cite{Wall}. We expand on the former in chapter \ref{chap_BeyondCANEC}, which aims to extend the CANEC to background spacetimes and CFTs which are not covered by the proof in chapter \ref{chap_curveANEC}. Inspired by the discussion in chapter \ref{chap_BeyondCANEC}, chapter \ref{chap_LO} discusses light-ray operators for the non-minimally coupled but otherwise free real scalar field and demonstrates that many of them only have non-negative expectation values. Finally, we summarise our results and discuss possible future research in chapter \ref{chap_Conc}. The rest of this introduction establishes our notation and conventions; a brief introduction of relevant concepts from general relativity and QFT can be found in appendices \ref{app_GR} and \ref{app_QFT} respectively. 

\section{Notation and conventions}\label{sec_Intro_Conventions}
In this thesis, we will employ the abstract index notation from \cite{abstracte_indices}, in which Latin indices from the beginning of the alphabet indicate how many (co-)vectors a tensor can act upon; tensor operations are denoted in the way that is familiar from the usual index notation. For example, $R^a_{\ \,bcd}$ is a $(1,3)$-tensor; its components with respect to some explicit basis of (co-)vectors are $R^\mu_{\ \,\nu\rho\sigma}$, and acting on a vector $l^a$ with $R^a_{\ \,bcd}$ is denoted as e.g. $R^a_{\ \,bcd}l^b$. For a covector acting on a vector (and vice versa) we will also use a dot, e.g. $p\cdot x = p_ax^a$. The main benefit of abstract index notation is that it makes coordinate independence more manifest.

We define a spacetime as a connected $d$ dimensional manifold $M$ equipped with a smooth, symmetric metric tensor $g_{ab}$ of Lorentzian signature; we also demand that $M$ is time-orientable, so one can consistently assign a sense of `past' and `future' to any point in $M$. We use the mostly plus sign convention, so that the Lorentzian signature implies that the matrix of metric components $(g_{\mu\nu})$ has one negative eigenvalue. Due to this negative eigenvalue, the norm $g_{ab}X^aX^b$ of a vector $X^a$ can be positive, zero, or negative, which classifies $X^a$ as respectively spacelike, null, or timelike; null and timelike vectors will also be referred to as `causal', as explained in section \ref{sec_Causality}. Similarly, we classify a submanifold $N\subseteq M$ based on its normal vector: if $N$ has an everywhere timelike, spacelike, or null normal vector, $N$ is referred to as spacelike, timelike, or null respectively. If $N$ is $d-n$ dimensional, we say that it has codimension $n$; a codimension 1 submanifold is also known as a hypersurface.

For the covariant derivative, we make use of the Levi-Civita connection $\Gamma^\mu_{\rho\sigma}$. The Riemann tensor, which measures the local curvature, is defined as
\begin{align}
    [\nabla_c,\nabla_d]X^a \alis R^a_{\ \,bcd}X^b\ .
\end{align}
The components of $R^a_{\ \,bcd}$ with respect to the coordinate basis can be computed as
\begin{align}
    R^\mu_{\ \,\nu\rho\sigma} = \partial_\rho\Gamma^\mu_{\nu\sigma} - \partial_\sigma\Gamma^\mu_{\nu\rho} + \Gamma^\mu_{\rho\lambda}\Gamma^\lambda_{\nu\sigma} - \Gamma^\mu_{\sigma\lambda}\Gamma^\lambda_{\nu\rho} \ , \label{H1_RiemannComponenten}
\end{align}
and we define the Ricci tensor and scalar as $R_{ab} = R^c_{\ \,acb}$ and $R = g^{ab}R_{ab}$ respectively. Given an action $S_\mathrm{mat}$ for the fields describing matter, we define the energy-momentum tensor as
\begin{align}
    T_{ab} = -\frac{2}{\sqrt{-g}}\funci{S_\mathrm{mat}}{g^{ab}}\ ,
\end{align}
where $g = \mathrm{det}(g_{\mu\nu})$. Classical general relativity relates $T_{ab}$ to the geometry through Einstein's field equations \eqref{H1_EFE}; one could add a cosmological constant term $+\Lambda g_{ab}$ to the left-hand side of \eqref{H1_EFE}, but we absorb it into $T_{ab}$. Semi-classically, we replace $T_{ab}$ by its expectation value $\mean{T_{ab}}_\psi = \bra{\psi}T_{ab}\ket{\psi}$, where $\ket{\psi}$ is the state of the quantised matter fields, to obtain \eqref{H1_SemiclassEFE}.

In Minkowski spacetime, with metric $g_{ab} = \eta_{ab}$, we will often use Cartesian coordinates $x^\mu = (x^0,\ldots,x^{d-1}) \equiv (x^0,\ve{x})$, for which the line element is
\begin{align}
    \d s^2 = \eta_{\mu\nu}\d x^\mu \d x^\nu = -(\d x^0)^2 + (\d\ve{x})^2 = -(\d x^0)^2 + (\d x^1)^2 + \ldots + (\d x^{d-1})^2\ .
\end{align}
The description of a null geodesic can be simplified by introducing null coordinates $x^\pm = x^0 \pm x^1$; the remaining coordinates are $\ve{x}^\perp = (x^2,\ldots,x^{d-1})$. The line element then becomes
\begin{align}
    \d s^2 = -\frac{1}{2}\roha{\d x^+\d x^- + \d x^-\d x^+} + (\d\ve{x}^\perp)^2\ .
\end{align}
Null indices are lowered as $x_\pm = -\frac{1}{2}x^\mp$, and from the product $\eta_{\mu\nu}p^\mu x^\nu$, the null components of an arbitrary vector $p^a$ are found to be $p^\pm = p^0 \pm p^{d-1}$. In Minkowski spacetime we usually consider null geodesics with $x^+ = \lambda$ and constant $x^-$ and $\ve{x}^\perp$, where $\lambda$ is the affine parameter.

\chapter{Energy conditions in classical gravity}\label{chap_ClassEC}

In this chapter, we review some energy conditions in the context of classical general relativity in order to motivate our later focus on the averaged null energy condition. To facilitate our discussion, section \ref{sec_Causality} introduces several concepts related to causality in general relativity based on the treatments of \cite{HawkingEllis} and \cite{Witten_causality}. In section \ref{sec_classEC}, we discuss four important energy conditions, before concluding this chapter with a brief review of one of their most important consequences, Penrose's singularity theorem, in section \ref{sec_Penrose}.

\section{Causality in general relativity}\label{sec_Causality}
The goal of much work in physics is to formulate and solve initial value problems: given initial data about a physical system, what will the state of the system be at some point in its future? The amount of initial data required to make this prediction is limited by the causal structure of spacetime, which this section introduces briefly based on \cite{HawkingEllis} and \cite{Witten_causality}.

In a spacetime $M$, the amount of information required to predict the future state of a system from initial data is restricted by the postulate of local causality: if $\U\subseteq M$ is a convex normal neighbourhood (i.e. any two points in \U\ are connected by a unique geodesic $\gamma\subset\U$), then a causal signal can only be sent between $p,q\in\U$ if there exists a curve between them with an everywhere either timelike or null tangent vector. Such a curve is called causal. It follows that events at $p\in M$ only affect events at $q\in M$ if $q$ can be reached from $p$ with a future-directed causal curve; after all, any curve connecting $p$ to $q$ can be covered with convex normal neighbourhoods and should be causal in each of these neighbourhoods to allow a signal to propagate along it. This leads us to define the causal future of a subset $W\subset M$ as follows:
\begin{definition}[Causal future]
    Let $M$ be a spacetime, and $W\subset M$ a subset. The causal future of $W$, denoted $J^+(W)$, is the set of points that can be reached from at least one point in $W$ by a future-directed causal curve; we let $W\subset J^+(W)$. The boundary of $J^+(W)$, denoted $\partial J^+(W)$, is formed by those $p\in J^+(W)$ for which no open region $\U\subset J^+(W)$ with $p\in\U$ exists.
\end{definition}
The causal past of $W$, denoted $J^-(W)$, is defined similarly by replacing `future' by `past'.

\subsection{Prompt curves and the boundary of the causal future}

Closely tied to the boundary of the causal future is the notion of promptness. A curve from a subset $W\subset M$ to any $p\in M$ is called prompt if it is future-directed and causal, and there are no future-directed causal curves from $W$ to any point in the interior of $J^-(p)$. Thus, if $p\in\partial J^+(W)$, all causal curves from $W$ to $p$ are prompt; if they weren't, then either $p\notin J^+(W)$, contradicting $p\in\partial J^+(W)$, or there would be an $r\in J^+(W)$ with $p$ in the interior of $J^+(r)$, implying the existence of an open region $\U\subset J^+(r)\subseteq J^+(W)$ containing $p$, also contradicting $p\in\partial J^+(W)$. Conversely, if a prompt curve from $W$ to $q\in M$ exists, then $q\in J^+(W)$ since prompt curves are causal and future-directed; furthermore, any open region $\U\subset M$ containing $q$ has a non-empty intersection with the interior of $J^-(q)$, implying that $\U\cap J^+(W)\neq\U$ since the curve from $W$ to $q$ is prompt. Hence, there is no open $\U\subset J^+(W)$ which contains $q$, so $q\in\partial J^+(W)$. In summary, $p\in \partial J^+(W)$ if and only if there is a prompt curve from $W$ to $p$.

\begin{wrapfigure}{r}{0.4\textwidth}
  \begin{center}
    \includegraphics[width=0.35\textwidth]{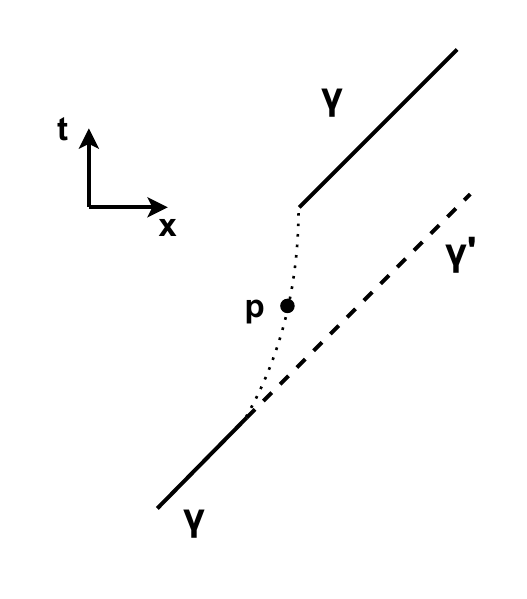}
  \end{center}
  \caption{A spacetime diagram, with time running vertically. A curve $\gamma$ starts out as a null geodesic (drawn as a solid line), has a timelike segment around some point $p$ (marked by a dotted line) and then continues as a null geodesic (represented with a solid line). The geometry near $p$ can be approximated by Minkowski spacetime, meaning that we can start a null geodesic $\gamma'$, drawn as a dashed line, from somewhere on the timelike segment of $\gamma$ and have it stay to the past of $\gamma$ everywhere.}
  \label{fig:Causaliteit}
\end{wrapfigure}
Besides the fact that they connect $W$ to points in $\partial J^+(W)$, prompt curves have other useful properties. For example, they are (segments of) null geodesics. That they are null curves follows from considering a causal curve $\gamma$ which is timelike near $p\in\gamma$. There exists a neighbourhood \U\ of $p$ where the metric is well-approximated by the Minkowski metric\footnote{Following \cite{Carroll_2019}, this can be seen in the Riemann normal coordinates introduced at the end of section \ref{sec_GeodCurv}. These are defined in a normal neighbourhood $\N_p$ of $p$ and allow the coordinates along any geodesic to be parametrised as $x^\mu(\lambda) = \lambda k^\mu$ for constant $k^\mu$. Through the geodesic equation \eqref{AppA_geodeet} with $V^\mu = X^\mu = k^\mu$, we find that $\Gamma^\rho_{\mu\nu} = 0$ at $p$, which by metric compatibility means that $\partial_\sigma g_{\mu\nu} = 0$ at $p$. Since we can choose the basis vectors such that $g_{\mu\nu} = \eta_{\mu\nu}$ at $p$, there must be some neighbourhood of $p$ that is well-described by Minkowski spacetime.}; in \U, it is evident that a null curve $\gamma'$ exists which coincides with $\gamma$ initially, deviates somewhere on the timelike section, and remains to the past of $\gamma$ afterwards, as sketched in figure \ref{fig:Causaliteit}. Hence, $\gamma$ was not prompt. That a prompt curve should be a geodesic follows by considering a null curve $\gamma$ which has a bend at $p\in\gamma$. By again looking at a neighbourhood \U\ of $p$ in which the metric is well-approximated by the Minkowski metric, we find that a null curve $\gamma'$ exists which initially coincides with $\gamma$, makes the same bend as $\gamma$ but earlier, and is afterwards to the past of $\gamma$, meaning that $\gamma$ was not prompt.

The above reasoning suggests another necessary property of a prompt curve starting at a set of points $W$: it cannot contain a caustic, defined as a point that can be reached from $W$ via two distinct null geodesics (if we have a null congruence emanating from $W$, a caustic can be recognised as a point where the Jacobi field $\eta$ vanishes). After all, if some $p\in M$ could be reached from $W$ via the null geodesics $\gamma_1$ and $\gamma_2$, a third null curve $\gamma_3$ could be constructed by combining the section of $\gamma_1$ between $W$ and $p$ with the section of $\gamma_2$ beyond $p$. $\gamma_3$ bends at $p$, so like before we can construct a null curve which is to the past of $\gamma_3$ beyond $p$; since $\gamma_3$ coincides with $\gamma_2$ here, we conclude that $\gamma_2$ is not prompt beyond $p$ either.

A third property of prompt curves is that a null geodesic $\gamma$ from a submanifold $W$ to a point $p$ can only be prompt if it intersects $W$ at a right angle. If it didn't, there would be another curve $\gamma'$ which comes closer to intersecting $W$ at a right angle and hence would need to traverse less spatial distance to reach $p$. $\gamma'$ could therefore be timelike, putting $p$ in the interior of $J^+(W)$, contradicting the promptness of $\gamma$. Also note that a null geodesic can only be perpendicular to submanifolds with a codimension of at least 2.

Finally, the fact that $\partial J^+(W)$ is spanned by prompt curves allows us to conclude that $\partial J^+(W)$ is an achronal hypersurface, i.e. a hypersurface that does not contain timelike separated points. First, consider that any prompt curve must be achronal. After all, if a null geodesic $\gamma$ is not achronal, there exist timelike separated $p,q\in\gamma$ and we can construct a new curve $\gamma'$ by replacing $\gamma_{pq}$ (the part of $\gamma$ connecting $p$ and $q$) with a timelike curve. Since $\gamma'$ is not a null geodesic, it is not prompt, and because $\gamma'$ coincides with $\gamma$ outside of $\gamma_{pq}$, we conclude that $\gamma$ is not prompt if it is not achronal. Because $\partial J^+(W)$ is spanned by prompt curves, this implies that $\partial J^+(W)$ is achronal.

We prove that $\partial J^+(W)$ is also a hypersurface by choosing any $p\in\partial J^+(W)$ and considering the convex normal neighbourhood $\N_p$ of $p$. In $\N_p$, we choose coordinates $x^\mu$ (with $x^i$ for $i\in\cuha{1,\ldots,d-1}$ the spacelike coordinates) such that the curves $\gamma(\ve{a}) = \cuha{q\in\N_p:x^i(q) = a^i}$ intersect the interiors of both $J^+(p)$ and $J^-(p)$. Because the interior of $J^+(p)$ is contained in $J^+(W)$ and that of $J^-(p)$ is not, $\gamma(\ve{a})$ intersects $\partial J^+(W)$ and does so exactly once since $\partial J^+(W)$ is achronal. The point $r$ where this happens is uniquely labelled by $\ve{a}$, so the map $\phi_p:\N_p\cap\partial J^+(W) \rightarrow \mathbb{R}^{d-1}$ with $\phi_p(r) = \ve{a}$ is injective. The combination $(\phi_p,\N_p)$ therefore defines a chart, which can be extended to an atlas of $\partial J^+(W)$ by repeating the above process at sufficiently many $p\in\partial J^+(W)$. If $M$ has no edge, $\partial J^+(W)$ has no edge either (if $\partial J^+(W)$ had an edge $\Gamma$, $\partial J^+(W)$ could be extended with the relevant part of $\partial J^+(\Gamma)$), so we conclude that $\partial J^+(W)$ is indeed an achronal codimension 1 submanifold or hypersurface.

\subsection{Initial value surfaces and globally hyperbolic spacetimes}

Having studied the boundary of the causal future in some detail, we now return to the relation between the causal structure of spacetime and initial value problems. Let us assume that we specify initial data for some theory on a subset $\Sigma\subset M$. If we can then predict what happens at $p\in J^+(\Sigma)$ based on this data, there should be no past-inextendible causal curves (causal curves that either stretch into the asymptotic past or originate at the past edge of spacetime) through $p$ that did not pass $\Sigma$ at some point, since these would carry influences that we cannot predict based on the data on $\Sigma$. This leads us to define the domain of dependence:
\begin{definition}[Domain of dependence]
    Let $M$ be a spacetime. The domain of dependence of a subset $\Sigma\subset M$, denoted $D(\Sigma)$, is the set of all points $p\in M$ such that every inextendible causal curve through $p$ intersects with $\Sigma$. $D(\Sigma)$ can be divided into the future domain of dependence $D^+(\Sigma) = D(\Sigma)\cap J^+(\Sigma)$ and the past domain of dependence $D^-(\Sigma) = D(\Sigma)\cap J^-(\Sigma)$. The edge of $D(\Sigma)$, denoted $\partial D(\Sigma)$, is called the Cauchy horizon of $\Sigma$; it too can be divided into a past and future Cauchy horizon, respectively $\partial D^-(\Sigma) = \partial D(\Sigma)\cap J^-(\Sigma)$ and $\partial D^+(\Sigma) = \partial D(\Sigma)\cap J^+(\Sigma)$.
\end{definition}
Note that by this definition, $\Sigma\subset D(\Sigma)$. Any $\Sigma$ has a domain of dependence, but formulating an initial value problem calls for a relativistic analogue to the Newtonian notion of an instant of time at which one defines the initial data. We will thus always define initial data on a $\Sigma$ that is a spacelike, achronal hypersurface; such a $\Sigma$ will be called an initial value hypersurface. The rationale behind defining initial data only on spacelike achronal hypersurfaces is that they do not contain points that can be connected by a causal curve, guaranteeing that the initial data cannot influence itself.

If one is interested in the global properties of spacetime, it becomes relevant to consider whether all physical processes in a spacetime could be formulated as an initial value problem; that is, whether there exists an initial value hypersurface $\Sigma$ such that $D(\Sigma) = M$. As both \cite{HawkingEllis} and \cite{Witten_causality} point out, there are no obvious grounds to assume this is the case in the actual universe and there are known solutions to general relativity that do not allow for such an initial value hypersurface (most notably the anti-de Sitter and Reissner-Nordström solutions); however, treating physical calculations as initial value problems has historically worked quite well, and we have no indications yet that the universe as a whole will deviate from that trend. We therefore assume that the universe is globally hyperbolic, which we define as follows:
\begin{definition}[Cauchy surface and global hyperbolicity]
    Let $M$ be a spacetime, and $\Sigma\subset M$ a spacelike, achronal hypersurface. If $D(\Sigma) = M$, $\Sigma$ is called a Cauchy surface, and a spacetime which contains a Cauchy surface is called globally hyperbolic.
\end{definition}
A useful fact about Cauchy surfaces is that any achronal hypersurface $S$ (which may be another Cauchy surface) can be mapped onto them injectively, because one can construct a future-directed timelike vector field (the following construction comes from \cite{Witten_causality}). This can be done by choosing a coordinate system for an open region containing any $p\in M$ and finding an eigenvector $V^a$ of $\roha{g_{\mu\nu}}$ at $p$ with a negative eigenvalue. $V^a$ is timelike, and hence can be chosen to be future-directed and normalised as $V_aV^a = -1$. Repeating this process at all points in $M$ extends $V^a$ to a future-directed timelike vector field, which can act as the tangent vector field for a set of inextendible curves known as the integral curves of $V^a$. These curves are everywhere timelike and hence intersect any achronal hypersurface at most once, and a Cauchy surface exactly once. Thus, given an achronal hypersurface $S$ and a Cauchy surface $\Sigma$, we can define a map $\varphi:S\rightarrow\Sigma$ which sends $p\in S$ to the point $q\in\Sigma$ where the integral curve through $p$ intersects $\Sigma$. $\varphi$ is injective, since every $p\in S$ lies on exactly one integral curve (the integral curves cannot intersect, since $V^a$ assigns exactly one vector to every point) but not all integral curves must intersect with $S$.

\section{Classical energy conditions and their interpretation}\label{sec_classEC}

As described in chapter \ref{chap_Intro}, energy conditions are often conceived of as constraints on the energy-momentum tensor $T_{ab}$ of a gravitating system, imposed to rule out geometries with pathological features. The simplest way to formulate such a constraint is as a pointwise inequality, providing a bound on some component of $T_{ab}$ at every point in spacetime; because we are usually interested in gravitating systems, we also impose \eqref{H1_EFE}, which translates a bound on $T_{ab}$ to a bound on $R_{ab}$, $R$, and $g_{ab}$. Four commonly used energy conditions are:
\begin{enumerate}
    \item The weak energy condition (WEC), which states that for any timelike vector $t^a$
    \begin{align}
        T_{ab}t^at^b \geq 0\ . \label{classEC_WEC}
    \end{align}
    The equivalent constraint on geometric quantities is straightforward to read off from \eqref{H1_EFE}, but due to difficulties in interpreting it \cite{Kontou_review}, it is not often used.
    \item The strong energy condition (SEC), according to which for any timelike $t^a$
    \begin{align}
        \roha{T_{ab} - \frac{T^c_{\ \,c}}{d-2}g_{ab}}t^at^b \geq 0\ . \label{classEC_SEC}
    \end{align}
    For the SEC, we will see that it is actually the geometric constraint $R_{ab}t^at^b\geq0$ which allows a more straightforward interpretation.
    \item The dominant energy condition (DEC) requires that, for any future-directed, timelike $t^a$, $-T^{ab}t_a$ is future-directed and causal (i.e. timelike or null). In turn, this means that for any two future-directed timelike vectors $t^a$ and $p^a$,
    \begin{align}
        T_{ab}t^ap^b \geq 0\ . \label{classEC_DEC}
    \end{align}
    Similarly to the WEC, the geometric bound this implies through \eqref{H1_EFE} is straightforward to read off, but difficult to interpret and hence rarely used.
    \item The null energy condition (NEC) claims that for any null vector $l^a$ we must have
    \begin{align}
        T_{ab}l^al^b\geq0\ . \label{classEC_NEC}
    \end{align}
    As with the SEC, the geometric version $R_{ab}l^al^b\geq 0$ will turn out to be easier to interpret.
\end{enumerate}
Before interpreting these energy conditions, we use a continuity argument from \cite{Kontou_review} to demonstrate that the NEC is implied by the other three conditions. We begin by observing that the WEC, SEC, and DEC can be written as $S_{ab}p^aq^b\geq0$ for co-oriented timelike $p^a$ and $q^a$ and a symmetric tensor $S_{ab}$. Let us consider a weakened statement, namely $S_{ab}p^aq^b\geq B$ for some fixed $B\inR$ and all $p^a$ and $q^a$ with $p_ap^a = q_aq^a = -1$ (without the restriction to unit vectors we could arbitrarily rescale $p^a$ and $q^a$, so that the bound would only hold for the rescaled vectors if $B\geq0$, which doesn't weaken the original statement). We then choose two null vectors $l^a$ and $k^a$ with $l_ak^a = -\frac{1}{2}$, and construct two timelike unit vectors from them:
\begin{align}
    p^a = \frac{1}{r}l^a + rk^a& &\text{and}& &q^a = \frac{1}{s}l^a + sk^a\ ,
\end{align}
for any $r,s>0$. With these choices of $p^a$ and $q^a$, the bound on $S_{ab}$ implies
\begin{align}
    S_{ab}p^aq^b = \frac{1}{rs}S_{ab}l^al^b + \roha{\frac{s}{r}l^ak^b + \frac{r}{s}k^al^b}S_{ab} + rsS_{ab}k^ak^b \geq&\ B \non
    S_{ab}l^al^b + \roha{s^2l^ak^b + r^2k^al^b}S_{ab} + r^2s^2S_{ab}k^ak^b \geq&\ rsB\ .
\end{align}
From the limit $r,s\downarrow0$ we see that $S_{ab}p^aq^b\geq B$ implies $S_{ab}l^al^b\geq0$ for any null vector $l^a$. For the WEC, SEC, and DEC we see that $S_{ab}l^al^b = T_{ab}l^al^b$, so all three conditions imply the NEC. The inverse obviously does not hold.

\subsection{Interpreting the classical conditions}

Let us now give a physical interpretation of the four energy conditions. For the WEC and DEC, this is relatively straightforward, as we can interpret them directly in terms of the energy-momentum tensor. The WEC simply states that any observer, travelling on a timelike curve with tangent vector $t^a$, must measure a positive energy density \cite{Kontou_review}. Note however \cite{Curiel2017}, which points out that negative energies arise quite naturally in the ergosphere of Kerr black holes \cite{Carroll_2019}. However, these energies are defined through the Killing vectors of the Kerr metric, and hence are not the energies any observer would naturally measure; nevertheless, they complicate the simplest interpretation of the WEC.

The statement of the DEC is that an observer on a timelike curve with tangent vector $t^a$ must always observe a causal flux $-T^{ab}t_a$ of energy-momentum \cite{Kontou_review}. This is sometimes said to imply that energy-momentum can only propagate causally (e.g. in \cite{HawkingEllis,Senovilla_1998,QEI_2D}), supported by a theorem which states that if a covariantly conserved $T_{ab}$ obeying the DEC vanishes on a closed, achronal set, then it must vanish on the future domain of dependence of this set \cite{HawkingEllis}. However, as shown in \cite{Earman_2014}, the DEC cannot be necessary or sufficient to rule out acausal propagation of energy-momentum, as there are DEC-violating theories which have a well-defined initial value formulation (limiting their propagation speed to the speed of light) and DEC-satisfying theories which allow signals to propagate superluminally.

In order to interpret the SEC and NEC, the Raychaudhuri equation \eqref{AppA_Raychaudhuri2}, reproduced here for convenience, will prove to be of considerable help:
\begin{align}
    \frac{d-n}{\eta}\diff{^2\eta}{\lambda^2} = - \sigma_{ab}\sigma^{ab} + \omega_{ab}\omega^{ab} - R_{ab}l^al^b\ . \label{classEC_Raychaudhuri}
\end{align}
The Raychaudhuri equation describes, roughly speaking, how the distance between a geodesic with tangent vector $l^a$ and affine parameter $\lambda$ and its neighbours in a congruence evolves. This distance is characterised by the Jacobi field $\eta$, while $\sigma_{ab}$ and $\omega_{ab}$ are respectively the shear and vorticity of the congruence; finally, $n$ is the codimension of the congruence, i.e. $n=1$ for time- and spacelike congruences and $n=2$ for null congruences. While the shear and vorticity are primarily properties of the congruence, $R_{ab}l^al^b$ depends on the curvature of spacetime and is directly constrained by the SEC and NEC.

The SEC posits that $R_{ab}t^at^b\geq0$ for any timelike $t^a$. By \eqref{classEC_Raychaudhuri}, this implies that spacetime curvature will draw timelike geodesics together (or at least not push them apart). However, as \cite{Curiel2017} remarks, $\eta$ is an average of the separation between geodesics that are neighbouring in various directions; the fact that spacetime curvature negatively accelerates $\eta$ does not imply that any two bodies moving on timelike geodesics cannot be accelerated away from each other, only that on average they will not. The formulation of the SEC in terms of the energy-momentum tensor is less straightforward to interpret: the term constrained by \eqref{classEC_SEC} has been interpreted as an effective energy density \cite{EED_Kontou}, but it is unclear why it should be constrained.

Finally, the interpretation of the NEC follows a similar line of reasoning: since the NEC states that $R_{ab}l^al^b\geq0$ for all null $l^a$, \eqref{classEC_Raychaudhuri} implies that spacetime curvature will not push null geodesics apart. Note that this should not be seen as a claim about gravity being attractive, since in a theory which satisfies the NEC but violates the SEC (for example due to the presence of a cosmological constant), gravity could cause timelike geodesics to accelerate away from each other, while causing null geodesics to focus. The formulation in terms of the energy-momentum tensor is, once again, difficult to interpret; analogously to the WEC, we may consider the NEC to be the statement that observers traversing null geodesics must detect positive energy densities, but it is far from obvious what is meant by an observer traversing a null geodesic. 

Despite the difficulties in interpreting the meaning of the NEC, the rest of this thesis will be concerned with its implications and generalisations, for two reasons. Firstly, we have argued that the NEC is the weakest of the four energy conditions discussed here, so a violation of the NEC immediately implies a violation of all the classical energy conditions. Secondly, it has been argued that violations of the NEC lead to instabilities in the theory \cite{classNEC_stab1,classNEC_stab2,classNEC_stab3} and to violations of e.g. Hawking's area theorem \cite{HawkingA} and the generalised second law \cite{classGSL}; constructing a sensible classical theory thus seems to require satisfying the NEC. With this in mind, we explore one particular consequence of the NEC: Penrose's theorem.

\section{Penrose's singularity theorem}\label{sec_Penrose}
Fundamentally, Penrose's theorem is not about singularities, for which a universally accepted definition is lacking \cite{Senovilla_1998}; rather, it is a statement about geodesic completeness. Beyond the causality-related concepts introduced in section \ref{sec_Causality}, the statement and proof of this theorem requires the notion of a trapped surface, which we introduce first.

\subsection{Trapped surfaces}

In a spacetime $M$, consider a codimension 2 spacelike submanifold \T. Since \T\ has codimension 2, it has two normal vectors, one of which is timelike (since \T\ is spacelike); since $g_{ab}$ has Lorentzian signature, the other one is spacelike. If \T\ is orientable, we can choose a consistent orientation for this spacelike normal vector and define two future-directed null congruences that emanate orthogonally from \T, propagating either along or opposite to the spacelike normal vector. For example, if \T\ is a $d-2$ dimensional sphere, these congruences would initially be directed away from the centre of the sphere and towards it.

A useful fact about these congruences (which is proven at the end of section \ref{sec_CongRay}, or in more detail in \cite{Witten_causality}) is that since they are orthogonal to a codimension 2 spacelike submanifold, $\omega_{ab} = 0$ along each of their null geodesics, at least until the geodesic encounters a caustic (a point where $\eta\rightarrow0$). If we also assume that the NEC holds, the Raychaudhuri equation \eqref{classEC_Raychaudhuri} implies that along every geodesic from these two null congruences
\begin{align}
    \diff{^2\eta}{\lambda^2} = -\frac{\eta}{d-2}\roha{\sigma_{ab}\sigma^{ab} + R_{ab}l^al^b} \leq 0\ , \label{classEC_Penrose1}
\end{align}
where the inequality can be imposed because $\eta\geq0$ by definition, $\sigma_{ab}\sigma^{ab}\geq0$ as argued in section \ref{sec_CongRay}, and $R_{ab}l^al^b\geq0$ as a consequence of the NEC. We can integrate both sides of the inequality in \eqref{classEC_Penrose1} twice to find that the Jacobi field can only grow if it was growing at \T:
\begin{align}
    \eta(\lambda) \leq \diff{\eta}{\lambda}\biggr\rvert_\T\,\lambda + \eta(0)\ , \label{classEC_JacobiBound}
\end{align}
where we assumed that $\lambda=0$ at \T. Hence, if $\eta$ is initially decreasing on \T, the geodesic must encounter a caustic at a finite affine parameter $\lambda_0$, with 
\begin{align}
    \lambda_0 \leq -\eta(0)\roha{\diff{\eta}{\lambda}\biggr\rvert_\T}^{-1}\ . \label{classEC_CausticAffine}
\end{align}
The observation that a future-directed congruence emanating orthogonally from \T\ cannot defocus if it was initially focused, leads one to define a trapped surface \cite{Penrose_1965,HawkingEllis}:
\begin{definition}[Trapped surfaces]
    Let $M$ be a spacetime, and $\T\subset M$ a compact codimension 2 spacelike submanifold. \T\ is called a trapped surface if both future-directed null congruences emanating orthogonally from \T\ have an initially decreasing Jacobi field.
\end{definition}
A physical interpretation of a trapped surface is that its presence signals the existence of a region of spacetime where gravity is so strong that even outgoing null geodesics are focused. 

\subsection{Proving the theorem}

Penrose's theorem is now concerned with the possibility of a trapped surface in a spacetime which is geodesically complete, obeys any of the energy conditions from section \ref{sec_classEC} and hence obeys the NEC, and is globally hyperbolic with a non-compact Cauchy surface $\Sigma$. Physically, these assumptions imply that there is no edge of spacetime, that matter focuses null geodesics, and that processes throughout spacetime can be predicted based on initial data on $\Sigma$, respectively. The exact statement of Penrose's theorem is \cite{Penrose_1965,HawkingEllis,Witten_causality}:
\begin{theorem}[Penrose]\label{thm_Penrose}
    Let $M$ be a globally hyperbolic spacetime with non-compact Cauchy surface $\Sigma$. If $M$ satisfies the NEC and the field equations of general relativity (so that $R_{ab}l^al^b\geq0$ for all null vectors $l^a$) and contains a trapped surface $\T$, then $M$ cannot be geodesically complete.
\end{theorem}
To prove this, we first assume that $M$ is geodesically complete. Recall now that $\partial J^+(\T)$ is spanned by prompt curves, which we argued are the segments of null geodesics that emanate orthogonally from \T\ and do not contain caustics. Since these geodesics emanate orthogonally from \T, they are subject to \eqref{classEC_JacobiBound}, and because \T\ is a trapped surface, every one of them must encounter a caustic at some finite affine parameter $\lambda_0$, bounded by \eqref{classEC_CausticAffine} if we set $\lambda=0$ at \T; the fact that we can extend the geodesics forming $\partial J^+(\T)$ sufficiently far is guaranteed by our assumption of geodesic completeness.

Since only the part of these geodesics prior to their caustic is prompt, $\partial J^+(\T)$ must be compact. After all, any $p\in\partial J^+(\T)$ can be specified with the spatial orientation of the null geodesic it is on (i.e. whether $p$ is on a geodesic which propagates along or opposite to the spacelike normal vector of \T), the point $q\in\T$ at which this geodesic intersects \T, and the affine parameter $\lambda_p$ assigned to $p$. Since only the part of the geodesic prior to the caustic at $\lambda_0$ is part of $\partial J^+(\T)$, $\lambda_p$ is restricted to a compact interval $0\leq\lambda_p<\lambda_0$, and since $q$ lies on a compact submanifold, $\partial J^+(\T)$ is compact.

We now recall that $\partial J^+(\T)$ is an achronal hypersurface. Then, because $M$ is globally hyperbolic with Cauchy surface $\Sigma$, there exists an injective map $\varphi:\partial J^+(\T)\rightarrow\Sigma$; because $\partial J^+(\T)$ is compact, $\varphi\roha{\partial J^+(\T)}$ is compact as well. Since $\Sigma$ is not compact, this compactness implies that $\varphi\roha{\partial J^+(\T)}$ has a boundary in $\Sigma$. On the other hand, $\partial J^+(\T)$ is a submanifold, and therefore does not have a boundary that can be mapped to a boundary of $\varphi\roha{\partial J^+(\T)}$ in $\Sigma$. This contradiction means that our starting assumption that $M$ is geodesically complete must have been wrong, completing the proof.

\subsection{Closing remarks}

We conclude with two remarks regarding theorem \ref{thm_Penrose}. First, as has been pointed out by many authors \cite{HawkingEllis,Senovilla_1998,Penrose_1965,Witten_causality}, the argument given here actually proves that the assumptions that a spacetime has a non-compact Cauchy surface, satisfies the NEC and \eqref{H1_EFE}, contains a trapped surface, and is geodesically complete are incompatible. It does not indicate which assumption fails, and it certainly does not provide sufficient grounds to conclude that a singularity appears, a point which may be illuminated further by looking at de Sitter spacetime \cite{Senovilla_1998,Witten_causality}. This is an exact solution of \eqref{H1_EFE} for $T_{ab} = -\Lambda g_{ab}/(8\pi G_N)$ with $\Lambda\geq0$, which saturates the NEC; assuming $d\geq3$, one can define global coordinates $x^\mu$ and write the metric as \cite{Carroll_2019}
\begin{align}
    \d s^2 = g_{\mu\nu}\d x^\mu \d x^\nu = -\d t^2 + \alpha^2\cosh^2(t/\alpha)\viha{\d\chi^2 + \sin^2\chi\,\d\Omega_{d-2}^2}\ ,
\end{align}
where $\d\Omega_{d-2}^2$ is the metric of the $d-2$ sphere, $\alpha$ is a constant length scale, and $0\leq\chi\leq\pi$. De Sitter spacetime is globally hyperbolic and geodesically complete, and contains trapped surfaces (e.g. the spacelike $d-2$ spheres at $t<0$); by theorem \ref{thm_Penrose}, its Cauchy surface must then be compact, which it is: e.g. $t=0$ defines a Cauchy surface with the topology of a sphere. In a different set of coordinates $\hat{x}^\mu$, defined on a part of de Sitter spacetime referred to as the (time-reversed) planar patch or flat slicing, the metric can be represented as \cite{HawkingEllis}
\begin{align}
    \d s^2 = g_{\mu\nu}\d \hat{x}^\mu \d \hat{x}^\nu = -\d\hat{t}^2 + e^{-2\hat{t}/\alpha}\d\hat{\ve{x}}^2\ .
\end{align}
This describes a spacetime which has a non-compact Cauchy surface (e.g. defined by $\hat{t} = 0$) and contains trapped surfaces (e.g. any spacelike $d-2$ sphere). It must then be geodesically incomplete by theorem \ref{thm_Penrose}, and indeed: once the time-reversed planar patch is embedded in the global de Sitter spacetime, null geodesics can leave it, entering the rest of the global de Sitter spacetime. Note that the geodesic incompleteness predicted by theorem \ref{thm_Penrose} does not imply the appearance of a singularity in this case, but rather of a Cauchy horizon.

Our second and final remark is that the NEC was crucial to the proof of theorem \ref{thm_Penrose}: without it, the argument for the compactness of $\partial J^+(\T)$ as given here would not have held and we would not have arrived at a contradiction. However, the next chapter will show that quantum fields, which describe the matter fields we observe, routinely violate the NEC. Whether the conclusion of Penrose's theorem must be relinquished in a semi-classical context therefore depends on two things: whether any energy condition can be formulated for quantum fields, and if it can be, whether such an energy condition is sufficient to prove a form of Penrose's theorem. Although there has been progress in reformulating Penrose's theorem for weakened energy conditions \cite{weak_thms1,weak_thms2,weak_thms3}, this is beyond the scope of this thesis, and the following chapters will focus on identifying possible energy conditions for quantum fields.

\chapter{The averaged null energy condition}\label{chap_ANEC}

Chapter \ref{chap_ClassEC} introduced energy conditions as pointwise restrictions on a contraction of the energy-momentum tensor, and demonstrated their profound implications for the global properties of spacetime by proving Penrose's singularity theorem from the null energy condition (NEC). In turn, the NEC was argued to be a sensible constraint due to its close relation to the matter theory's stability \cite{classNEC_stab1,classNEC_stab2,classNEC_stab3} and consistency with black hole thermodynamics \cite{HawkingA,classGSL}. However, these are results of classical field theory. A more realistic description of matter using quantum field theory (QFT) permits negative energy densities, e.g. in the Casimir effect \cite{CasimirEffect}, which makes it necessary to reconsider the validity of the NEC if the results of the classical theory are to be maintained in a semi-classical context.

As a first step towards an energy condition for semi-classically self-consistent theories, we will consider QFTs on fixed spacetimes, i.e. we specify the geometry of spacetime without imposing \eqref{H1_SemiclassEFE}. Section \ref{sec_FlatSpacetime} considers the simple case of a QFT in Minkowski spacetime. In this setting, it is shown that the NEC cannot hold as a statement about the expectation value of the energy-momentum tensor (i.e. we cannot demand that $\mean{T_{ab}}_\psi l^al^b\geq0$ for all null vectors $l^a$ and states $\ket{\psi}$), after which we prove that violations of the NEC cannot be arbitrarily large and long-lasting by arguing that the NEC must hold on average along a complete null geodesic. We consider this averaged NEC (ANEC) in some more detail in section \ref{sec_CurvedSpacetimes}, where several examples demonstrate that the ANEC cannot hold for QFTs on curved spacetimes.

In this chapter, we consider a QFT with Hilbert space \calH\ and energy-momentum tensor $T_{ab}$. We denote the expectation value of this operator by $\bra{\psi}T_{ab}\ket{\psi} = \mean{T_{ab}}_\psi$ for any $\ket{\psi}\in\calH$.

\section{Energy conditions in flat spacetime}\label{sec_FlatSpacetime}
A straightforward way to implement the energy conditions \eqref{classEC_WEC}-\eqref{classEC_NEC} in a QFT is to demand that $\mean{T}_\psi \geq 0$ for any $\ket{\psi}\in\calH$, where $T$ is the relevant contraction of $T_{ab}$ (e.g. the NEC for a QFT would have $T = T_{ab}l^al^b$ for any null vector $l^a$). In this section, we consider a QFT in a Minkowski spacetime $M$ and show that it necessarily violates a bound like $\mean{T}_\psi\geq0$ if $T$ is a local operator \cite{EGJ_NietPuntsgewijs}. It is then proven that the average of $\mean{T_{ab}}_\psi l^al^b$ along a null geodesic with tangent vector $l^a$ will be non-negative \cite{ANEC_Entropy}.

\subsection{Pointwise energy conditions in QFT}

To prove that energy conditions cannot be formulated as $\mean{T}_\psi\geq0$ for a local operator $T$, we require the Reeh-Schlieder theorem \cite{Kontou_review,Haag_1996,Witten_ReehSchlieder} (\cite{Reeh_Schlieder_1961} is the German original):
\begin{theorem}[Reeh-Schlieder]
    Consider a QFT defined on a Minkowski spacetime $M$, with Hilbert space \calH\ and Minkowski vacuum $\ket{0}$. Let $\calH_0\subseteq\calH$ be the set of states that can be created by acting on $\ket{0}$ with a local field operator, and let $\A(\U)$ be the set of local operators of the QFT localised in an open region $\U\subset M$. Then the set $\A(\U)\ket{0}$ is dense in $\calH_0$, i.e. any $\ket{\psi}\in\calH_0$ can be approximated to arbitrary accuracy as $\hat{a}\ket{0}$ for $\hat{a}\in\A(\U)$.
\end{theorem}
An important corollary follows by considering two spacelike separated regions $\U,\U'\subset M$. Due to the spacelike separation, all $\hat{a}\in\A(\U)$ must (anti-)commute with all $\hat{a}'\in\A(\U')$ to preserve causality. Hence, if $\hat{a}\ket{0} = 0$, then $\hat{a}\hat{a}'\ket{0} = 0$ for all $\hat{a}'\in\A(\U')$. However, since $\A(\U')\ket{0}$ is dense in $\calH_0$, this implies that $\hat{a}$ annihilates all $\ket{\psi}\in\calH_0$, which in turn means that $\hat{a}$ must be identically zero. 

By combining this corollary with the Reeh-Schlieder theorem itself, the following theorem is immediate \cite{Kontou_review,EGJ_NietPuntsgewijs}:
\begin{theorem}[No pointwise constraints]\label{thm_NoPointwise}
    Let $T\in\A(\U)$ be a self-adjoint local operator with $\mean{T}_\psi\geq0$ for all $\ket{\psi}\in\calH_0$. If $T$ is renormalised such that $\mean{T}_0 = 0$, then $T = 0$.
\end{theorem}
The proof is as follows. Since $T$ is a non-negative operator, we can formally write the vacuum expectation value of $T$ as a norm: $\mean{T}_0 = \abs{T^{1/2}\ket{0}}^2 = 0$. This implies that $T^{1/2}$ annihilates $\ket{0}$, which by the Reeh-Schlieder theorem means that $T^{1/2} = 0$. We therefore conclude that $T=0$ as well, as claimed.

This clearly implies that we cannot impose the energy conditions \eqref{classEC_WEC}-\eqref{classEC_NEC} as constraints on the pointwise expectation value of the energy-momentum tensor. This claim can be sharpened slightly. After all, a realistic measurement of the energy density will always have a finite resolution, and thus it makes sense study weighted averages of $T$, defined as
\begin{align}
    T(f) = \int\d^dx\,f(x)T(x)\ ,
\end{align}
where we take $f:M\rightarrow\mathbb{R}$ to be smooth, positive, and supported only on a compact region $\U$ (i.e. $f$ is a test function). Since we still have $T(f)\in\A(\U)$, theorem \ref{thm_NoPointwise} applies and we conclude that the energy conditions from section \ref{sec_classEC} also cannot hold as constraints on expectation values of $T_{ab}$ averaged over some compact region.

\subsection{An information-theoretical proof of the ANEC}

A loophole presents itself at this point: if we average $T_{ab}$ and its contractions over a non-compact region, the expectation value of this operator may still be bounded. An example of such a bound is the ANEC, which is the statement that for a QFT in Minkowski spacetime
\begin{align}
    \int_\gamma\d\lambda\, \mean{T_{ab}}_\psi l^al^b \geq0\ . \label{thANEC_ANEC}
\end{align}
Here, $\gamma\subset M$ is an inextendible null geodesic with tangent vector $l^a$ and affine parameter $\lambda$. In order to prove the ANEC quite generally, \cite{ANEC_Entropy} uses the monotonicity of the relative entropy. In this subsection, we will briefly go over the information-theoretical part of their argument, which was also treated in \cite{BlancoCasini_2013}.

The first step is to split the Hilbert space. To that end, consider a Cauchy surface $\Sigma\subset M$; because $\Sigma$ is a Cauchy surface, any state in \calH\ can be reconstructed from the state of the fields on $\Sigma$. We denote the set of possible states on $\Sigma$ by $\calH_\Sigma$. This set can be further split up by dividing $\Sigma$ into two regions, $A$ and $A^c$, such that $A\cup A^c=\Sigma$ and $A\cap A^c = \emptyset$; if the QFT is local, any state on $\Sigma$ can be split into a state on $A$ and a state on $A^c$, since these regions are spacelike separated. Hence, $\calH_\Sigma$ factorises as $\calH_\Sigma = \calH_A\otimes\calH_{A^c}$.

As the next step, we recall from section \ref{sec_DensityMatrices} that a density matrix can be assigned to the field configuration on $\Sigma$ as
\begin{align}
    \rho^\psi = \sum_\alpha c^\psi_\alpha\smallket{\psi_\alpha}\smallbra{\psi_\alpha}\ ,
\end{align}
where $\ket{\psi_\alpha}\in\calH_\Sigma$ are unit vectors, and the $c^\psi_\alpha$ are restricted by $0\leq c^\psi_\alpha\leq1$ and $\sum_\alpha c^\psi_\alpha = 1$ \cite{Griffiths_QM}. Because $\calH_\Sigma$ factorises, we can define a reduced density matrix $\rho^\psi_A$ on $A$, as
\begin{align}
    \rho^\psi_A = \Tr_{A^c}(\rho^\psi) = \sum_\alpha c^\psi_\alpha\smallket{\psi^A_\alpha}\smallbra{\psi^A_\alpha}\ ,
\end{align}
where $\Tr_{A^c}$ is defined in \eqref{AppB_ReducedDensity} and the $\smallket{\psi^A_\alpha}\in\calH_A$ are again unit vectors. With the reduced density matrix, expectation values of operators $\order_A$ that act only on the field configuration on $A$ can be denoted as $\mean{\order_A}_\psi = \Tr_A(\rho^\psi_A\order_A)$. Furthermore, it can be used to define two quantities of great importance to quantum information theory. The first is the entanglement entropy across the entangling surface $\partial A$:
\begin{align}
    S_e[\psi,A] = -\Tr_A(\rho^\psi_A\ln\rho^\psi_A)\ . \label{thANEC_EntanglementEntropy}
\end{align}
The other quantity is the modular Hamiltonian:
\begin{align}
    K^\psi_A = -\ln\rho^\psi_A\ .
\end{align}
Of course, a similar modular Hamiltonian $K^\psi_{A^c}$ can be defined for the field configuration on $A^c$, and they can be combined into the full modular Hamiltonian as
\begin{align}
    \hat{K}^\psi_A = K^\psi_A\otimes\mathbb{1}_{A^c} - \mathbb{1}_A\otimes K^\psi_{A^c}\ .
\end{align}
Both the entanglement entropy and modular Hamiltonian play a role in the relative entropy. This quantity, which indicates the degree to which states $\ket{\psi},\ket{\phi}\in\calH_\Sigma$ can be distinguished based on measurements restricted to $A$, is defined as
\begin{align}
    S(\rho^\psi_A|\rho^\phi_A) \equiv&\ \Tr_A(\rho^\psi_A\ln\rho^\psi_A) - \Tr_A(\rho^\psi_A\ln\rho^\phi_A) \non
    \alis \Tr_A(\rho^\psi_A\ln\rho^\psi_A) - \Tr_A(\rho^\phi_A\ln\rho^\phi_A) -\Tr_A(\rho^\phi_AK^\phi_A) + \Tr_A(\rho^\psi_A K^\phi_A) \non
    \alis -S_e[\psi,A] + S_e[\phi,A] - \smallmean{K^\phi_A}_\phi + \smallmean{K^\phi_A}_\psi \non
    \equiv&\ -\Delta S_e[A] + \Delta \smallmean{K^\phi_A}\ . \label{thANEC_RelEnt}
\end{align}
The relative entropy has two important properties: it is non-negative and monotonous \cite{Witten_ReehSchlieder}. The latter property is the statement that
\begin{equation}
    S(\rho^\psi_a|\rho^\phi_a) \leq S(\rho^\psi_A|\rho^\phi_A)\ , \label{thANEC_RelEntMon}
\end{equation}
if we choose a region $a$ such that $D(a)\subset D(A)$; in essence, this property means that it is more difficult to distinguish states by taking measurements in a smaller region. By combining \eqref{thANEC_RelEnt} and \eqref{thANEC_RelEntMon}, it follows that
\begin{align}
    \Delta \smallmean{K^\phi_a} - \Delta \smallmean{K^\phi_A} + \Delta S_e[A] - \Delta S_e[a] \leq 0\ . \label{thANEC_ApplyMonotonic1}
\end{align}
In particular, choose a second Cauchy surface $\Sigma'$ with $a\subset\Sigma'$, and call $a^c=\Sigma'\setminus a$. The equivalent of \eqref{thANEC_ApplyMonotonic1} for $a^c$ and $A^c$ must take into account that $D(A^c)\subset D(a^c)$, leading to
\begin{align}
    \Delta \smallmean{K^\phi_{A^c}} - \Delta \smallmean{K^\phi_{a^c}} + \Delta S_e[a^c] - \Delta S_e[A^c] \leq 0\ . \label{thANEC_ApplyMonotonic2}
\end{align}
Adding \eqref{thANEC_ApplyMonotonic1} and \eqref{thANEC_ApplyMonotonic2} together yields
\begin{align}
    \Delta\smallmean{\hat{K}^\phi_a} - \Delta\smallmean{\hat{K}^\phi_A} + \Delta S_e[A] - \Delta S_e[A^c] - \Delta S_e[a] + \Delta S_e[a^c] \leq 0\ . \label{thANEC_ApplyMonotonic3}
\end{align}
The final step taken in \cite{ANEC_Entropy} is to choose two specific field configurations; following them, we choose $\phi$ to be the vacuum $\ket{0}$ and let $\psi$ denote a pure state $\ket{\psi}\in\calH_\Sigma$ ($\ket{\psi}$ can then be evolved to $\Sigma'$ through the equations of motion to obtain the corresponding state in $\calH_{\Sigma'}$). Since these are both pure states, all terms in \eqref{thANEC_ApplyMonotonic3} that involve the entanglement entropy must vanish. To see this, consider for example $S_e[\psi,A]$ for a pure state $\ket{\psi}\in\calH_\Sigma$; the other terms can be treated similarly. The reduced density matrix for this state is
\begin{align}
    \rho^\psi_A = \Tr_{A^c}(\rho^\psi) = \sum_i\langle\psi^{A^c}_i|\psi\rangle\langle\psi|\psi^{A^c}_i\rangle = \smallket{\psi^A}\smallbra{\psi^A}\ ,
\end{align}
since $\ket{\psi} = \smallket{\psi^A}\smallket{\psi^{A^c}}$ with $\smallket{\psi^{A}}\in\calH_{A}$ and $\smallket{\psi^{A^c}}\in\calH_{A^c}$ both unit vectors. Importantly, $\rho^\psi_A$ also describes a pure state, which means that $(\rho^\psi_A)^2 = \rho^\psi_A$. We can then evaluate the following quantity (using a trick from \cite{Kiritsis}):
\begin{align}
    \rho^\psi_A\ln\rho^\psi_A = \lim_{n\rightarrow1}\parti{(\rho^\psi_A)^n}{n} = \lim_{n\rightarrow1}\parti{\rho^\psi_A}{n} = 0\ .
\end{align}
Hence, the entanglement entropy of a pure state vanishes, leaving only the terms involving modular Hamiltonians in \eqref{thANEC_ApplyMonotonic3}. Since the full modular Hamiltonian annihilates the vacuum \cite{BlancoCasini_2013}, \eqref{thANEC_ApplyMonotonic3} reduces to
\begin{align}
    \smallmean{\hat{K}^0_A}_\psi - \smallmean{\hat{K}^0_a}_\psi \geq 0\ , \label{thANEC_ModHamPos}
\end{align}
where the superscript indicates that this is the modular Hamiltonian of the vacuum. It is difficult to calculate these modular Hamiltonians in general, but the difference in \eqref{thANEC_ModHamPos} can be calculated perturbatively if $a$ is an infinitesimal deformation of $A$. If one chooses $A = \cuha{x\in M:x^0 = 0,\,x^1>0}$ and picks an infinitesimal vector field $\zeta^a$ with, in null coordinates, $\zeta^\mu = (\zeta^+,\zeta^-,\ve{0}^\perp) = \zeta^\mu(\ve{x}^\perp)$ and $\pm\zeta^\pm>0$, then \cite{ANEC_Entropy} proved that
\begin{align}
    \smallmean{\hat{K}^0_A}_\psi - \smallmean{\hat{K}^0_a}_\psi = 2\pi\int\d^{d-2}\ve{x}^\perp\roha{\zeta^+\int_{L_+(\ve{x}^\perp)}\d x^+\,\smallmean{T_{++}}_\psi - \zeta^-\int_{L_-(\ve{x}^\perp)}\d x^-\,\smallmean{T_{--}}_\psi}\ . \label{thANEC_ModHamANEC}
\end{align}
In this expression, $L_+(\ve{x}^\perp)$ is a complete null geodesic with affine parameter $\lambda$ such that $L_+(\ve{x}^\perp) = \cuha{x\in M:x^\mu = (x^+ = \lambda,x^- = 0,\ve{x}^\perp)}$. Similarly, the coordinates along $L_-(\ve{x}^\perp)$ can be written as $x^\mu(\lambda) = (x^+=0,x^-=\lambda,\ve{x}^\perp)$. Combining \eqref{thANEC_ModHamPos} and \eqref{thANEC_ModHamANEC} for $\zeta^+\gg-\zeta^-$ yields
\begin{align}
    \int\d^{d-2}\ve{x}^\perp\,\zeta^+\int_{L_+(\ve{x}^\perp)}\d x^+\,\smallmean{T_{++}}_\psi \geq 0\ .
\end{align}
The ANEC \eqref{thANEC_ANEC} follows because this should hold for any choice of $\zeta^+ = \zeta^+(\ve{x}^\perp)$.

\section{The ANEC in curved spacetimes}\label{sec_CurvedSpacetimes}
In the previous section, it was shown that the ANEC holds for general QFTs in Minkowski spacetime. The situation is more complicated in other spacetimes, so in this section several examples of violations of the ANEC will be given.

A first counterexample to the ANEC can be given in a Minkowski spacetime in which one spatial direction is compactified \cite{Klinkhammer}. As a result, a phenomenon closely related to the Casimir effect \cite{CasimirEffect} occurs, and the ANEC is violated on any null geodesic that moves along the compactified direction. Another class of counterexamples can be found for QFTs that would be conformally invariant field theories (CFTs) in flat spacetime. These theories are invariant under Weyl transformations \cite{WeylVsConformal}, that is, local rescalings of the metric as given in \eqref{AppA_Weyl}:
\begin{align}
    \tilde{g}_{ab}(x)\rightarrow g_{ab}(x) = \Omega^2(x)\tilde{g}_{ab}(x) \ ,
\end{align}
for a smooth, non-zero function $\Omega:M\rightarrow\mathbb{R}$ called the conformal factor. If $\tilde{T}_{ab}$ and $T_{ab}$ are the energy-momentum tensors of the CFT in the original and transformed spacetime respectively, they are related as in \eqref{AppB_CFTEMT}:
\begin{align}
    U\tilde{T}_{ab}U^\dagger = \Omega^{d-2}(x)\roha{T_{ab} - X_{ab}}\ , \label{thANEC_TransformEMT}
\end{align}
where $U:\tilde{\calH}\rightarrow\calH$ is a unitary map between states in the original and transformed spacetime, and $X_{ab}$ is a purely geometrical conformal anomaly \cite{Conforme_anomalie}. We may choose $\tilde{g}_{ab} = \eta_{ab}$; then the ANEC integral in the transformed spacetime can be related to an integral in Minkowski spacetime, since
\begin{align}
    \int_\gamma \mean{T_{ab}}_\psi l^al^b\d\lambda = \int_\gamma \Omega^{-d}(x)\smallmean{\tilde{T}_{ab}}_{\tilde{\psi}}\tilde{l}^a\tilde{l}^b + X_{ab}\tilde{l}^a\tilde{l}^b\,\d\tilde{\lambda}\ , \label{thANEC_TransformANEC}
\end{align}
where we used the fact that Weyl transformations leave null geodesics invariant, defined $\smallket{\tilde{\psi}} = U^\dagger\ket{\psi}$, and used \eqref{AppA_WeylAffine}. It may happen that this integral is still non-negative; this is the case for e.g. (anti-)de Sitter spacetime, for which $\Omega$ is constant along $\gamma$ and $X_{ab}\tilde{l}^a\tilde{l}^b=0$ \cite{Rosso_2020}. In these instances, the ANEC continues to hold, but for most choices of $\Omega$, \eqref{thANEC_TransformANEC} indicates that the ANEC can be violated in two distinct ways: the conformal factor $\Omega$ can enhance the contribution of a segment of $\gamma$ along which the null energy condition is violated, or the conformal anomaly violates the ANEC \cite{ANEC_violation_old,ANEC_violation_new,ANEC_violation_VeryOld}.

Finally, one may consider spacetimes that are unrelated to Minkowski spacetime. An example of this is provided by a QFT in Schwarzschild spacetime, where the conformally coupled (but otherwise free) massless scalar field violates the ANEC for any null geodesic that does not cross the event horizon if one considers the Boulware vacuum \cite{ANECviolation_Schwarzschild}. This state, which is the vacuum as seen by a stationary observer outside the event horizon, is singular at the event horizon \cite{Carroll_2019}. This may indicate that it is not physically relevant. Nevertheless, the ANEC is violated on null geodesics that never cross the event horizon, and thus it should be possible to violate the ANEC with a state that agrees with the Boulware vacuum well outside the black hole but smooths out the divergence.

In summary, section \ref{sec_FlatSpacetime} showed that QFTs routinely violate the pointwise null energy condition and this section demonstrated that they also violate the averaged version when considered in curved spacetimes. However, the success of the ANEC in Minkowski spacetime hints at the possibility of a generally valid energy condition. For this, we would have to modify the ANEC to a form suitable for various geometries. Two approaches to this modification are discussed in the next chapter: the conformally invariant ANEC and the self-consistent achronal ANEC.

\chapter{Generalisations of the ANEC}\label{chap_curveANEC}
As the previous chapter showed, quantum field theories (QFTs) in Minkowski spacetime obey the averaged null energy condition (ANEC), which states that
\begin{align}
    \int_\gamma\d\lambda\,\mean{T_{ab}}_\psi l^al^b\geq0
\end{align}
for any complete null geodesic $\gamma$ with affine parameter $\lambda$ and tangent vector $l^a$ and any state $\ket{\psi}$. However, several examples showed that QFTs fail to obey this condition when considered in curved spacetimes, which is a relevant setting if one is interested in the role of energy conditions in (semi-classical) gravity. 

This chapter therefore examines two generalisations of the ANEC to curved spacetimes. To facilitate this discussion, section \ref{sec_AdS/CFT} very briefly introduces the AdS/CFT correspondence, which is used in section \ref{sec_CANEC} to prove, for holographic field theories in a class of fixed curved spacetimes, the $d=3$ conformally invariant ANEC \eqref{intro_CANEC} or CANEC. Although this condition is secure against violations by Weyl transformations, its proof only holds for a relatively restricted range of situations. Thus, section \ref{sec_AANEC} goes beyond the framework of QFT on fixed spacetime and considers a geometry determined by the quantum field. A perturbative proof by Wall \cite{Wall} of the resulting self-consistent achronal ANEC (AANEC) is presented.

\section{The AdS/CFT correspondence}\label{sec_AdS/CFT}
As chapter \ref{chap_Intro} remarked, general relativity is a classical theory, while realistic matter and its interactions are described by QFTs. Often, this discrepancy can be dealt with using semi-classical gravity \cite{Rosenfeld_1963}, but this is not a quantum theory of gravity, which remains elusive. An example of progress towards such a theory is the AdS/CFT correspondence, first proposed in \cite{Maldacena}, which we introduce very briefly in this section; a more extensive introduction can be found in e.g. \cite{Kiritsis} or \cite{Polchinski_AdS/CFT} and references therein.

\subsection{Anti-de Sitter spacetime}

Anti-de Sitter (AdS) spacetime is the maximally symmetric spacetime with constant Ricci scalar $R < 0$ (i.e. the AdS metric is invariant under the maximum number of independent diffeomorphisms and the AdS manifold is negatively curved); as such, it contains no matter and has a negative cosmological constant $\Lambda < 0$. In coordinates that cover an entire $d+1$ dimensional AdS manifold, the line element is
\begin{align}
    \d s^2 = \roha{-\cosh^2(\rho/L)\d t^2 + \d\rho^2 + L^2\sinh^2(\rho/L)\d\Omega_{d-1}^2}\ , \label{curveANEC_AdSGlobal}
\end{align}
where $t\inR$ and $\rho\geq0$, $L$ is a length scale related to $\Lambda$ as $\Lambda = -d(d-1)/(2L^2)$, and $\d\Omega^2_{d-1}$ is the metric of the $d-1$ dimensional unit sphere. An important property of AdS spacetime is that it has a timelike conformal boundary, which becomes apparent upon defining new coordinates $\chi$ and $\tau$ with $\cos\chi = 1/\cosh(\rho/L)$ and $\tau = t/L$, so that the metric reads
\begin{align}
    \d s^2 = \frac{L^2}{\cos^2\chi}\roha{-\d \tau^2 + \d\chi^2 + \sin^2\chi\,\d\Omega_{d-1}^2}\ . \label{curveANEC_AdSSphere}
\end{align}
This shows that AdS spacetime is conformally related to the Einstein static universe (which has the geometry described by the line element in brackets), restricted to $0\leq\chi<\frac{\pi}{2}$. Therefore, AdS spacetime has a timelike boundary at $\chi = \frac{\pi}{2}$ and is not globally hyperbolic, since null geodesics can terminate on the boundary in finite coordinate time.

If a spacetime $M$ has $\Lambda < 0$ and contains matter fields that decay sufficiently quickly towards the conformal boundary $\partial M$, then the causal structure of $M$ asymptotically approaches that of AdS spacetime near $\partial M$. A convenient way to represent this is with the Fefferman-Graham gauge \cite{FeffermanGraham}, which is a set of coordinates $y^\alpha = (x^\mu,z)$ with $\alpha\in\cuha{0,\ldots,d}$ and $\mu\in\cuha{0,\ldots,d-1}$ such that the metric near $\partial M$ can be written as
\begin{align}
    \d s^2 = g_{\alpha\beta}(y)\d y^\alpha \d y^\beta = \frac{1}{z^2}\roha{\d z^2 + \hat{g}_{\mu\nu}(x,z)\d x^\mu\d x^\nu}\ . \label{CurveANEC_FGgauge}
\end{align}
If no matter is present in $M$, $\hat{g}_{\mu\nu} = \eta_{\mu\nu}$ so that the metric in \eqref{CurveANEC_FGgauge} reduces to that of the Poincaré patch, which covers half of AdS spacetime. On the other hand, $\hat{g}_{\mu\nu}$ has a (formal) power series expansion \cite{FeffermanGraham} in the presence of matter:
\begin{align}
    \hat{g}_{\mu\nu}(x,z) = \sum_{n=0}^\infty z^n\hat{g}^{(n)}_{\mu\nu}(x) + z^d\ln(z^2)h_{\mu\nu}(x)\ , \label{curveANEC_FGexpansie}
\end{align}
where $h_{\mu\nu}=0$ for odd $d$ and $\hat{g}_{\mu\nu}^{(2n+1)} = 0$ for $2n+1<d$ \cite{holographic_EMT}. $\partial M$, the boundary of $M$, is located at $z\rightarrow0$, and forms another spacetime with metric $\lim_{z\rightarrow0}z^2g_{\mu\nu}(z,x) = \hat{g}^{(0)}_{\mu\nu}$. As such, knowledge of $\hat{g}^{(0)}_{\mu\nu}$ can be used as boundary conditions for the (semi-)classical Einstein field equations near the boundary, where only the cosmological constant contributes to $T_{ab}$; in fact, $\hat{g}^{(n)}_{\mu\nu}$ can be expressed purely in terms of $\hat{g}^{(0)}_{\mu\nu}$ for all $n<d$. For example, $\hat{g}^{(2)}_{\mu\nu}$ is \cite{holographic_EMT}
\begin{align}
    \hat{g}^{(2)}_{\mu\nu} = -\frac{1}{d-2}\roha{\hat{R}_{\mu\nu} - \frac{\hat{R}}{2(d-1)}\hat{g}_{\mu\nu}^{(0)}}\ , \label{CurveANEC_G2}
\end{align}
with $\hat{R}_{\mu\nu}$ and $\hat{R}$ respectively the Ricci tensor and scalar associated with the boundary metric $\hat{g}^{(0)}_{\mu\nu}$. Note the sign difference between \eqref{CurveANEC_G2} and \cite{holographic_EMT}; this is due to a different sign convention for geometric quantities. We follow \cite{CANEC_odd,CANEC_even}.

\subsection{The correspondence in a nutshell}

The relationship between a spacetime $M$ and its boundary $\partial M$ goes even deeper according to the AdS/CFT correspondence, which states that there is a correspondence between a string theory describing quantum gravity in $M$ (the bulk) and a non-gravitational CFT on $\partial M$ (the boundary). In particular, it says that every CFT state or correlator maps to a string theory state or correlator respectively. This is particularly useful for holographic CFTs, which correspond to semi-classical general relativity in their low-energy limit, since we then understand both the gravitational and the field theoretical side of the correspondence. For example, one can relate the energy-momentum tensor $T_{\mu\nu}$ of a holographic CFT to geometric quantities in the bulk \cite{holographic_EMT}:
\begin{align}
    \mean{T_{\mu\nu}}_\psi = \frac{L^{d-1}d}{16\pi G_N}\hat{g}^{(d)}_{\mu\nu} + X_{\mu\nu}\ , \label{curveANEC_HolographicEMt}
\end{align}
where $X_{\mu\nu}$ is the conformal anomaly, which can be expressed in terms of $\hat{g}^{(n<d)}_{\mu\nu}$ and vanishes for odd $d$. Another important consequence of the AdS/CFT correspondence is that signals between points $p,q\in\partial M$ may travel on two types of curves: those that stay in the boundary, and those that move into the bulk. If the boundary CFT is a sensible theory on its own, it should not allow signals to travel between points that cannot be connected by a causal curve in the boundary. It makes sense, then, to demand that there is no causal bulk curve connecting points in $\partial M$ that cannot be causally connected by a boundary curve. A spacetime $M\cup\partial M$ with this property is said to have the no-bulk-shortcut property.

\section{The \texorpdfstring{$d=3$}{d=3} conformally invariant ANEC}\label{sec_CANEC}
In the previous section, we have briefly seen how the AdS/CFT correspondence relates gravitational dynamics in a $d+1$ dimensional asymptotically AdS spacetime $M$ to the dynamics of a CFT in the $d$ dimensional boundary spacetime $\partial M$. Following \cite{CANEC_odd}, we will use this correspondence and the related no-bulk-shortcut property to demonstrate that, at least for $d=3$, holographic CFTs obey the conformally invariant ANEC (CANEC). The main idea of this proof of CANEC is that we can choose two points in $\partial M$ that cannot be connected by a causal boundary curve, and then attempt to construct a causal bulk curve which does connect them. Insisting on the no-bulk-shortcut property leads to the CANEC.

\subsection{Constructing a bulk curve}

We begin with general $d$ and choose a point $p\in\partial M$. Now let $\partial J^+(p)\subset\partial M$ be the boundary of the causal future of $p$ in $\partial M$, and choose $q\in \partial J^+(p)$; this means that there is a null geodesic $\gamma\subset\partial M$ connecting $p$ to $q$. Assuming $d\geq3$, we can choose a codimension 2 spacelike submanifold $W\subset\partial M$ such that $\gamma$ intersects $W$ orthogonally in $p$. This submanifold can be used to choose Gaussian normal coordinates $x^\mu = (u,v,x^I)$, with $I\in\cuha{2,d-1}$, as the boundary coordinates $x^\mu$ in a region containing $p$ and $q$. These coordinates are described in more detail in section \ref{sec_CongRay} (based on \cite{Witten_causality}), but their most important feature is that curves with fixed $v$ and $x^I$ describe null geodesics with affine parameter $u$ that emanate orthogonally from $W$. As a consequence, the components of the boundary metric $\hat{g}^{(0)}_{\mu\nu}$ can be read off from
\begin{align}
    \d s^2_{\partial M} = -e^q\roha{\d u\d v + \d v\d u} + \hat{g}_{vv}^{(0)}\d v^2 + \hat{g}_{IJ}^{(0)}\d x^I\d x^J + \hat{g}_{vI}^{(0)}\roha{\d v\d x^I + \d x^I \d v}\ . \label{CurveANEC_GaussianNormal}
\end{align}
Here, $q = q(v,x^I)$ while $\hat{g}_{vv}^{(0)}$, $\hat{g}_{IJ}^{(0)}$ and $\hat{g}_{vI}^{(0)}$ can be functions of all boundary coordinates; note that $\hat{g}^{(0)}_{uv} = -e^q < 0$. 

Returning to the task at hand, we may choose $v = x^I = 0$ for points along $\gamma$, and set $u = \lambda_-$ at $p$ and $u = \lambda_+$ at $q$. We further choose a point $r\in\partial M$ which also has $u = \lambda_+$ and $x^I = 0$; if there exists a causal boundary curve from $p$ to $r$, then the choice of signs in \eqref{CurveANEC_GaussianNormal} and the fact that $v=0$ along part of $\partial J^+(p)$ imply that $r$ must have $v\geq0$. The same must then be true if $M\cup\partial M$ has the no-bulk-shortcut property and there exists a causal bulk curve from $p$ to $r$. To analyse the consequences of this restriction, we construct a causal bulk curve $\Gamma\subset M\cup\partial M$ which is a perturbation of $\gamma$ and connects $p$ to an $r$ with minimal $v$. The coordinates along $\Gamma$ can be parametrised as
\begin{align}
    y^\alpha = \roha{u,v,x^I,z} = \roha{\lambda,v(\lambda),0^I,z(\lambda)}\ ,
\end{align}
where $\lambda\in[\lambda_-,\lambda_+]$ is the parameter and $0^I$ indicates that all $x^I$ vanish; because $p,r\in\partial M$, we require $z(\lambda_\pm) = 0$. Since we assume that $\Gamma$ is a perturbation of $\gamma$, we can choose a small parameter $0<\varepsilon\ll1$ and expand $v(\lambda)$ and $z(\lambda)$ as respectively
\begin{align}
    v(\lambda) = \sum_{n=1}^\infty\varepsilon^nv_n(\lambda)& &\text{and}& &z(\lambda) = \sum_{n=1}^\infty\varepsilon^nz_n(\lambda)\ . \label{curveANEC_vzExpansie}
\end{align}
Here, $v_n$ and $z_n$ are arbitrary sufficiently smooth functions. Because $\hat{g}^{(1)}_{\mu\nu} = 0$ \cite{holographic_EMT}, the norm of the tangent vector to $\Gamma$, with components $K^\alpha$, is
\begin{align}
    z^2g_{\alpha\beta}K^\alpha K^\beta \alis \hat{g}_{uu} + \hat{g}_{vv}\dot{v}^2 + \dot{z}^2 + 2\hat{g}_{uv}\dot{v} \non
    \alis \roha{\varepsilon^2z_1^2 + 2\varepsilon^3z_1z_2}\hat{g}^{(2)}_{uu} + \varepsilon^3z_1^3\hat{g}^{(3)}_{uu} + \roha{\varepsilon^2\dot{v}_1^2 + 2\varepsilon^3\dot{v}_1\dot{v}_2}\hat{g}^{(0)}_{vv} + \varepsilon^2\dot{z}_1^2 + 2\varepsilon^3\dot{z}_1\dot{z}_2 \non
    &\qquad\ \qquad\ \qquad\ + 2\roha{\varepsilon\dot{v}_1 + \varepsilon^2\dot{v}_2 + \varepsilon^3\dot{v}_3}\hat{g}_{uv}^{(0)} + 2\varepsilon^3\dot{v}_1z_1^2\hat{g}^{(2)}_{uv} + \order(\varepsilon^4)\ , \label{curveANEC_expansie}
\end{align}
where a dot signifies a derivative with respect to $\lambda$. Using \eqref{curveANEC_vzExpansie}, we can do a similar expansion for the $v$-coordinate of $r$:
\begin{align}
    v(\lambda_+) = \int_{\lambda_-}^{\lambda_+}\dot{v}\d\lambda = \varepsilon\int_{\lambda_-}^{\lambda_+}\dot{v}_1\d\lambda + \varepsilon^2\int_{\lambda_-}^{\lambda_+}\dot{v}_2\d\lambda + \varepsilon^3\int_{\lambda_-}^{\lambda_+}\dot{v}_3\d\lambda + \order(\varepsilon^4)\ .
\end{align}
Two facts can now be combined: first, that $\Gamma$ is causal, implying that $g_{\alpha\beta}K^\alpha K^\beta\leq0$, and second, that $\Gamma$ satisfies the no-bulk-shortcut property, meaning that $v(\lambda_+)\geq0$. We impose these inequalities order-by-order, in the sense that the leading non-vanishing order in the $\varepsilon$-expansion should obey them. At first order, the causal nature of $\Gamma$ implies that
\begin{align}
    \hat{g}_{uv}^{(0)}\dot{v}_1\leq0\ .
\end{align}
Recalling from \eqref{CurveANEC_GaussianNormal} that $\hat{g}^{(0)}_{uv}<0$, we conclude that $\dot{v}_1\geq0$, which means that the no-bulk-shortcut property holds identically at first order in $\varepsilon$. However, we may minimise $v(\lambda_+)$ by choosing $\dot{v}_1 = 0$, so that the signs of $g_{\alpha\beta}K^\alpha K^\beta$ and $v(\lambda_+)$ are determined at second order in $\varepsilon$. At this order and setting $\dot{v}_1 = 0$, a causal $\Gamma$ requires
\begin{align}
    \dot{v}_2\hat{g}^{(0)}_{uv} \leq -z_1^2\hat{g}_{uu}^{(2)} - \dot{z}_1^2\ .
\end{align}
Again using that $\hat{g}^{(0)}_{uv} < 0$, we find that this bounds the second order term in the expansion of $v(\lambda_+)$ from below as
\begin{align}
    \int_{\lambda_-}^{\lambda_+}\dot{v}_2\d\lambda \geq -\frac{1}{\hat{g}^{(0)}_{uv}}\int_{\lambda_-}^{\lambda_+}\dot{z}_1^2 - \frac{\hat{R}_{uu}}{d-2}z_1^2\,\d\lambda \equiv -\frac{1}{\hat{g}^{(0)}_{uv}}I_2[z_1]\ , \label{curveANEC_V2}
\end{align}
where we used the fact that $\hat{g}^{(0)}_{uv}$ is independent of $u = \lambda$ and can therefore be taken out of the integral, and rewrote $\hat{g}^{(2)}_{uu}$ in terms of the boundary geometry via \eqref{CurveANEC_G2}. Whether the no-bulk-shortcut property continues to hold is now determined by the minimum value of the integral on the right-hand side, referred to as $I_2$; this is a typical problem from functional analysis \cite{vanBrunt}. 

\subsection{Determining the minimum value of \texorpdfstring{$I_2$}{I2}}

The first step to determining the minimum of $I_2$ is to extremise it, which means that we impose the Euler-Lagrange equation on the integrand. For $I_2$, this equation reads
\begin{align}
    \ddot{z}_1 = -\frac{\hat{R}_{uu}}{d-2}z_1\ . \label{curveANEC_EulerLagrange}
\end{align}
If $z_1 = s_1(\lambda)$ solves this equation, one can see that $I_2[s_1] = 0$ by integrating by parts and using that $z_1(\lambda_\pm) = 0$. This cannot be a maximum of $I_2$, since we can always choose a rapidly oscillating $z_1$ to make $I_2$ arbitrarily big, but to distinguish a minimum from a saddle point one should consider the second variation of $I_2$ around $s_1$. To formulate a sufficient condition for a minimum of $I_2$, we denote the integrand in the definition of $I_2$ by $f$ (i.e. $I_2[z_1] = \int_{\lambda_-}^{\lambda_+}f(\lambda,z_1,\dot{z}_1)\d\lambda$) and consider the Jacobi accessory equation:
\begin{align}
    \diff{}{\lambda}\roha{\dot{\xi}\partial_{\dot{z}_1}^2f} - \roha{\partial_{z_1}^2f - \diff{}{\lambda}\roha{\partial_{z_1}\partial_{\dot{z}_1}f}}\xi = 0\ , \label{curveANEC_JacobiAccessory}
\end{align}
where $\xi$ is some sufficiently smooth function and all derivatives of $f$ are evaluated at a solution $z_1 = s_1(\lambda)$ to \eqref{curveANEC_EulerLagrange}. It can be shown that $I_2[s_1]$ is a minimum of $I_2$ if $\partial_{\dot{z}_1}^2f > 0$ and there exists no non-trivial solution $\xi$ to \eqref{curveANEC_JacobiAccessory} which vanishes anywhere in $\langle\lambda_-,\lambda_+]$ if $\xi(\lambda_-) = 0$ \cite{vanBrunt}.

In our present case, the Jacobi accessory equation implies that $\xi$ should satisfy \eqref{curveANEC_EulerLagrange}. Since we impose $z_1(\lambda_\pm) = 0$, the sufficient condition for a minimum of $I_2$ cannot be satisfied unless $z_1 = 0$ is the only solution to \eqref{curveANEC_EulerLagrange} that is consistent with the boundary conditions; this happens, for example, when $\hat{R}_{uu}\leq0$ for all $\lambda\in[\lambda_-,\lambda_+]$. However, we may demand that no solution to \eqref{curveANEC_EulerLagrange} should exist which vanishes for a $\lambda\in\langle\lambda_-,\lambda_+\rangle$ when $z_1(\lambda_-) = 0$, which means that $I_2$ satisfies the Jacobi necessary condition \cite{vanBrunt}. This condition states that, for $I_2$ to have a minimum, it is necessary but not sufficient that $\partial_{\dot{z}_1}^2f > 0$ and there exists no solution $\xi$ to \eqref{curveANEC_JacobiAccessory} which vanishes anywhere in $\langle\lambda_-,\lambda_+\rangle$ if $\xi(\lambda_-) = 0$.

This is about as much as can be said about the nature of the extremum of $I_2$ on general grounds: it is not a maximum and satisfies the necessary conditions to be a minimum, but unless $\hat{R}_{uu}\leq0$ everywhere along $\Gamma$, it cannot be concluded generally that the extremum is indeed a minimum. An important explicit case is the case of a constant and positive $\hat{R}_{uu}$, because one can then use the Poincaré inequality \cite{Poincare}:
\begin{align}
    \int_{\lambda_-}^{\lambda_+}z_1^2\d\lambda \leq \roha{\frac{\lambda_+ - \lambda_-}{\pi}}^2\int_{\lambda_-}^{\lambda_+}\dot{z}_1^2\d\lambda\ .
\end{align}
Using this inequality in the case of a constant $\hat{R}_{uu}>0$, $I_2$ can be directly estimated as
\begin{align}
    I_2 \geq \roha{1 - \frac{\hat{R}_{uu}}{d-2}\roha{\frac{\lambda_+ - \lambda_-}{\pi}}^2}\int_{\lambda_-}^{\lambda_+}\dot{z}_1^2\d\lambda\ .
\end{align}
By choosing $\lambda_+ = \lambda_- + \pi\sqrt{(d-2)/\hat{R}_{uu}}$, $I_2 = 0$ is certain to be a minimum. Another important class of spacetimes is formed by the conformally flat spacetimes, which have $\hat{g}_{ab}^{(0)} = \Omega^2(x)\eta_{ab}$. In that case, $\hat{R}_{uu}$ can be calculated directly from \eqref{AppA_ConformalRicciT}, leading to
\begin{align}
    I_2[z_1] \alis \int_{\lambda_-}^{\lambda_+}\dot{z}_1^2 + \frac{\ddot{\Omega}}{\Omega}z_1^2\d\lambda \non
    \alis \int_{\lambda_-}^{\lambda_+}\roha{\dot{z}_1 - \frac{z_1\dot{\Omega}}{\Omega}}^2\d\lambda + \viha{\frac{\dot{\Omega}}{\Omega}z_1^2}^{\lambda_+}_{\lambda_-}\ .
\end{align}
If one not only demands that $z_1^2$ vanishes as $\lambda\rightarrow\lambda_\pm$ but also that it does so faster than $\Omega$, the boundary terms vanish and $I_2$ is manifestly non-negative.

\subsection{Finding the energy condition}

To proceed, let us assume that $\hat{R}_{uu}$ is such that $I_2 = 0$ is a minimum. According to \eqref{curveANEC_V2}, this means that the no-bulk-shortcut property continues to hold at second order in $\varepsilon$, although this contribution can be made to vanish by choosing $z_1$ to obey \eqref{curveANEC_EulerLagrange} and $\dot{v}_2 = 0$ as well. We can then consider the third order contribution to $z^2g_{\alpha\beta}K^\alpha K^\beta$ in \eqref{curveANEC_expansie}, setting $\dot{v}_1 = \dot{v}_2 = 0$:
\begin{align}
    \dot{v}_3\hat{g}^{(0)}_{uv} \leq -z_1z_2\hat{g}^{(2)}_{uu} - \frac{1}{2}z_1^3\hat{g}^{(3)}_{uu} - \dot{z}_1\dot{z}_2\ .
\end{align}
This shows that the third-order contribution to $v(\lambda_+)$ is limited as
\begin{align}
    \int_{\lambda_-}^{\lambda_+}\dot{v}_3\d\lambda \geq -\frac{1}{\hat{g}^{(0)}_{uv}}\int_{\lambda_-}^{\lambda_+}\roha{\dot{z}_1\dot{z}_2 - \frac{\hat{R}_{uu}}{d-2}z_1z_2} + \frac{z_1^3}{2}\hat{g}^{(3)}_{uu}\,\d\lambda\ .
\end{align}
By integrating by parts and imposing \eqref{curveANEC_EulerLagrange}, the terms vanish. The remaining term on the right-hand side, proportional to $\hat{g}^{(3)}_{uu}$, vanishes trivially if the only solution to \eqref{curveANEC_EulerLagrange} with $z_1(\lambda_\pm) = 0$ is $z_1 = 0$, or if $d > 3$, which implies that $\hat{g}^{(3)}_{uu} = 0$. However, in the $d=3$ spacetimes for which $z_1$ is non-trivial and $I_2\geq0$, there is no further constraint on this final contribution. Instead, we invert the logic and insist upon the no-bulk-shortcut property, meaning that we must have
\begin{align}
    \int_{\lambda_-}^{\lambda_+}z_1^3\hat{g}^{(3)}_{uu}\d\lambda \geq 0\ . \label{curveANEC_protoCANEC}
\end{align}
In particular, note that $\hat{g}^{(3)}_{uu} = \hat{g}^{(3)}_{\mu\nu}l^\mu l^\nu$, where $l^\mu$ are the components of the tangent vector to $\gamma$; we also recall \eqref{curveANEC_HolographicEMt}, which for $d=3$ allows us to replace $\hat{g}^{(3)}_{\mu\nu}$ by (a positive multiple of) the energy-momentum tensor of the holographic CFT. To completely formulate \eqref{curveANEC_protoCANEC} in terms related to the boundary theory, we compare \eqref{curveANEC_EulerLagrange} to \eqref{classEC_Raychaudhuri} for a null congruence (which has $n=2$). The similarity is manifest if $\gamma$ can be embedded in a shear-free and irrotational congruence. We have already arranged the latter condition implicitly, when we embedded $p$ in a codimension 2 spacelike submanifold $W$ such that $\gamma\subset\partial J^+(W)$; after all, $\partial J^+(W)$ forms a null congruence which emanates orthogonally from a codimension 2 spacelike submanifold, which by \eqref{AppA_NoVort} means that it is irrotational. It will not, typically, be shear-free, but in the cases that it is we can identify $z_1$ with the Jacobi field. This finally establishes the $d=3$ CANEC:
\begin{align}
    \int_{\lambda_-}^{\lambda_+}\eta^3\mean{T_{ab}}_\psi l^al^b\d\lambda \geq 0\ . \label{curveANEC_CANEC}
\end{align}
To confirm the Weyl invariance of this condition, we perform a Weyl transformation $\hat{g}^{(0)}_{\mu\nu} \rightarrow g'_{\mu\nu} = \Omega^2(x)\hat{g}_{\mu\nu}^{(0)}$. By \eqref{thANEC_TransformEMT}, \eqref{AppA_WeylAffine}, and \eqref{AppA_WeylJacobi}, the CANEC integral transforms as
\begin{align}
    \int_{\lambda_-}^{\lambda_+}\eta^3\mean{T_{ab}}_\psi l^al^b\d\lambda \alis \int_{\lambda'_-}^{\lambda_+'}\roha{\Omega^{-1}\eta'}^3\roha{\Omega\mean{T'_{ab}}_{\psi'}}\Omega^4l'^a l'^b \Omega^{-2}\d\lambda' \non
    \alis \int_{\lambda'_-}^{\lambda'_+}\eta'^3\mean{T'_{ab}}_{\psi'} l'^al'^b\d\lambda'\ ,
\end{align}
where primes indicate transformed quantities. Evidently, if the CANEC holds in the geometry described by $\hat{g}^{(0)}_{\mu\nu}$, then it will hold in all conformally related geometries as well.

The calculation leading to \eqref{curveANEC_CANEC} can be repeated for different values of $d$, as has been done in \cite{CANEC_odd} and \cite{CANEC_even} for a class of spacetimes. The general picture that emerges from these calculations is that the energy-momentum tensor is bounded as
\begin{align}
    \int_{\lambda_-}^{\lambda_+}\eta^d\mean{T_{ab}}_\psi l^a l^b\d\lambda \geq B\ , \label{curveANEC_CANECform}
\end{align}
where $\lambda_\pm$ are the affine parameters of caustics and $B$ is a purely geometrical quantity that is closely related to the conformal anomaly (and hence vanishes for odd $d$). However, this bound has only been confirmed for $d=3,4,5$, in rather specific backgrounds (e.g. the $d=5$ calculation from \cite{CANEC_odd} assumes a static deformation of the static Einstein universe), and for holographic theories. In chapter \ref{chap_BeyondCANEC}, we will attempt to broaden the context to which the CANEC applies, but first \ref{sec_AANEC} will discuss another way of generalising the ANEC.

\section{The self-consistent achronal ANEC}\label{sec_AANEC}
Until now, we have formulated the ANEC and CANEC as energy conditions for QFTs in fixed spacetimes. These are not the theories we are ultimately interested in: as stated in chapter \ref{chap_Intro}, energy conditions aim to constrain dynamical geometries, which are determined by quantum fields through the semi-classical Einstein field equations \eqref{H1_SemiclassEFE}. Such theories might obey the self-consistent achronal ANEC (AANEC) \cite{AANEC}, which posits that the ANEC \eqref{thANEC_ANEC} holds on complete achronal null geodesics (which do not connect timelike separated points) in spacetimes that are self-consistent in the sense that they obey semiclassical gravity.

Note that achronality of the null geodesic and self-consistency of the spacetime are both required to avoid the violations of the ANEC mentioned in section \ref{sec_CurvedSpacetimes}. After all, there exist Weyl transformations that do not change the completeness or achronality of a null geodesic and nevertheless lead to violations of the ANEC \cite{ANEC_violation_new}; these are expected to be ruled out because the required Weyl factor cannot be semi-classically supported \cite{AANEC}. On the other hand, there exist potentially self-consistent spacetimes, such as Schwarzschild spacetime, in which the ANEC can be violated \cite{ANECviolation_Schwarzschild}. These spacetimes often do not have null geodesics that are both complete and achronal \cite{AANEC}; for Schwarzschild spacetime in particular, the radial null geodesics are achronal but incomplete while the non-radial geodesics are complete but not achronal.

Of course, avoiding these counterexamples does not constitute a proof of the AANEC. Formulating such a proof is a considerable undertaking due to the highly non-linear nature of the semi-classical Einstein field equations, but there are several reasons to believe the AANEC holds quite generally. Firstly, it was shown in \cite{FlanaganWald} that the massless, minimally coupled free real scalar field obeys the AANEC when it is allowed to perturb Minkowski spacetime. Secondly, \cite{Wall} demonstrated that the AANEC can be derived perturbatively from the generalised second law of thermodynamics, which states that the sum of the entropy associated with a black hole horizon and with matter outside this horizon cannot decrease over time. The aim of this section is to go over this second proof in some detail.

\subsection{The generalised second law of thermodynamics}

To state the proof of the AANEC in \cite{Wall}, it is first necessary to understand the generalised second law (GSL) of thermodynamics. The usual second law states that the entropy $S$ of an isolated system is non-decreasing with time. This law can be violated in the presence of a black hole by letting part of the system fall into the hole. However, it was noticed that this usually increases the surface area of the black hole \cite{Bekenstein_1973}, which led to the definition of the generalised entropy as
\begin{align}
    S_\mathrm{gen} = \frac{A}{4G_N} + S_\mathrm{out}\ , \label{curveANEC_GeneralisedS}
\end{align}
with $A$ the area of the event horizon and $S_\mathrm{out}$ the entanglement entropy of the matter outside the black hole, defined analogously to \eqref{thANEC_EntanglementEntropy} with the exterior and interior of the black hole as the regions $A$ and $A^c$ from \eqref{thANEC_EntanglementEntropy}, respectively. The GSL then states that $S_\mathrm{gen}$ cannot decrease \cite{Bekenstein_1973,Hawking_entropie}. In fact, similar statements hold for horizons in spacetimes without black holes (e.g. \cite{dS_entropie_hawking,dS_entropie,Friedmann_entropie}), motivating a GSL for all causal horizons $H_\mathrm{fut} = \partial J^-(W_\mathrm{fut})$, where $W_\mathrm{fut}$ is a future-complete timelike curve (i.e. its proper time can increase without limit) \cite{SecondLaw_CausalHorizon}. Such horizons can be used to specify $S_\mathrm{gen}$ by choosing a spacelike hypersurface $T$ and defining $A$ as the area of $H_\mathrm{fut}\cap T$ and $S_\mathrm{out}$ as the entropy in $J^-(W_\mathrm{fut})\cap T$ \cite{Wall}. The GSL then states that $S_\mathrm{gen}[T_2]\geq S_\mathrm{gen}[T_1]$ if $T_2$ is nowhere to the past of $T_1$.

The GSL can be supplemented by its time reversal, the anti-GSL, for which one defines a past horizon $H_\mathrm{past} = \partial J^+(W_\mathrm{past})$ with $W_\mathrm{past}$ a past-complete timelike curve. Then $A$ is the area of $H_\mathrm{past}\cap T$ for a spacelike hypersurface $T$, and $S_\mathrm{out}$ is the entropy in $J^+(W_\mathrm{past})\cap T$; the anti-GSL then states that $S_\mathrm{gen}[T_2] \leq S_\mathrm{gen}[T_1]$ if $T_1$ is everywhere to the past of $T_2$. The GSL and the anti-GSL can hold at the same time since they hold at different horizons, and they still lead to a thermodynamic arrow of time because the GSL only holds on future horizons. Furthermore, the anti-GSL should hold if the theory is to be CPT invariant.

\subsection{The background spacetime}\label{sec_AANEC_Background}

The last step to set the stage for a perturbative proof of the AANEC is to specify the classical background that the quantum fields will perturb. Following \cite{Wall}, we define it as a spacetime $M$ which satisfies the classical Einstein equations \eqref{H1_EFE} sourced by minimally coupled classical fields. As argued in section \ref{sec_classEC}, such fields can reasonably be assumed to satisfy the null energy condition. Furthermore, to prove the AANEC, $M$ is required to contain at least one complete achronal null geodesic $N$. This geodesic can be embedded in a causal horizon, namely an achronal null congruence $H\subseteq \partial J^+(N)\cap\partial J^-(N)$ with vanishing expansion, shear, and vorticity (i.e. $\theta = \sigma_{ab} = \omega_{ab} = 0$; see appendix \ref{sec_CongRay} for detailed definitions) \cite{Galloway_2000}. By the Raychaudhuri equation \eqref{AppA_RaychaudhuriExpansion}, these properties imply that $8\pi G_N T_{ab}l^al^b = R_{ab}l^al^b = 0$ where $l^a$ is the tangent vector to $N$. Finally, they imply that GSL is classically saturated: $A$ must be constant because $\theta$, which measures the relative change in the horizon area, vanishes along $N$, and any entropy flux must be supported by an energy flux which is impossible because $T_{ab}l^al^b$, which measures the energy density along $N$, vanishes everywhere along $N$. 

\subsection{A perturbative proof of the AANEC}

The proof of the AANEC in \cite{Wall} now starts by observing that the (anti-)GSL implies a restriction on the energy-momentum tensor of the matter fields: if $T_{ab}$ were completely free, one could violate the GSL by sending a negative energy flux across the future horizon, decreasing its area without necessarily lowering the entropy on the outside. This restriction becomes especially clear if one considers, following \cite{Wall}, a perturbative expansion around a classical spacetime $M$ as described in the previous paragraph. This expansion is done by restoring factors of $\hbar$ and then using them as the small expansion parameter. This $\hbar$-expansion can often be interpreted as an expansion in the number of loops in the Feynman diagrams that contribute to an expectation value \cite{hbar_expansion}, which necessitates finding a way of applying this expansion to geometric quantities. The method of choice is to quantise the geometry as well, and impose
\begin{align}
    \mean{R_{ab}} - \frac{1}{2}\mean{Rg_{ab}} = 8\pi G_N\mean{T_{ab}}\ ,\label{curveANEC_EhrenfestGR}
\end{align}
where it should be emphasised that \eqref{curveANEC_EhrenfestGR} does not describe semi-classical gravity but something reminiscent of the Ehrenfest theorem \cite{Griffiths_QM} for quantum gravity. Nevertheless, \eqref{H1_SemiclassEFE} and \eqref{curveANEC_EhrenfestGR} approximate the same theory and should therefore obey similar constraints, justifying the further use of \eqref{curveANEC_EhrenfestGR}. Given this quantised geometry, the $\hbar$-expansion can be denoted as
\begin{align}
    \mean{T_{ab}} \alis\mean{T_{ab}}_{(0)} + \mean{T_{ab}}_{(1)} + \order(\hbar^{3/2}) \ ,\\
    \mean{g_{ab}} \alis g_{ab(0)} + \mean{g_{ab}}_{(1/2)} + \mean{g_{ab}}_{(1)} + \order(\hbar^{3/2})\ ,
\end{align}
and similarly for other quantities; in particular, the area $A$ in \eqref{curveANEC_GeneralisedS} should be replaced by its expectation value to allow such an expansion \cite{Wall_entropy}. The orders of $\hbar$ can be interpreted as follows:

\begin{itemize}
    \item The classical background enters the expansion as the zeroth order. Thus, $g_{ab(0)}$ is the classical metric, supported by the classical energy-momentum tensor $\mean{T_{ab}}_{(0)}$, which obeys $\mean{T_{ab}}_{(0)}l^a_{(0)}l^b_{(0)} = 0$ (with $l^a_{(0)}$ the tangent vector of $N$) on $H$, the horizon in the classical spacetime.
    \item Gravitons perturb the metric at half order, i.e. $\mean{g_{ab}}_{(1/2)}$ describes fluctuations in the metric due to free gravitational waves. After all, a graviton with wavelength $\lambda$ has an energy of $\sim\hbar/\lambda$, corresponding to the squared metric perturbation. There is no contribution to $\mean{T_{ab}}$ at this order, since gravitons don't have a local energy-momentum tensor.
    \item The first perturbations due to the matter fields appear at first order, where they perturb the energy-momentum tensor by $\mean{T_{ab}}_{(1)}$ and source the metric perturbation $\mean{g_{ab}}_{(1)}$.
\end{itemize}

Contributions to the expansion of \eqref{curveANEC_EhrenfestGR} that are of order $\order(\hbar^{3/2})$ are ignored. To similarly expand the generalised entropy, one must first define the horizon in the perturbed spacetime. This can be done through two assumptions: first, that the horizon persists in the perturbed spacetime, and second, that all perturbations are sufficiently localised so that there exists an identification of points between the classical and perturbed spacetimes in which $\theta(H^\pm) = \order(\hbar^{3/2})$, with $H^+$ the asymptotic future of $H$ and $H^-$ its asymptotic past. The future and past horizons are then defined as $H_\mathrm{fut} = \partial J^-(H^+)$ and $H_\mathrm{past} = \partial J^+(H^-)$; clearly, $H_\mathrm{fut} = H_\mathrm{past} = H$ at zeroth order in $\hbar$, and they typically separate at first order in $\hbar$. We assume that the three horizons also coincide at half order.

The next step is to find the generalised entropy at any spacelike hypersurface $T$. This can be done by recognising that the horizon $H$ splits the classical spacetime $M$ in three regions: $P = J^-(H)$, $F = J^+(H)$, and potentially $O = M \setminus (P\cup F)$. Because the GSL is saturated by the classical setup, the relevant term in the expansion is the first quantum correction, which due to the factor $\hbar^{-1}$ in the area term is given by the zeroth order:
\begin{align}
    S_{\mathrm{fut}(0)}[T] = S_{\mathrm{out}(0)}[P\cap T] + \frac{\mean{A}_{(1)}}{4G_N\hbar}[H_\mathrm{fut}\cap T]\ , \label{curveANEC_GenEntPer}
\end{align}
where we suppressed the dependence on fields and states and the subscript `fut' emphasises that \eqref{curveANEC_GenEntPer} gives the generalised entropy that is relevant to the GSL, not the anti-GSL. Note that in principle, $P\cap T$ is defined using $H$ and $T$ in the classical spacetime, while $H_\mathrm{fut}\cap T$ is defined in the perturbed spacetime; since the difference is of order $\hbar$, it is irrelevant to $S_{\mathrm{fut}(0)}$.

\begin{figure}[!htb]
    \centering
    \includegraphics[width=\textwidth]{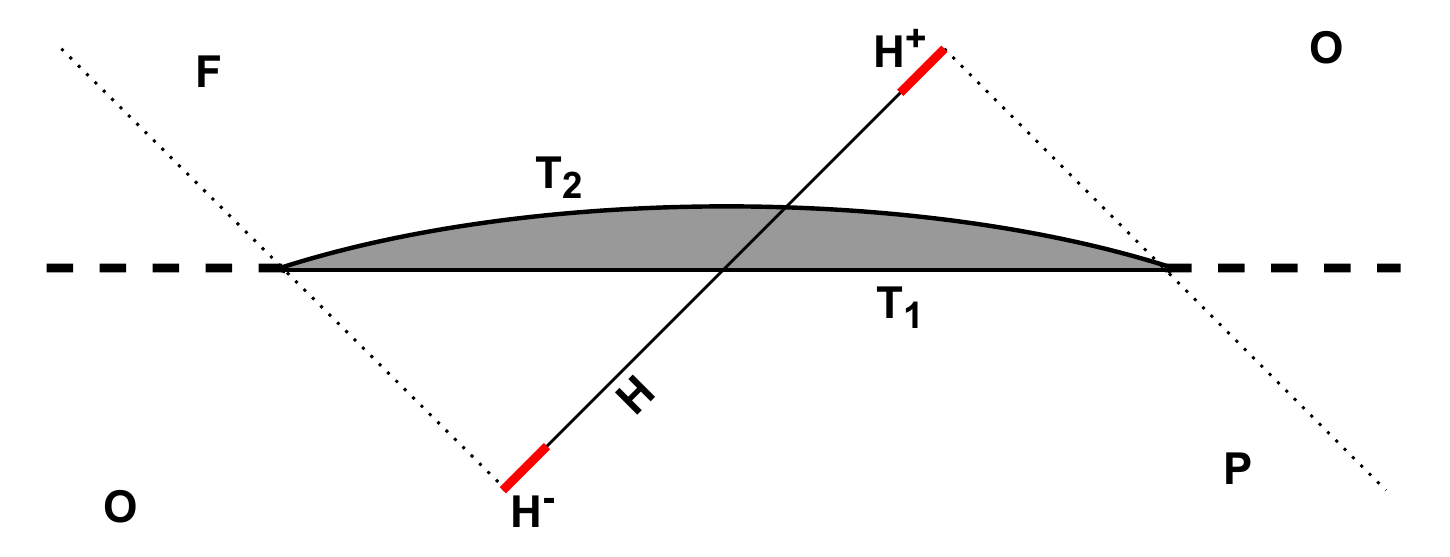}
    \caption{Penrose diagram of the classical spacetime, reproduced from \cite{Wall}. The horizon $H$ is drawn as a solid diagonal line, with its asymptotic past and future (respectively $H^-$ and $H^+$) represented as thicker red line segments. The dotted diagonal lines indicate $\partial J^+(H)$ and $\partial J^-(H)$ (on the left and on the right, respectively); the region above $\partial J^+(H)$ and $H$ is $F = J^+(H)$, and the region below $\partial J^-(H)$ and $H$ is $P = J^-(H)$. The spacelike hypersurfaces $T_1$ and $T_2$ are drawn as solid mostly horizontal lines; they coincide in the region $O$, which is the region outside both $F$ and $P$. In $O$, $T_1$ and $T_2$ are drawn as horizontal dashed lines. Finally, $\Delta T$ (the region between $T_1$ and $T_2$) is drawn as a shaded region.}
    \label{fig:wall_sketch}
\end{figure}

This allows us to state the GSL. Choose any two achronal spacelike hypersurfaces $T_1$ and $T_2$, with $T_1$ never to the future of $T_2$, and let them coincide in region $O$; the region between them is $\Delta T$. A sketch of this setup can be found in figure \ref{fig:wall_sketch}. If we define $\delta f[T] = f[T_2] - f[T_1]$, the GSL becomes
\begin{align}
    \delta S_{\mathrm{out}(0)}[P\cap T] + \frac{1}{4G_N\hbar}\delta\mean{A}_{(1)}[H_\mathrm{fut}\cap T] \geq 0\ . \label{curveANEC_PerturbedGSL}
\end{align}

By taking the time reversal of \eqref{curveANEC_PerturbedGSL}, the anti-GSL can be seen to yield
\begin{align}
    \delta S_{\mathrm{out}(0)}[F\cap T] + \frac{1}{4G_N\hbar}\delta\mean{A}_{(1)}[H_\mathrm{past}\cap T] \leq 0\ . \label{curveANEC_PerturbedAntiGSL}
\end{align}
From the difference between \eqref{curveANEC_PerturbedGSL} and \eqref{curveANEC_PerturbedAntiGSL}, it follows that
\begin{align}
    \frac{1}{4G_N\hbar}\roha{\delta\mean{A}_{(1)}[H_\mathrm{fut}\cap T] - \delta\mean{A}_{(1)}[H_\mathrm{past}\cap T]} \geq \delta S_{\mathrm{out}(0)}[F\cap T] - \delta S_{\mathrm{out}(0)}[P\cap T]\ . \label{curveANEC_GSLminAntiGLS}
\end{align}
The right-hand side of this inequality can be bounded by weak monotonicity \cite{Pippenger_2003}, which for three regions $R_1$, $R_2$, and $R_3$ states that the entropy $S_\mathrm{out}$ obeys
\begin{align}
    S_\mathrm{out}[R_1] + S_\mathrm{out}[R_2] \leq S_\mathrm{out}[R_1\cup R_3] + S_\mathrm{out}[R_2\cup R_3]\ . \label{curveANEC_WeakMonotonicity}
\end{align}
Here, we take $R_1 = F\cap T_1$, $R_2 = P\cap T_2$, and $R_3 = H\cap\Delta T$ and note that the evolution of a quantum state from $R_1 \cup R_3 = (T_1\cap F)\cup(H\cap\Delta T)$ to $T_2\cap F$ is unitary; this is because $F\cap T_1$ and $F\cap T_2$ are achronal hypersurfaces in $F = J^+(H)$ which coincide at $\partial J^+(H)\setminus H$, so $F\cap T_2\subset D^+(R_1\cup R_3)$ and $R_1\cup R_3 \subset D^-(F\cap T_2)$. See also figure \ref{fig:wall_sketch}. The evolution from $P\cap T_1$ to $(H\cap\Delta T)\cup(P\cap T_2) = R_3\cup R_2$ is unitary by the same logic. The entanglement entropy is invariant under such unitary evolution (this can be verified in \eqref{thANEC_EntanglementEntropy} by replacing $\rho^\psi\ln\rho^\psi \rightarrow U(\rho^\psi)U^\dagger U(\ln\rho^\psi) U^\dagger$ for a unitary $U$ and using the cyclic property of the trace), so $S_\mathrm{out}[R_1\cup R_3] = S_\mathrm{gen}[F\cap T_2]$ and $S_\mathrm{gen}[R_2\cup R_3] = S_\mathrm{gen}[P\cap T_1]$. Thus, \eqref{curveANEC_WeakMonotonicity} implies
\begin{align}
    S_{\mathrm{out}(0)}[F\cap T_1] + S_{\mathrm{out}(0)}[P\cap T_2] \leq&\ S_{\mathrm{out}(0)}[F\cap T_2] + S_{\mathrm{out}(0)}[P\cap T_1] \non
    \delta S_{\mathrm{out}(0)}[P\cap T] \leq&\ \delta S_{\mathrm{out}(0)}[F\cap T]\ , \label{curveANEC_WeakMonoImplied}
\end{align}
which demonstrates that the right-hand side of \eqref{curveANEC_GSLminAntiGLS}, and by the inequality the left-hand side as well, is non-negative. Thus, the area of the future horizon doesn't grow slower than that of the past horizon as one goes to $T$ that are further in the future. This can only be true if the expansion of the past horizon does not exceed that of the future horizon, so
\begin{align}
    \mean{\theta_\mathrm{fut}}_{(1)} - \mean{\theta_\mathrm{past}}_{(1)} \geq 0\ . \label{curveANEC_ExpansionInequality}
\end{align}
We can consider both quantities to be evaluated on $H$, since the difference between $H$ and $H_\mathrm{fut}$ ($H_\mathrm{past}$) would only affect $\mean{\theta_\mathrm{fut}}_{(1)}$ ($\mean{\theta_\mathrm{past}}_{(1)}$) at second order in $\hbar$. The evolution of the expansion parameters along this congruence is described by expectation value of the Raychaudhuri equation \eqref{AppA_RaychaudhuriExpansion}, which under the assumption that $\smallmean{\theta^2} + (d-2)\smallmean{\sigma_{ab}\sigma^{ab}} = \order(\hbar^{3/2})$ can be expanded to first order as:
\begin{align}
    \diff{\mean{\theta}_{(1)}}{\lambda} = -8\pi G_N\mean{T_{ab}}_{(1)}l^a_{(0)}l^b_{(0)}\ . \label{curveANEC_RaychaudhuriPerturbative1}
\end{align}
In principle, the tangent vectors on the right-hand side should be those of curves in $H_\mathrm{fut}$ and $H_\mathrm{past}$ for $\mean{\theta_\mathrm{fut}}_{(1)}$ and $\mean{\theta_\mathrm{past}}_{(1)}$ respectively, but as argued we only need the expansion parameters along a geodesic in $H$. Since we will primarily be interested in the expansion along $N$, the relevant tangent vector is $l^a_{(0)}$.

To finish the proof, we express \eqref{curveANEC_ExpansionInequality} at a point on $N$ with affine parameter $\lambda'$ in terms of integrals along $N$, fixing $\mean{\theta_\mathrm{fut}}_{(1)}(\lambda\rightarrow\infty) = \theta(H^+) = 0$ and $\mean{\theta_\mathrm{past}}_{(1)}(\lambda\rightarrow-\infty) = \theta(H^-) = 0$:
\begin{align}
    \mean{\theta_\mathrm{fut}}_{(1)} - \mean{\theta_\mathrm{past}}_{(1)} \alis -8\pi G_N\int_\infty^{\lambda'}\mean{T_{ab}}_{(1)}l^a_{(0)}l^b_{(0)}\d\lambda + 8\pi G_N \int_{-\infty}^{\lambda'}\mean{T_{ab}}_{(1)}l^a_{(0)}l^b_{(0)}\d\lambda \non
    \alis 8\pi G_N\int_N \mean{T_{ab}}_{(1)}l^a_{(0)}l^b_{(0)}\d\lambda\ . \label{curveANEC_AANECintegraal}
\end{align}
Since this must be non-negative by \eqref{curveANEC_ExpansionInequality}, the AANEC must hold at first order in $\hbar$.

Several issues have been glossed over in this derivation. The most pressing of these is the fact that the entanglement entropy usually diverges, necessitating a renormalisation procedure. Such a renormalisation procedure, \cite{Wall} argues, must exist to make the GSL a meaningful statement in the first place; furthermore, it should subtract the same divergence from all entropies in \eqref{curveANEC_WeakMonoImplied}. This means that the renormalisation does not invalidate the conclusion of \eqref{curveANEC_WeakMonoImplied}, and hence the derivation of the AANEC should continue to hold.

Another issue is that first-order contributions from $\smallmean{\sigma_{ab}\sigma^{ab}}$ have been neglected, while these could describe graviton fluctuations (the shear tensor $\sigma_{ab}$ describes the influence of distant matter on a geodesic \cite{shearfree}). If one assumes that these fluctuations do not separate $H_\mathrm{fut}$ from $H_\mathrm{past}$ at half order, their effect can be included while neglecting contributions from $\smallmean{\theta^2}$ (since $\smallmean{\theta^2} = \order(\hbar^2)$), and modify \eqref{curveANEC_RaychaudhuriPerturbative1} to
\begin{align}
    \diff{\mean{\theta}_{(1)}}{\lambda} = -8\pi G_N\mean{T_{ab}}_{(1)}l^a_{(0)}l^b_{(0)} - \smallmean{\sigma_{ab}\sigma^{ab}}_{(1)}\ . \label{curveANEC_RaychaudhuriPerturbative2}
\end{align}
Then, instead of \eqref{curveANEC_AANECintegraal}, the preceding calculation would show that
\begin{align}
    \int_N 8\pi G_N\mean{T_{ab}}_{(1)}l^a_{(0)}l^b_{(0)} + \smallmean{\sigma_{ab}\sigma^{ab}}_{(1)}\,\d\lambda \geq 0\ . \label{curveANEC_ShearInclusive}
\end{align}
This conjectured condition is referred to as the `shear-inclusive' AANEC in \cite{Wall}; it implies the usual AANEC, since one could imagine putting $n$ non-interacting copies of the fields in the same state. This replaces $\mean{T_{ab}}_{(1)}$ by $n\mean{T_{ab}}_{(1)}$ in \eqref{curveANEC_ShearInclusive}, with $\mean{T_{ab}}_{(1)}$ the energy-momentum tensor of a single copy. If one then chooses $n$ big enough to let $n\mean{T_{ab}}_{(1)}$ dominate over $\smallmean{\sigma_{ab}\sigma^{ab}}_{(1)}$, but small enough to keep $n\mean{T_{ab}}_{(1)}$ at first order in $\hbar$, it follows that the AANEC must also be satisfied.

\chapter{Extending the CANEC}\label{chap_BeyondCANEC}

In the previous chapter, we established the conformally invariant ANEC (CANEC) for holographic conformal field theories (CFTs) in $d=3$ spacetime dimensions. In $d=3,4,5$ spacetime dimensions, the CANEC took the general form \cite{CANEC_odd,CANEC_even}
\begin{align}
    \int_{\lambda_-}^{\lambda_+}\eta^d\smallmean{T_{ab}}_\psi l^al^b\d\lambda \geq B\ , \label{H5_CANECform}
\end{align}
for any state $\ket{\psi}$, where the integral is taken along a null geodesic with affine parameter $\lambda$, tangent vector $l^a$, and Jacobi field $\eta$, with caustics (where $\eta=0$) occurring at $\lambda = \lambda_\pm$. The lower bound $B$ vanishes for odd $d$ and is related to the conformal anomaly in even $d$.

In this chapter, we pursue several ways of generalising this statement. First, section \ref{sec_RicciFlat} studies an extension of the CANEC \eqref{curveANEC_CANEC} to holographic CFTs in two classes of spacetimes that were not covered by the proof in \ref{sec_CANEC}. Then, \ref{sec_CFTinConFlat} extends the CANEC to non-holographic CFTs on conformally flat spacetimes by Weyl transforming the ANEC. Finally, an attempt to prove the self-consistent, achronal ANEC (AANEC) based on the properties of a CANEC-like integral is presented in section \ref{sec_FailedProof}, and the reasons for the failure of this proof are discussed.

\section{Holographic CFTs in spacetimes with \texorpdfstring{$\hat{R}_{uu} = 0$}{Ruu = 0}}\label{sec_RicciFlat}
In section \ref{sec_CANEC}, we considered a $d+1$ dimensional asymptotically anti-de Sitter (AdS) spacetime $M$ with a $d$ dimensional boundary $\partial M$ that could (as described in section \ref{sec_CongRay}) be equipped with Gaussian normal coordinates $x^\mu = (u,v,x^I)$ with $I\in\cuha{2,d-1}$. By the AdS/CFT correspondence, the gravitational dynamics in $M$ can be described by a CFT in $\partial M$. For $d=3$, it was shown that this CFT should obey the CANEC \eqref{curveANEC_CANEC} \cite{CANEC_odd} by attempting to construct a bulk shortcut for $\gamma$, a boundary null geodesic with $u = \lambda$, $v = x^I = 0$. However, it was mentioned that the CANEC is trivially satisfied if $\hat{R}_{uu}$, the $uu$-component of the boundary Ricci tensor, is non-negative everywhere along $\gamma$. In this section, we aim to slightly loosen this constraint by considering two classes of spacetimes with $\hat{R}_{uu} = 0$: maximally symmetric spacetimes and a weak-field approximation to Schwarzschild spacetime. We will use a method from \cite{holographic_ANEC,Rosso_2020}, and slightly extend their results as they considered holographic CFTs on Minkowski and de Sitter (dS) spacetime.

\subsection{Holographic CFTs in maximally symmetric spacetimes}\label{sec_MaxSym}
For a $d$ dimensional spacetime, there can be at most $d(d+1)/2$ independent diffeomorphisms that leave its metric invariant, and if a metric actually is invariant under this maximum number of independent diffeomorphisms, the spacetime it is associated with is called `maximally symmetric' \cite{Carroll_2019}. The three maximally symmetric spacetimes with a metric with Lorentzian signature are Minkowski spacetime, AdS spacetime, and dS spacetime which is exactly like AdS spacetime but with $\Lambda > 0$. All three of these spacetimes are conformally related to the static Einstein universe, with metrics that can be written as
\begin{align}
    \d s^2 \alis \frac{L^2}{f_\Lambda^2(x)}\roha{-\d\tau^2 + \d\chi^2 + \sin^2\chi\,\d\Omega_{d-2}^2}\ , \label{H5_MaxSym} \\
    f_\Lambda^2(x) \alis \begin{cases}
        \cos^2\tau & \Lambda > 0 \\
        \cos^2\roha{\frac{\tau + \chi}{2}}\cos^2\roha{\frac{\tau - \chi}{2}} & \Lambda = 0 \\
        \cos^2\chi & \Lambda < 0
    \end{cases}\ , \label{H5_MaxSymFuncties}
\end{align}
where $-\frac{\pi}{2}<\tau<\frac{\pi}{2}$ and $0\leq\chi\leq\pi$ cover the entire manifold for $\Lambda > 0$, $-\pi + \chi < \tau < \pi - \chi$ and $0\leq\chi<\pi$ do so for $\Lambda = 0$, and the whole manifold is covered by $-\infty<\tau<\infty$ and $0\leq\chi<\frac{\pi}{2}$ for $\Lambda < 0$. Finally, $L$ is a length scale which is defined as $L^2 = (d-1)(d-2)/(2\abs{\Lambda})$ for $\Lambda\neq0$ and freely chosen for $\Lambda = 0$.

One can now place a quantum field theory on any of the maximally symmetric spacetimes. If this theory is a holographic CFT, it will give rise to a gravitational theory in a bulk spacetime $M$, making the maximally symmetric spacetime a boundary spacetime $\partial M$. As argued in section \ref{sec_AdS/CFT}, $M\cup\partial M$ should possess the no-bulk-shortcut property to ensure that the holographic CFT is sensible, and hence one may hope to apply the same techniques as in section \ref{sec_CANEC} to impose a CANEC on the boundary CFT. However, for a maximally symmetric spacetime with metric $\hat{g}^{(0)}_{ab}$, the Riemann tensor can be expressed as \cite{Carroll_2019}
\begin{align}
    \hat{R}_{abcd} = \frac{2\Lambda}{(d-2)(d-1)}\roha{\hat{g}_{ac}^{(0)}\hat{g}_{bd}^{(0)} - \hat{g}^{(0)}_{ad}\hat{g}^{(0)}_{bc}}\ , \label{H5_RiemannTensor}
\end{align}
which implies that any geometric $(0,2)$ tensor will be proportional to the metric. By defining $u = (\tau + \chi)/2$ and $v = (\tau - \chi)/2$ in \eqref{H5_MaxSym}, it becomes clear that in a maximally symmetric spacetime $\hat{g}_{uu}^{(0)} = \hat{g}^{(0)}_{vv} = 0$. This means that the argument as presented in \ref{sec_CANEC} will yield a condition which is trivially satisfied.

Therefore, we will instead follow the argument from \cite{holographic_ANEC} and \cite{Rosso_2020} (who made the argument in Minkowski and dS spacetime, respectively). The essence of this argument is to take a boundary null geodesic $\gamma$ and a particular causal bulk curve $\Gamma$, and then show that the no-bulk-shortcut property implies an energy condition in the limit that $\Gamma$ approaches $\gamma$. Our first step is therefore to claim that an inextendable null geodesic $\gamma$ can be described by the following coordinates (with $\mu\in\cuha{0,\ldots,d-1}$ and $u = (\tau + \chi)/2$ and $v = (\tau - \chi)/2$):
\begin{align}
    x^\mu = (u,v,\ve{x}^\perp) = \begin{cases}
        (0,\arctan\lambda,\ve{n}^\perp) & -\infty<\lambda\leq0 \\
        (\arctan\lambda,0,-\ve{n}^\perp) & 0<\lambda<\infty
    \end{cases}\ , \label{H5_MaxSymBoundaryCurve}
\end{align}
where $\ve{n}^\perp$ is a constant $d-2$ vector which indicates the direction of $\gamma$ on the $d-2$ dimensional sphere; its sign flips at $\lambda = 0$ because $\gamma$ passes through the origin there. It can be verified that \eqref{H5_MaxSymBoundaryCurve} describes a null geodesic by inserting it into \eqref{AppA_geodeet}; also note that $\gamma$ is inextendable since $\lambda\rightarrow-\infty$ corresponds to $\chi = -\tau = \frac{\pi}{2}$ and the point with $\lambda\rightarrow\infty$ is located at $\chi = \tau = \frac{\pi}{2}$. Both of these are points on the conformal edge of $\partial M$.

Our next step is to define a causal bulk curve $\Gamma$, which begins at a point on $\gamma$, moves into the bulk, and then returns to $\partial M$. We describe this path with coordinates $y^\alpha = (x^\mu,z)$, for $\alpha\in\cuha{0,\ldots,d}$, by setting \cite{Rosso_2020}
\begin{align}
    y^\alpha = (u,v,\ve{x}^\perp,z) = \begin{cases}
        \roha{f_+(\lambda),\arctan\lambda,\ve{n}^\perp,f_z(\lambda)} & -\lambda_0\leq\lambda\leq\lambda_1 \\
        \roha{\arctan\lambda,f_-(\lambda),-\ve{n}^\perp,f_z(\lambda)} & \lambda_1<\lambda\leq\lambda_0
    \end{cases}\ , \label{H5_MaxSymBulkCurve}
\end{align}
where $0<\lambda_0<\infty$ and $\lambda_1$ is the parameter at which $\chi = 0$; it therefore has $\abs{\lambda_1} < \lambda_0$. $f_\pm$ and $f_z$ are functions which are constrained by
\begin{align}
    f_+(-\lambda_0) = f_z(\pm\lambda_0) \alis 0\ , \label{H5_Voorwaarde1} \\
    f_+(\lambda_1) = f_-(\lambda_1) \alis \arctan\lambda_1\ , \label{H5_Voorwaarde2} \\
    f_-(\lambda_0) \geq&\ 0\ . \label{H5_Voorwaarde3}
\end{align}
Here, \eqref{H5_Voorwaarde1} expresses the requirement that $\Gamma$ starts on $\gamma$ and ends in $\partial M$, while \eqref{H5_Voorwaarde2} imposes continuity at $\lambda_1$; \eqref{H5_Voorwaarde3} is the statement of the no-bulk-shortcut property. We now pick particular functions for $f_z$ and $f_\pm$ to construct a $\Gamma$ which extends into the bulk perturbatively. We let $0<\varepsilon\ll1$ and choose
\begin{align}
    f_z(\lambda) \alis \begin{cases}
        \varepsilon L\roha{1 + \frac{\lambda}{\lambda_0}} & \lambda_0\leq\lambda\leq\lambda_1 \\
        \varepsilon L\roha{1 + \frac{2\lambda_1}{\lambda_0 - \lambda_1} - \frac{\lambda_0 + \lambda_1}{\lambda_0 - \lambda_1}\frac{\lambda}{\lambda_0}} & \lambda_1<\lambda\leq\lambda_0
    \end{cases}\ , \\
    f_+(\lambda) \alis \varepsilon^dL^{d-2}\int_{-\lambda_0}^\lambda p(\lambda)\hat{g}^{(d)}_{vv}\d\lambda + \varepsilon^{d+\delta}P(\lambda)\ , \label{H5_f+} \\
    f_-(\lambda) \alis \varepsilon^dL^{d-2}\int_{\lambda_1}^\lambda q(\lambda)\hat{g}^{(d)}_{uu}\d\lambda + \varepsilon^{d+\delta}Q(\lambda) + \arctan\lambda_1\ . \label{H5_f-}
\end{align}
Here, $p$ and $q$ are sufficiently smooth functions while $P$ and $Q$ obey $P(-\lambda_0) = Q(\lambda_1) = 0$; all four are of order 1, as opposed to $\delta$ for which $0<\delta\ll1$. The integrals in \eqref{H5_f+} and \eqref{H5_f-} are taken along $\gamma$, and finally, we fix $\lambda_0 = \varepsilon^{1-d}$, so that $\Gamma$ becomes inextendable for $\varepsilon\rightarrow0$. This also ensures that $f_z$ is of order $\varepsilon$, while $\dot{f}_z$ and $\dot{f}_\pm$ are of order $\varepsilon^d$.

Our third step will be to examine the consequences of the demand that $\Gamma$ is causal. To do this, we must calculate the norm of its tangent vector, which requires knowledge of the bulk metric. Since $\Gamma$ extends perturbatively into the bulk, this metric can be described using the Fefferman-Graham gauge and expansion from \eqref{CurveANEC_FGgauge} and \eqref{curveANEC_FGexpansie}. As a reminder, in this gauge the metric of an asymptotically AdS spacetime near the boundary can be expanded as \cite{FeffermanGraham}
\begin{align}
    \d s^2 = \frac{1}{z^2}\roha{\d z^2 + \hat{g}_{\mu\nu}\d x^\mu \d x^\nu} = \frac{1}{z^2}\roha{\d z^2 + \roha{\sum_{n=0}^\infty z^n\hat{g}^{(n)}_{\mu\nu} + z^d\ln(z^2)h_{\mu\nu}}\d x^\mu \d x^\nu}\ .
\end{align}
Here, the coefficients $\hat{g}^{(n)}_{\mu\nu}$ and $h_{\mu\nu}$ are functions of the boundary geometry described by $\hat{g}^{(0)}_{\mu\nu}$ for $n<d$, while $\hat{g}^{(d)}_{\mu\nu}$ is related to the energy-momentum tensor of the boundary CFT by \eqref{curveANEC_HolographicEMt} \cite{holographic_EMT}. However, as was pointed out based on \eqref{H5_RiemannTensor}, geometric $(0,2)$ tensors will be proportional to the metric in a maximally symmetric spacetime. Therefore, in our current setup, the Fefferman-Graham expansion simplifies to
\begin{align}
    \hat{g}_{\mu\nu} \alis \roha{\sum_{n=0}^{d-1} z^n m_n + z^d\ln(z^2)m_d}\hat{g}^{(0)}_{\mu\nu} + z^d\hat{g}^{(d)}_{\mu\nu} + \order(z^{d+1}) \non
    \equiv&\ m(z)\hat{g}^{(0)}_{\mu\nu} + z^d\hat{g}^{(d)}_{\mu\nu} + \order(z^{d+1})\ ,
\end{align}
where the $m_n$ are the proportionality constants between $\hat{g}^{(n)}_{\mu\nu}$ and $\hat{g}^{(0)}_{\mu\nu}$ \cite{Rosso_2020}. Let us now calculate the norm of the tangent vector of $\Gamma$ up to the first non-trivial order in $\varepsilon$ for $\lambda \leq \lambda_1$:
\begin{align}
    z^2g_{\alpha\beta}\dot{y}^\alpha\dot{y}^\beta \alis (\dot{f}_z)^2 - \frac{4L^2m(f_z(\lambda))}{f^2_\Lambda(\lambda)}\frac{\dot{f}_+(\lambda)}{1 + \lambda^2} + f_z^d(\lambda)\roha{\hat{g}^{(d)}_{uu}\dot{f}_+^2 + \hat{g}^{(d)}_{vv}\dot{v}^2 + 2\hat{g}^{(d)}_{uv}\frac{\dot{f}_+}{1+\lambda^2}} + \order(\varepsilon^{d+1}) \non
    \alis -4L^2\roha{\varepsilon^dL^{d-2}p(\lambda)\hat{g}^{(d)}_{vv} + \varepsilon^{d+\delta}\dot{P}} + \varepsilon^dL^d\hat{g}^{(d)}_{vv}\dot{v}^2 + \order(\varepsilon^{d+1}) \non
    \alis -4L^2\varepsilon^{d+\delta}\dot{P} + \varepsilon^dL^d\roha{\dot{v}^2 - 4p(\lambda)}\hat{g}^{(d)}_{vv} + \order(\varepsilon^{d+1})\ , \label{H5_BulkCausaal1}
\end{align}
where to go from the first to the second line, we expanded $f^2_\Lambda(\lambda)$ assuming that $\arctan{\lambda}\gg f_+(\lambda) \sim\order(\varepsilon^d)$ to cancel a factor of $(1+\lambda^2)^{-1}$; for very small $\abs{\lambda}$, both $f_\Lambda$ and $1 + \lambda^2$ tend to 1 and the combination still yields 1. We now choose $p(\lambda) = \dot{v}^2/4$; it follows that $\Gamma$ is causal if
\begin{align}
    z^2g_{\alpha\beta}\dot{y}^\alpha\dot{y}^\beta = -L^2\varepsilon^{d+\delta}\dot{P} + \order(\varepsilon^{d+1}) \leq& 0 \ .
\end{align}
This can only hold if at least $\dot{P}\geq0$, so $P$ cannot decrease as $\lambda$ increases. We can do essentially the same calculation for $\lambda>\lambda_1$, which if we choose $q(\lambda) = \dot{u}^2/4$ leads to $\dot{Q}\geq0$.

The penultimate step in this holographic calculation is to impose the no-bulk-shortcut property \eqref{H5_Voorwaarde3}, which together with \eqref{H5_Voorwaarde2}, \eqref{H5_f+}, and \eqref{H5_f-} leads to
\begin{align}
    f_-(\lambda_0) = \frac{\varepsilon^dL^{d-2}}{4}\int_{\lambda_1}^{\lambda_0}\dot{u}^2\hat{g}^{(d)}_{uu}\d\lambda + \varepsilon^{d+\delta}Q(\lambda_0) + \arctan\lambda_1 \geq&\ 0 \non
    \frac{\varepsilon^dL^{d-2}}{4}\int_{\lambda_1}^{\lambda_0}\dot{u}^2\hat{g}^{(d)}_{uu}\d\lambda + \frac{\varepsilon^dL^{d-2}}{4}\int_{-\lambda_0}^{\lambda_1} \dot{v}^2\hat{g}^{(d)}_{vv}\d\lambda + \varepsilon^{d+\delta}Q(\lambda_0) + \varepsilon^{d+\delta}P(\lambda_1) \geq&\ 0 \non
    L^{d-2}\int_{-\lambda_0}^{\lambda_0}\hat{g}^{(d)}_{ab}l^al^b\,\d\lambda + 4\varepsilon^\delta\roha{Q(\lambda_0) + P(\lambda_1)} + L^{d-2}\int_0^{\lambda_1}\roha{\dot{v}^2\hat{g}^{(d)}_{vv} - \dot{u}^2\hat{g}^{(d)}_{uu}}\d\lambda \geq&\ 0\ , \label{H5_MaxSymProtoCANEC}
\end{align}
where we have used the fact that the two integrals in the second line are evaluated along $\gamma$ to rewrite them as the first integral in \eqref{H5_MaxSymProtoCANEC}, which is taken along a segment of $\gamma$ and involves its tangent vector $l^a$. Because $\hat{g}^{(d)}_{ab}l^al^b$ jumps from $\hat{g}^{(d)}_{vv}\dot{v}^2$ to $\hat{g}^{(d)}_{uu}\dot{u}^2$ at $\lambda=0$, this rewriting required adding and subtracting a term, which introduced the last integral in \eqref{H5_MaxSymProtoCANEC}. However, since $\lambda_1 = \tan\roha{f_+(\lambda_1)} \sim \order(\varepsilon^d)$ this integral is negligible compared to the other terms in \eqref{H5_MaxSymProtoCANEC}.

To make the final step and interpret \eqref{H5_MaxSymProtoCANEC} as an energy condition, we therefore neglect its second integral and recall that since $\dot{P},\dot{Q}\geq0$, we must have $P(\lambda_1)\geq P(-\lambda_0) = 0$ and $Q(\lambda_0) \geq Q(\lambda_1) = 0$. We thus have three options for both $P(\lambda_1)$ and $Q(\lambda_0)$: they diverge to $+\infty$ faster than $\varepsilon^{-\delta}$ in the limit $\varepsilon\rightarrow0$, they diverge slower than $\varepsilon^{-\delta}$ in this limit, or they converge to some positive value in this limit. In the first case, \eqref{H5_MaxSymProtoCANEC} reduces to a trivial constraint on the first integral, but the second and third option lead, in the limit $\varepsilon\rightarrow0$, to a non-trivial constraint:
\begin{align}
    \int_{-\infty}^{\infty}\hat{g}^{(d)}_{ab}l^al^b\,\d\lambda \geq 0\ . \label{H5_MaxSymProtoCANEC2}
\end{align}
This can finally be interpreted as an energy condition by recalling \eqref{curveANEC_HolographicEMt}, which relates $\hat{g}^{(d)}_{ab}$ to $\mean{T_{ab}}_\psi$ for a holographic CFT, and allows us to rewrite \eqref{H5_MaxSymProtoCANEC2} as the ANEC:
\begin{align}
    \int_{-\infty}^\infty\mean{T_{ab}}_\psi l^al^b\d\lambda \geq 0\ .
\end{align}
Note the absence of the conformal anomaly $X_{ab}$; it is purely geometric and therefore proportional to the metric in a maximally symmetric spacetime, implying $X_{ab}l^al^b = 0$.

One may ask whether finding the ANEC is consistent with our expectations based on the CANEC. To see that this is the case, we define the following coordinates:
\begin{align}
    \frac{t\pm r}{2L} = \tan\roha{\frac{\tau\pm\chi}{2}}\ .
\end{align}
These coordinates are only well-defined for $-\pi+\chi<\tau<\pi-\chi$, but this is sufficient to include $\gamma$ completely. Applying this coordinate transformation to \eqref{H5_MaxSym}, we find that the maximally symmetric spacetimes are all conformally flat:
\begin{align}
    \d s^2 \alis \frac{1}{f_\Lambda^2(x)}\cos^2\roha{\frac{\tau + \chi}{2}}\cos^2\roha{\frac{\tau - \chi}{2}}\roha{-\d t^2 + \d r^2 + r^2\d\Omega_{d-2}^2} \non
    \equiv&\ \Omega^2(x)\roha{-\d t^2 + \d r^2 + r^2\d\Omega_{d-2}^2}\ ,
\end{align}
where we consider $\tau$ and $\chi$ as functions of $t$ and $r$. $\gamma$ can be described by $\chi = \abs{\tau}$ for $-\frac{\pi}{2}<\tau<\frac{\pi}{2}$, and it is straightforward to check using \eqref{H5_MaxSymFuncties} that this means that along $\gamma$, we have $\Omega^2 = 1$. This is relevant because, as explained in more detail in section \ref{sec_CFTinConFlat} or \ref{sec_Weyl}, a Weyl transformation is a local rescaling of the metric, for example $\eta_{ab} \rightarrow \Omega^2\eta_{ab}$, which maps null congruences with Jacobi field $\tilde{\eta}$ to null congruences with Jacobi field $\eta = \Omega\tilde{\eta}$. Thus, if $\gamma$ in Minkowski spacetime is part of a null plane (a congruence consisting of parallel null geodesics, for which $\tilde{\eta}$ is a constant), then the Jacobi field along $\gamma$ in a maximally symmetric spacetime would be $\eta = \Omega^2\tilde{\eta} = \tilde{\eta}$, a constant as well. We would therefore indeed expect the CANEC \eqref{H5_CANECform} to reduce to the ANEC in a maximally symmetric spacetime.

\subsection{Holographic CFTs in a non-conformally flat spacetime}\label{sec_HolographicCFTinSchwarz}
For a maximally symmetric boundary spacetime, the calculation from section \ref{sec_CANEC} does not yield an energy condition because the boundary Ricci tensor is proportional to $\hat{g}^{(0)}_{ab}$, meaning that $\hat{R}_{uu} = 0$. Another class of spacetimes in which this calculation does not yield an energy condition is formed by the Ricci flat spacetimes, which are defined by their vanishing Ricci tensor. A well-known non-trivial example of such a spacetime is Schwarzschild spacetime, described by \cite{Carroll_2019}
\begin{align}
    \d s^2 = -\roha{1 - \frac{2G_NM}{r}}\d t^2 + \roha{1 - \frac{2G_NM}{r}}^{-1}\d r^2 + r^2\d\Omega_{d-2}^2\ ,
\end{align}
which is the metric around a static point source with mass $M$, or outside a static spherically symmetric body with the same mass. Since this metric has a vanishing Ricci tensor wherever it applies, the calculation from section \ref{sec_CANEC} will yield an identically satisfied condition. However, one might hope that the techniques from \cite{holographic_ANEC,Rosso_2020} as applied in the previous subsection will lead to a non-trivial energy condition. The goal of this subsection is therefore to apply the procedure from section \ref{sec_MaxSym} in a spacetime with approximately the same geometry as Schwarzschild spacetime to derive an energy condition for holographic CFTs. That is, we construct an explicit causal bulk curve $\Gamma$ and demonstrate that the no-bulk-shortcut property implies a non-trivial condition in the limit that $\Gamma$ becomes a boundary null geodesic.

For simplicity, we do not study the exact $d$ dimensional Schwarzschild spacetime; rather, we consider a $d=4$ Minkowski spacetime perturbed by a static Newtonian source (i.e. pressureless dust producing a weak gravitational field). The metric describing this spacetime is \cite{Carroll_2019}
\begin{align}
    \d s^2 = -(1 + 2\Phi)\d t^2 + (1 - 2\Phi)\roha{\d x^2 + \d y^2 + \d z^2}\ , \label{H5_PerturbedMinkowski}
\end{align}
where $\Phi = \Phi(x,y,z)$ is the Newtonian potential obeying the Poisson equation $(\partial_x^2 + \partial_y^2 + \partial_z^2)\Phi = 4\pi G_N\rho$ ($\rho$ is the density of the Newtonian source) and the distance into the bulk will be denoted by $Z$ instead of the $z$ used in \eqref{CurveANEC_FGgauge}. We always work far away from the source, allowing us to set $(\partial_x^2 + \partial_y^2 + \partial_z^2)\Phi = 0$ and use $\Phi$ as a perturbative parameter. Finally, we assume that the Newtonian source responsible for $\Phi$ is localised around $x = y = z = 0$.

Let us now consider a holographic CFT living in a boundary spacetime $\partial M$ whose metric is given by \eqref{H5_PerturbedMinkowski}. Using the Fefferman-Graham gauge \eqref{CurveANEC_FGgauge} and the corresponding power series expansion \eqref{curveANEC_FGexpansie} \cite{FeffermanGraham}, the metric of the five-dimensional, asymptotically AdS bulk spacetime $M$ can be written as
\begin{align}
    \d s^2 = g_{\alpha\beta}\d y^\alpha\d y^\beta = \frac{1}{Z^2}\roha{\d Z^2 + \roha{\sum_{n=0}^\infty Z^n\hat{g}^{(n)}_{\mu\nu} + Z^4\ln(Z^2)h_{\mu\nu}}\d x^\mu\d x^\nu}\ , \label{H5_FGexpansie}
\end{align}
where $\hat{g}^{(0)}_{\mu\nu}$ can be read off from \eqref{H5_PerturbedMinkowski}. An important remark is that the spacetime described by \eqref{H5_PerturbedMinkowski} has $\hat{R}_{\mu\nu} = \order(\Phi^2)$. By \eqref{CurveANEC_G2}, it follows that $\hat{g}^{(2)}_{\mu\nu} = \order(\Phi^2)$, and since $h_{\mu\nu}$ is constructed from $\hat{g}^{(2)}_{\mu\nu}$ for $d=4$ \cite{holographic_EMT} we also have $h_{\mu\nu} = \order(\Phi^2)$.

\begin{figure}[!htb]
    \centering
    \includegraphics[width=\textwidth]{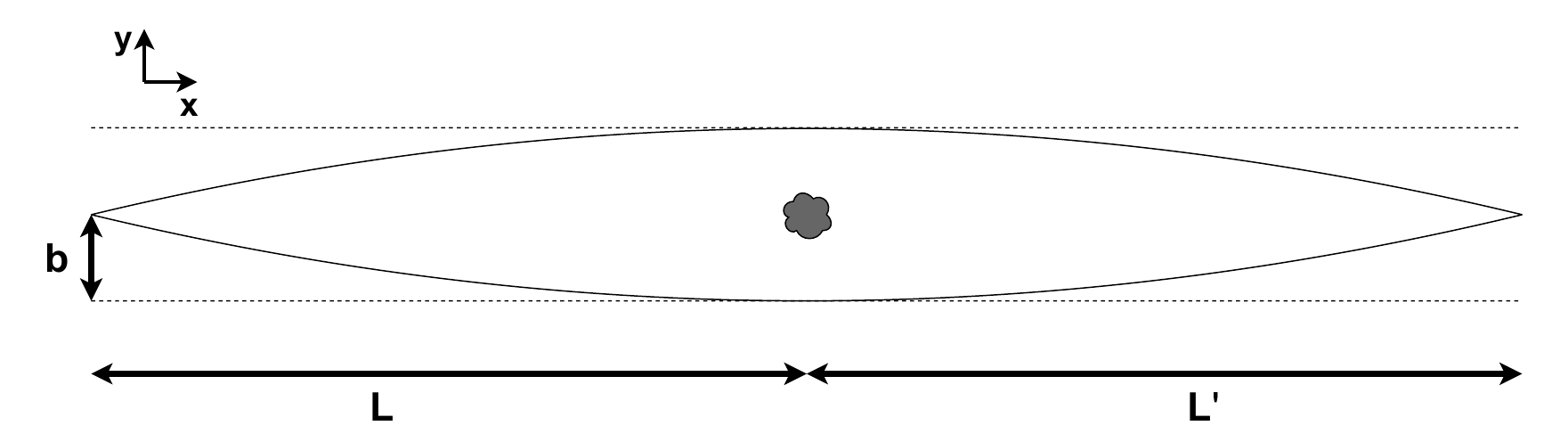}
    \caption{A sketch of the spatial trajectories of two null geodesics from the congruence in which $\gamma$ is embedded. The grey blob in the centre represents the Newtonian source creating $\Phi$, the dashed lines represent two unperturbed null geodesics (the upper one may for example represent $x^\mu_{(0)}$), and the solid curved lines represent two perturbed geodesics (again, the upper one may represent $x^\mu_{(0)} + x^\mu_{(1)}$). The bold arrows indicate three length scales, $L,L'\gg b$, all of which are assumed to be much larger than the size of the gravitating source.}
    \label{fig:SchwarzschildTraject}
\end{figure}

As in section \ref{sec_MaxSym}, our first step towards an energy condition will be to describe the relevant boundary null geodesic $\gamma$. Following \cite{Carroll_2019}, we do this perturbatively with $\Phi$ as the perturbative parameter. The coordinates along $\gamma$ are then\footnote{Although the notation is quite similar, there should be no confusion about what kind of expansion a perturbative quantity belongs to: the metric is only ever expanded in the sense of the Fefferman-Graham expansion, never in $\Phi$, and vice versa for other quantities.} $x^\mu(\lambda) = x^\mu_{(0)}(\lambda) + x^\mu_{(1)}(\lambda) + \order(\Phi^2)$, where $x^\mu_{(0)}$ describes a null geodesic in Minkowski spacetime and $x^\mu_{(1)}$ is a small correction due to $\Phi$, where `small' is meant in the sense that $x^i_{(1)}\partial_i\Phi \ll \Phi$. Our specific choice for $x^\mu_{(0)}$ is
\begin{align}
    x^\mu_{(0)}(\lambda) = \roha{t_{(0)},x_{(0)},y_{(0)},z_{(0)}} = \roha{\lambda,\lambda,b,0}\ . \label{H5_PerMinNuldeGeodeet}
\end{align}
We will not need explicit expressions for the first-order correction $x^\mu_{(1)}$, but we can describe its qualitative effect by considering the congruence in which we embed $\gamma$. The unperturbed congruence is the Minkowski null plane, described by choosing different constant values of $y_{(0)}$ and $z_{(0)}$ in \eqref{H5_PerMinNuldeGeodeet}. Once the perturbation is taken into account, the congruence in which we wish to embed $\gamma$ is sketched in figure \ref{fig:SchwarzschildTraject}. It is quite similar to the scenario of gravitational lensing: the null geodesics start at some point at a large distance $L$ from the gravitating source (to which we assign affine parameter $\lambda_-$ along $\gamma$), pass by the source at a large distance $b$, and are gravitationally deflected to refocus at a large distance $L'$ from the source (at a point to which we assign affine parameter $\lambda_+$ along $\gamma$). `Large' in this context is relative to the size of the source, and to justify the perturbative expansion of $x^\mu(\lambda)$ we assume that $L,L'\gg b$.

To simplify the subsequent discussion, we define $u = (t + x)/2$ and $v = (t - x)/2$. These are not lightcone coordinates; for example, the boundary spacetime metric \eqref{H5_PerturbedMinkowski} reads
\begin{align}
    \d s^2_{\partial M} = -2(\d u\d v + \d v\d u) + \d y^2 + \d z^2 - 2\Phi\roha{2\d u^2 + 2\d v^2 + \d y^2 + \d z^2}\ .
\end{align}
Nevertheless, when using them there are only two non-zero terms in $x^\mu_{(0)}$, namely $u_{(0)} = \lambda$ and $y_{(0)} = b$. Finally, the tangent vector of $\gamma$ can also be expanded:
\begin{align}
    l^\mu = l^\mu_{(0)} + l^\mu_{(1)} + \order(\Phi^2) \alis \roha{\dot{u}_{(0)},\dot{v}_{(0)},\dot{y}_{(0)},\dot{z}_{(0)}} + \roha{\dot{u}_{(1)},\dot{v}_{(1)},\dot{y}_{(1)},\dot{z}_{(1)}} + \order(\Phi^2) \non
    \alis \roha{1,0,0,0} + \roha{\dot{u}_{(1)},\dot{v}_{(1)},\dot{y}_{(1)},\dot{z}_{(1)}} + \order(\Phi^2)\ .
\end{align}
At first order in $\Phi$, the only constraint from $\hat{g}^{(0)}_{\mu\nu}l^\mu l^\nu = 0$ is that $\dot{v}_{(1)} = -\Phi$.

Having chosen a boundary geodesic, we now construct a bulk curve $\Gamma$. Analogously to \eqref{H5_MaxSymBulkCurve}, we describe the coordinates along $\Gamma$ as
\begin{align}
    y^\alpha = \roha{u,v,y,z,Z} = \roha{u_{(0)} + u_{(1)}, v_{(1)} + f_v(\lambda), b + y_{(1)},z_{(1)},f_Z(\lambda)} + \order(\Phi^2)\ ,
\end{align}
that is, we choose $\Gamma$ such that it shares its $u$-, $y$-, and $z$-coordinate with $\gamma$. Since we want $\Gamma$ to start at the same point as $\gamma$ and to end in the boundary, we demand $f_v(\lambda_-) = f_Z(\lambda_\pm) = 0$, where we remind the reader that $\gamma$ begins and ends at $\lambda_-$ and $\lambda_+$ respectively.

We proceed by choosing specific functions for $f_v$ and $f_Z$. For the latter, we recall that $\gamma$ begins and ends at a caustic, where neighbouring geodesics in the congruence intersect. Therefore, along $\gamma$ the Jacobi field satisfies $\eta(\lambda_\pm) = 0$, making it a useful ingredient for $f_Z$. It becomes even more appealing once we realise that the unperturbed boundary congruence is the null plane, with a constant Jacobi field. This implies that any variability in $\eta$ is due to $\Phi$, and hence $\dot{\eta}=\order(\Phi)$, which is a helpful property for reasons that will soon become apparent. Our choice of $f_v$ is analogous to \eqref{H5_f+}, so we set
\begin{align}
    f_v(\lambda) \alis \varepsilon^4\int_{\lambda_-}^\lambda p(\lambda)\hat{g}^{(4)}_{ab}l^al^b\,\d\lambda + \varepsilon^{4+\delta}P(\lambda)\ , \\
    f_Z(\lambda) \alis \varepsilon\eta^n(\lambda)\ ,
\end{align}
where we demand $P(\lambda_-) = 0$, we have $0<\varepsilon,\delta\ll1$ and $n > 0$, and the integral in $f_v$ is taken along $\gamma$. With these explicit forms of $f_v$ and $f_Z$, the norm of the tangent vector of $\Gamma$ can be expanded up to leading order in both $\varepsilon$ and $\Phi$ using \eqref{H5_FGexpansie}:
\begin{align}
    Z^2g_{\alpha\beta}\dot{y}^\alpha\dot{y}^\beta \alis (\dot{f}_Z)^2 + 2\hat{g}^{(0)}_{uv}l^u\dot{f}_v + \varepsilon^4\eta^{4n}\hat{g}^{(4)}_{\mu\nu}l^\mu l^\nu + \order(\Phi^2,\varepsilon^5) \non
    \alis n^2\varepsilon^2\eta^{2n-2}\dot{\eta}^2 - 4\roha{\dot{u}_{(0)} + \dot{u}_{(1)}}\roha{\varepsilon^4p\hat{g}^{(4)}_{\mu\nu}l^\mu l^\nu + \varepsilon^{4+\delta}\dot{P}} + \varepsilon^4\eta^{4n}\hat{g}^{(4)}_{\mu\nu}l^\mu l^\nu + \order(\Phi^2,\varepsilon^5) \non
    \alis \varepsilon^4\roha{\eta^{4n} - (\dot{u}_{(0)} + \dot{u}_{(1)})p}\hat{g}^{(4)}_{\mu\nu}l^\mu l^\nu - 4\varepsilon^{4+\delta}(\dot{u}_{(0)} + \dot{u}_{(1)})\dot{P} + \order(\Phi^2,\varepsilon^5)\ .
\end{align}
Here, we have made use of the observation following \eqref{H5_FGexpansie} that $\hat{g}^{(2)}_{\mu\nu},h_{\mu\nu}=\order(\Phi^2)$ and the facts that $\hat{g}^{(0)}_{\mu\nu}l^\mu l^\nu = 0$, $\dot{f}_v=\order(\varepsilon^4)$, and $\dot{\eta} = \order(\Phi)$. We can then fix $p$ as $p(\lambda) = (\dot{u}_{(0)} + \dot{u}_{(1)})^{-1}\eta^{4n}$, which is allowed because $\dot{u}_{(0)} + \dot{u}_{(1)} = 1 + \order(\Phi) > 0$. This implies that $\Gamma$ is a causal curve if
\begin{align}
    Z^2g_{\alpha\beta}\dot{y}^\alpha\dot{y}^\beta = -4\varepsilon^{4+\delta}(\dot{u}_{(0)} + \dot{u}_{(1)})\dot{P} + \order(\Phi^2,\varepsilon^5) \leq 0\ . \label{H5_NonFlatCausal}
\end{align}
If this inequality is to hold even in the limit $\varepsilon\rightarrow0$, we must have $\dot{P}\geq0$ (the prefactor of $\dot{P}$ plays no role, since $\dot{u}_{(0)} + \dot{u}_{(1)}$ is positive). 

As in section \ref{sec_MaxSym}, this property of $P$ plays an important role once we consider the no-bulk-shortcut property. Since $\gamma$ is prompt (it contains no caustics), it should be impossible to causally connect its starting point to the interior of the causal past of its endpoint. Therefore, a causal bulk curve starting at the same point as $\gamma$ should not be able to reach points with $v < v(\lambda_+)$ for $u\leq u(\lambda_+)$, where $v(\lambda_+)$ and $u(\lambda_+)$ are coordinates of the endpoint of $\gamma$. In particular, for $\Gamma$ we should demand that
\begin{align}
    f_v(\lambda_+) = \varepsilon^4\int_{\lambda_-}^{\lambda_+}\frac{\eta^{4n}}{\dot{u}_{(0)} + \dot{u}_{(1)}}\hat{g}^{(4)}_{ab}l^al^b\,\d\lambda + \varepsilon^{4+\delta}P(\lambda_+) \geq 0\ . \label{H5_NonFlatNoBulk}
\end{align}
Because we found that $\dot{P}\geq0$, we must have that $P(\lambda_+)\geq P(\lambda_-) = 0$. As argued at the end of section \ref{sec_MaxSym}, this leaves to options: either $P(\lambda_+)$ diverges to $+\infty$, meaning that \eqref{H5_NonFlatNoBulk} is trivially satisfied, or $P(\lambda_+)$ is finite, meaning that the $\varepsilon\rightarrow0$ limit of \eqref{H5_NonFlatNoBulk} implies that
\begin{align}
    \int_{\lambda_-}^{\lambda_+}\frac{\eta^{4n}}{\dot{u}_{(0)} + \dot{u}_{(1)}}\hat{g}^{(4)}_{ab}l^al^b\,\d\lambda \geq 0\ .
\end{align}
Recalling \eqref{curveANEC_HolographicEMt}, this leads us to an energy condition:
\begin{align}
    \int_{\lambda_-}^{\lambda_+}\frac{\eta^{4n}}{\dot{u}_{(0)} + \dot{u}_{(1)}}\roha{\mean{T_{ab}}_\psi - X_{ab}} l^al^b\,\d\lambda \geq 0 \ , \label{H5_NonFlatBijna}
\end{align}
where $X_{ab}$ is the conformal anomaly and $\ket{\psi}$ is any state of the holographic CFT. Our final step will be to ensure the Weyl invariance of this bound by fixing $n$. This can be done by considering a spacetime with metric $g'_{ab} = \Omega^2\hat{g}_{ab}^{(0)}$ for an appropriate function $\Omega$. Transforming $\mean{T_{ab}}_\psi - X_{ab}$ as in \eqref{thANEC_TransformEMT}, $\lambda$ as in \eqref{AppA_WeylAffine}, and $\eta$ as in \eqref{AppA_WeylJacobi}, we find that the left-hand side of \eqref{H5_NonFlatBijna} is related to a quantity in the geometry described by $g'_{ab} = \Omega^2\hat{g}_{ab}^{(4)}$ as
\begin{align}
    \int_{\lambda_-}^{\lambda_+}\frac{\eta^{4n}}{\diff{u_{(0)}}{\lambda} + \diff{u_{(1)}}{\lambda}} \roha{\smallmean{T_{ab}}_\psi - X_{ab}} l^al^b\,\d\lambda \alis \int_{\lambda'_-}^{\lambda'_+}\frac{\Omega^{2-4n}(\eta')^{4n}}{\diff{u_{(0)}}{\lambda'} + \diff{u_{(1)}}{\lambda'}}\roha{\smallmean{T'_{ab}}_{\psi'} - X'_{ab}} l'^al'^b\,\d\lambda'\ , \label{H5_NonFlatTransform}
\end{align}
where primes denote quantities defined in the spacetime with metric $g'_{ab}$. To ensure that the right-hand side of \eqref{H5_NonFlatTransform} does not explicitly depend on $\Omega$, we choose $2 - 4n = 0$. Thus, a Weyl invariant energy condition for holographic CFTs in the geometry described by \eqref{H5_PerturbedMinkowski} is
\begin{align}
    \int_{\lambda_-}^{\lambda_+}\frac{\eta^2}{\dot{u}_{(0)} + \dot{u}_{(1)}}\roha{\mean{T_{ab}}_\psi - X_{ab}} l^al^b\,\d\lambda \geq 0\ . \label{H5_NonFlatCANEC}
\end{align}
Here, the integral is taken along a geodesic which can be seen as a small perturbation of a null geodesic in Minkowski spacetime, between points where $\eta(\lambda_\pm) = 0$; $u$ is a coordinate which acts as an affine parameter for the Minkowski spacetime geodesic. Comparing \eqref{H5_NonFlatCANEC} to the general form \eqref{H5_CANECform} of the CANEC, we note that there exists a discrepancy between the two. A possible cause of this discrepancy is that \eqref{H5_CANECform} was derived for shear-free congruences, whereas the congruence to which \eqref{H5_NonFlatCANEC} applies has a non-vanishing shear which communicates the influence of the distant Newtonian perturbation to the null geodesics \cite{shearfree}. However, further testing of this hypothesis is beyond the scope of this thesis.

\section{CFTs in conformally flat spacetime}\label{sec_CFTinConFlat}
So far, we have only considered the CANEC for holographic CFTs, for which we can employ the no-bulk-shortcut property to derive an energy condition. However, there are many non-holographic CFTs, for which the corresponding gravitational theory is intractable, and one may ask whether these theories still obey something similar to the CANEC. To explore this question, we will apply a Weyl transformation to the Minkowski spacetime ANEC for a general CFT. We then argue that the resulting bound can be interpreted as the CANEC for a particular null congruence in a conformally flat spacetime.

Consider two spacetimes, $\tilde{M}$ equipped with metric $\tilde{g}_{ab}$ and $M$ equipped with metric $g_{ab}$, which are related by a Weyl transformation; tildes will be used to distinguish quantities and concepts in $\tilde{M}$ from those in $M$. The fact that $\tilde{M}$ and $M$ are related by a Weyl transformation means that wherever points in $\tilde{M}$ can be identified with points in $M$ by a diffeomorphism, their metrics are related as $g_{ab}(x) = \Omega^2(x)\tilde{g}_{ab}(x)$ for a non-negative function $\Omega$. As argued in section \ref{sec_Weyl}, $\Omega$ can only diverge at $\partial M$, the edge of $M$, and only vanish at $\partial \tilde{M}$, the edge of $\tilde{M}$, assuming that the Weyl transformation is invertible.

Now consider using $\tilde{M}$ and $M$ as backgrounds for quantum field theory (QFT) which would be a CFT if it were placed in Minkowski spacetime; we will simply refer to this theory as a CFT. The field operators in $\tilde{M}$ and $M$ must be related due to the Weyl invariance of the CFT \cite{WeylVsConformal}, and hence the Hilbert spaces $\tilde{\calH}$ and \calH\ should be related as well. Assuming that the Weyl transformation is invertible, we can follow \cite{Rosso_2020} and represent this relation with a unitary map $U:\tilde{\calH}\rightarrow\calH$. In particular, the energy-momentum tensors are related as in \eqref{AppB_CFTEMT}:
\begin{align}
    U\tilde{T}_{ab}U^\dagger = \Omega^{d-2}\roha{T_{ab} - X_{ab}}\ , \label{H5_emtTrans}
\end{align}
where the conformal anomaly $X_{ab}$ depends only on the geometry of spacetime and vanishes for odd $d$ \cite{Conforme_anomalie,holographic_EMT}; it arises due to counterterms that are introduced to make the action of the CFT finite in a general spacetime \cite{holographic_EMT,duff_anomaly}. The scaling with $\Omega$ can be derived classically for a theory with a Weyl invariant action $S$ by realising that a variation $\delta \tilde{g}^{ab}$ induces $\delta g^{ab} = \Omega^{-2}\delta \tilde{g}^{ab}$. Hence, the energy-momentum tensors in $M$ and $\tilde{M}$ are classically related as
\begin{align}
    T_{ab} = -\frac{2}{\sqrt{-g}}\funci{S}{g^{ab}} = -\Omega^{2-d}\frac{2}{\sqrt{-\tilde{g}}}\funci{S}{\tilde{g}^{ab}} = \Omega^{2-d}\tilde{T}_{ab}\ ,
\end{align}
confirming the non-anomalous part of \eqref{H5_emtTrans}. Under a Weyl transformation $\tilde{g}_{ab}\rightarrow g_{ab}$, null geodesics will be left invariant, although an affine parameter $\tilde{\lambda}$ is no longer affine. Instead, \eqref{AppA_WeylAffine} relates an affine parameter $\lambda$ to $\tilde{\lambda}$:
\begin{equation}
    \diff{\tilde{\lambda}}{\lambda} = \Omega^{-2}\ ,
\end{equation}
which also implies that the tangent vector $\tilde{l}^a$ to such a geodesic transforms to $l^a = \Omega^{-2}\tilde{l}^a$. Contracting both sides of \eqref{H5_emtTrans} with $\tilde{l}^a$ twice and integrating along the null geodesic $\gamma$ with $\tilde{l}^a$ as tangent vector leads to
\begin{equation}
    \int_\gamma U\tilde{T}_{ab}\tilde{l}^a\tilde{l}^bU^\dagger\d\tilde{\lambda} = \int_\gamma \Omega^d\roha{T_{ab} - X_{ab}}l^al^b\,\d\lambda\ . \label{H5_transformANEC}
\end{equation}
Note that this equality is only sensible if $\gamma\subset M$ after the transformation. For example, if $\tilde{M}$ is Minkowski spacetime and $M$ is the Poincaré patch of AdS spacetime, with line element $\d s^2 = (-\d t^2 + \d z^2 + \d \ve{x}^2)/z^2$ and therefore an edge at $z=0$, it is unreasonable to consider a null geodesic which crosses $z=0$ in $\tilde{M}$, as part of it is not mapped to $M$. In general, we require $\gamma$ not to intersect $\partial\tilde{M}$ or $\partial M$ except possibly at its conformal ends, where $\tilde{\lambda}\rightarrow\pm\infty$; except at these ends, we therefore require $\Omega$ to be positive and finite along $\gamma$. 

Let us now specialise to the case where $\tilde{M}$ is Minkowski spacetime and $\gamma$ a complete null geodesic, implying $0<\Omega<\infty$ everywhere along $\gamma$ except possibly at its conformal ends. Then $\tilde{T}_{ab}$ obeys the ANEC for any $\smallket{\tilde{\psi}} = U^\dagger\smallket{\psi}\in\tilde{\calH}$ with $\smallket{\psi}\in\calH$ and we recover a result from \cite{Rosso_2020}:
\begin{align}
    0 \leq \int_\gamma \smallbra{\tilde{\psi}}\tilde{T}_{ab}\tilde{l}^a\tilde{l}^b\smallket{\tilde{\psi}}\d\tilde{\lambda} \alis \int_\gamma\Omega^d\bra{\psi}T_{ab}l^al^b\ket{\psi} - \Omega^dX_{ab}l^al^b\,\d\lambda \ ,
\end{align}
Hence, the conformal anomaly provides a lower bound for a weighted average of the null-null component of the energy-momentum tensor:
\begin{align}
    \int_\gamma\Omega^d\smallmean{T_{ab}}_\psi l^al^b\d\lambda \geq \int_\gamma\Omega^dX_{ab}l^al^b\,\d\lambda\ . \label{H5_CANECconformFlat}
\end{align}
The similarity to \eqref{H5_CANECform} is already evident, including the property that the lower bound is geometric and vanishes in odd dimensions. To make this correspondence even more concrete, let us consider the Ricci tensor of $M$, given in \eqref{AppA_ConformalRicciT}:
\begin{align}
    R_{ab} \alis \tilde{R}_{ab} - (d-2)\roha{\tilde{\nabla}_a\tilde{\nabla}_b\ln\Omega - \tilde{\nabla}_a\ln\Omega\tilde{\nabla}_b\ln\Omega} \non
    &\qquad\ \qquad\ \qquad\ \qquad\ \qquad\ - \tilde{g}^{cd}\roha{\tilde{\nabla}_c\tilde{\nabla}_d\ln\Omega + (d-2)\tilde{\nabla}_c\ln\Omega\tilde{\nabla}_d\ln\Omega}\tilde{g}_{ab}\ , \label{H5_RicciTransform}
\end{align}
where $\tilde{\nabla}_a$ is the covariant derivative with regard to $\tilde{g}_{ab}$. We can contract \eqref{H5_RicciTransform} with $l^a$ twice, use that $\tilde{l}^a\tilde{\nabla}_a = \DIFF{}{\tilde{\lambda}}$, and note that $\d\tilde{\lambda} = \Omega^{-2}\d\lambda$ to obtain
\begin{align}
    R_{ab}l^al^b \alis \tilde{R}_{ab}l^al^b - \frac{d-2}{\Omega^4}\roha{\diff{^2\ln\Omega}{\tilde{\lambda}^2} - \roha{\diff{\ln\Omega}{\tilde{\lambda}}}^2} \non
    \alis \Omega^{-4}\tilde{R}_{ab}\tilde{l}^a\tilde{l}^b -\frac{d-2}{\Omega^4}\roha{\Omega^2\diff{}{\lambda}\roha{\Omega\diff{\Omega}{\lambda}} - \roha{\Omega\diff{\Omega}{\lambda}}^2} \non
    \alis \Omega^{-4}\tilde{R}_{ab}\tilde{l}^a\tilde{l}^b -\frac{d-2}{\Omega}\diff{^2\Omega}{\lambda^2} \ . \label{H5_RicciContractie}
\end{align}
Now $\tilde{R}_{ab} = 0$ since $\tilde{M}$ is Minkowski spacetime, and hence \eqref{H5_RicciContractie} may be recognised as the Raychaudhuri equation for a shear-free, irrotational null congruence if the relevant Jacobi field is $\eta = \alpha\Omega$ with constant $\alpha\inR$. Such a shear-free, irrotational congruence can always be constructed for $\gamma$ in Minkowski spacetime (any null geodesic can be embedded in a null plane), and a short calculation shows that this congruence will not gain shear or vorticity through a Weyl transformation \cite{CANEC_odd} (the general case is considered in section \ref{sec_Weyl}). To see this, consider the Jacobi field $\tilde{\eta}$ of a null congruence with $\sigma_{ab} = \omega_{ab} = 0$ in $\tilde{M}$. By \eqref{AppA_WeylJacobi}, the Jacobi field of the corresponding congruence in $M$ is $\eta = \Omega\tilde{\eta}$, so the Raychaudhuri equation \eqref{classEC_Raychaudhuri} states that
\begin{align}
    0 \alis \tilde{\eta}\tilde{R}_{ab}\tilde{l}^a\tilde{l}^b + (d-2)\diff{^2\tilde{\eta}}{\tilde{\lambda}^2} \non
    \alis \roha{R_{ab}l^al^b + \frac{d-2}{\Omega}\diff{^2\Omega}{\lambda^2}}\Omega^3\eta + (d-2)\Omega^2\diff{}{\lambda}\roha{\Omega^2\diff{\Omega^{-1}\eta}{\lambda}} \non
    \alis \roha{\eta R_{ab}l^al^b + (d-2)\diff{^2\eta}{\lambda^2}}\Omega^3\ , \label{H5_MappedRaychaudhuri}
\end{align}
where we applied \eqref{H5_RicciContractie} to go to the second line. Thus, as long as $\Omega > 0$ (which we assume to be true everywhere on $\tilde{M}$), \eqref{H5_MappedRaychaudhuri} shows that the mapped congruence is shear-free and irrotational as well. However, since $\eta = \Omega\tilde{\eta}$, the Weyl transformation may introduce caustics.

Therefore, by interpreting the ANEC to hold on null geodesics that are part of a null plane with constant Jacobi field $\tilde{\eta} = \eta_0>0$, the factor $\Omega^d$ in \eqref{H5_CANECconformFlat} may be identified as a power of the Jacobi field $\eta = \Omega\eta_0$ of the Weyl transformed congruence. Attempts to describe $\gamma$ with less explicit reference to Minkowski spacetime run into the problem that it will always be necessary to specify that $\gamma$ begins and ends at $\partial\tilde{M}$, so we make no such attempt. Nevertheless, one may identify \eqref{H5_CANECconformFlat} as a CANEC-type bound which holds for general CFTs in a conformally flat spacetime:
\begin{align}
    \int_\gamma \eta^d\smallmean{T_{ab}}_\psi l^al^b\d\lambda \geq \int_\gamma\eta^dX_{ab}l^al^b\d\lambda\ , \label{H5_CANECextended}
\end{align}
where $\gamma$ is a null geodesic which was complete in Minkowski spacetime, and $\eta$ is the Jacobi field of the Weyl transformed null plane. Note that the bound is not exactly Weyl invariant: different choices of $\Omega$ will typically lead to different conformal anomalies $X_{ab}$. Nevertheless, the form of the bound remains the same after a Weyl transformation, and therefore we will still refer to \eqref{H5_CANECextended} as the CANEC.

To finish this section, let us consider more closely to which conformally flat spacetimes and CFTs it applies. With regard to constraints on the spacetime, we have already remarked that $\Omega$ must be positive and finite everywhere except possibly on $\partial\tilde{M}$ and $\partial M$, where it may respectively vanish and diverge. Furthermore, to make sense of \eqref{H5_transformANEC} we require $\gamma\subset M$ after the Weyl transformation. This means that $\Omega$ must also be positive and finite along $\gamma$, except possibly at its conformal end points, where it may vanish or diverge.

On the other hand, the CFT is constrained by its behaviour under a Weyl transformation. For example, a primary scalar operator $\tilde{\Phi}$ transforms as \cite{WeylVsConformal}
\begin{align}
    \tilde{\Phi}(x) \rightarrow \Phi(x) = \Omega^{-\Delta_{\tilde{\Phi}}}(x)\tilde{\Phi}(x)\ , \label{H5_WeylTransformField}
\end{align}
for a Weyl transformation $\tilde{g}_{ab}\rightarrow g_{ab}$; $\Delta_{\tilde{\Phi}}$ is the scaling dimension of $\tilde{\Phi}$. For many physically relevant fields, $\Delta_{\tilde{\Phi}}$ is strictly positive in $d>2$ spacetime dimensions (e.g. $\Delta_{\tilde{\Phi}} = \frac{1}{2}d - 1$ for the free scalar field), meaning that $\Omega^{-\Delta_{\tilde{\Phi}}}$ diverges as $\Omega\rightarrow0$. This can only occur at $\partial\tilde{M}$, so if $\partial\tilde{M}$ intersects the interior of $M$, \eqref{H5_WeylTransformField} indicates that we must demand that $\tilde{\Phi}$ vanishes sufficiently quickly as one approaches $\partial\tilde{M}$ in order to have a smooth $\Phi$ on $M$. Similarly, $\Omega^{-\Delta_{\tilde{\Phi}}}$ vanishes when $\Omega\rightarrow\infty$, which may happen at $\partial M$; by \eqref{H5_WeylTransformField}, demanding that $\tilde{\Phi}$ be smooth on $\tilde{M}$ even when $\partial M$ intersects the interior of $\tilde{M}$ requires $\Phi$ to vanish sufficiently quickly at $\partial M$. Thus, to have a CFT with smooth fields, it must be possible to impose Dirichlet boundary conditions in both $\tilde{M}$ and $M$.

In summary, in this section we have considered a conformally flat spacetime $M$ with metric $g_{ab} = \Omega^2\eta_{ab}$, where $\Omega$ can only vanish at the conformal boundary of Minkowski spacetime and diverge at the conformal boundary of $M$; otherwise, it is positive and finite. A CFT in such a spacetime obeys \eqref{H5_CANECextended} if there exists a complete null geodesic $\gamma$ in Minkowski spacetime such that $\Omega$ is positive and finite along $\gamma$ except possibly at the conformal ends (i.e. when the affine parameter in Minkowski spacetime diverges) and we can impose Dirichlet boundary conditions on the CFT, both in $M$ and in Minkowski spacetime.

\section{A failed proof of the achronal ANEC}\label{sec_FailedProof}
Section \ref{sec_CFTinConFlat} has demonstrated that for CFTs in conformally flat spacetimes, the CANEC is a generalisation of the flat spacetime ANEC. One might wonder whether it is possible to push this programme further, and to that end this section will consider an attempt to derive the self-consistent, achronal ANEC (AANEC) from the properties of a CANEC-like integral, using a perturbative method analogous to the one employed in \cite{Wall} (or section \ref{sec_AANEC}). Although we eventually argue that the assumptions that are required to formulate the proof undermine its usefulness, the way in which this happens may illuminate some aspects of other perturbative proofs of the AANEC, in particular the proof from \cite{Wall}.

We will be concerned with a $d\geq3$ dimensional semi-classically self-consistent spacetime, i.e. one on which the matter fields are quantised and related to the classical geometry by the semi-classical Einstein equation \eqref{H1_SemiclassEFE} \cite{Rosenfeld_1963}. In this spacetime, we choose a null congruence with tangent vector field $l^a$, Jacobi field $\eta$, shear $\sigma_{ab}$, and vorticity $\omega_{ab}$; from this congruence, we select one null geodesic $\gamma$ with affine parameter $\lambda$. For brevity, we will denote $\sigma_{ab}\sigma^{ab} = \sigma^2$ and $\omega_{ab}\omega^{ab} = \omega^2$. Inspired by the shear-inclusive AANEC \eqref{curveANEC_ShearInclusive} conjectured in \cite{Wall}, we consider the `shear-inclusive CANEC-integral':
\begin{align}
    I = \int_{\lambda_-}^{\lambda_+}\eta^d\roha{8\pi G_N\mean{T_{ab}}_\psi l^al^b + \sigma^2}\d\lambda\ , \label{NewCANEC_scCANEC_tussen0}
\end{align}
where the integral is taken along $\gamma$, $\lambda_\pm$ are (for now) the affine parameters of two arbitrary points on the geodesic with $\lambda_+ > \lambda_-$, and $\ket{\psi}$ is any state of the quantum fields. \eqref{H1_SemiclassEFE} implies that $8\pi G_N\mean{T_{ab}}_\psi l^al^b = R_{ab}l^al^b$, and $R_{ab}l^al^b$ can in turn be expressed in terms of properties of the congruence through the Raychaudhuri equation \eqref{classEC_Raychaudhuri}, which continues to hold because the geometry is treated classically. By imposing \eqref{H1_SemiclassEFE} and \eqref{classEC_Raychaudhuri} on \eqref{NewCANEC_scCANEC_tussen0}, $I$ can be rewritten to
\begin{align}
    I \alis \int_{\lambda_-}^{\lambda_+}\eta^d\roha{\omega^2 + (d-2)(d-1)\roha{\frac{1}{\eta}\diff{\eta}{\lambda}}^2}\d\lambda - (d-2)\viha{\eta^{d-1}\diff{\eta}{\lambda}}^{\lambda_+}_{\lambda_-}\ , \label{NewCANEC_scCANEC_tussen1}
\end{align}
where we immediately integrated the second derivative of $\eta$ in \eqref{classEC_Raychaudhuri} by parts. Observe that the integral is now positive definite: $\eta^d\geq0$ because $\eta$ is non-negative by definition, section \ref{sec_CongRay} argued that $\omega^2\geq0$, and $\frac{1}{\eta}\diff{\eta}{\lambda}$ is real and therefore squares to a non-negative quantity. Hence, $I$ is bounded from below by the boundary terms:
\begin{align}
    I = \int_{\lambda_-}^{\lambda_+}\eta^d\roha{8\pi G_N\mean{T_{ab}}l^al^b + \sigma^2}\d\lambda \geq - (d-2)\viha{\eta^{d-1}\diff{\eta}{\lambda}}^{\lambda_+}_{\lambda_-}\ . \label{NewCANEC_scCANEC_exact}
\end{align}
This does not provide information beyond pure geometry and semi-classical gravity, merely stating that a congruence which is focused must have encountered non-negative $8\pi G_N\mean{T_{ab}}_\psi l^al^b + \sigma^2$, whereas the same quantity may have been negative if the congruence is defocused (a congruence may also defocus due to vorticity or initial conditions).

As in section \ref{sec_AANEC} and following \cite{Wall}, our attempted proof will be perturbative, using an $\hbar$-expansion of $\mean{T_{ab}}_\psi$. We remind the reader that in this expansion, one restores factors of $\hbar$, expands the expectation value in powers of $\hbar$, and sets $\hbar=1$ again; this can be interpreted as an expansion in Feynman diagrams organised by the number of loops, since every loop brings in a factor of $\hbar$ \cite{hbar_expansion}. We denote it as
\begin{align}
    \smallmean{T_{ab}}_\psi = \smallmean{T_{ab}}_{(0)} + \smallmean{T_{ab}}_{(1)} + \order(\hbar^2)\ , \label{NewCANEC_HbarExpansion}
\end{align}
where $\smallmean{T_{ab}}_{(0)}$ corresponds to tree-level or classical contributions to $\smallmean{T_{ab}}_\psi$, while $\smallmean{T_{ab}}_{(1)}$ represents the one-loop contributions, which are pure quantum effects. In contrast with section \ref{sec_AANEC}, we do not quantise any geometric objects; instead, we assume that the expansion in \eqref{NewCANEC_HbarExpansion} is perturbative (i.e. that each term is small compared to previous non-vanishing terms). Then \eqref{H1_SemiclassEFE} can be solved iteratively, leading to a similar expansion of $g_{ab}$, in which the order indicates the highest contributing order from the expansion of $\smallmean{T_{ab}}_\psi$: $g_{ab(n)}$ is sourced by all $\smallmean{T_{ab}}_{(m)}$ with $m\leq n$. Thus, we have obtained a perturbative expansion of the geometry, but do note that the underlying manifold $M$ is always the same (one may equivalently refer to a manifold $M_{(n)}$ equipped with a metric $g_{ab(0)} + \ldots+g_{ab(n)}$ as a perturbed manifold if points on $M_{(n)}$ and $M_{(m)}$ for $n\neq m$ can be identified).

Given this perturbative expansion of the metric, the geodesic equation $l^a\nabla_al^b = 0$ can be solved iteratively as well, leading to expansions for the geodesic $\gamma = \gamma_{(0)} + \gamma_{(1)} + \order(\hbar^2)$ and its tangent vector $l^a = l^a_{(0)} + l^a_{(1)} + \order(\hbar^2)$; $l^a_{(0)} + l^a_{(1)}$ is the tangent vector of the curve $\gamma_{(0)} + \gamma_{(1)}$, which is a geodesic in a geometry described by $g_{ab(0)} + g_{ab(1)}$. Similar to section \ref{sec_AANEC}, we assume that these expansions are perturbative as well, in the sense that $\gamma_{(n+1)}$ is a small shift (in appropriate units) in the curve $\gamma_{(0)} + \ldots + \gamma_{(n)}$ and that $l^a_{(n+1)}$ is a small change in the tangent vector $l^a_{(0)} + \ldots + l^a_{(n)}$. Justifying the assumption that the expansion of $\gamma$ is perturbative is a delicate matter to which we return in due time. With this assumption the expansions of $l^a$ and $g_{ab}$ define perturbative expansions for $\sigma_{ab}$ and $\omega_{ab}$, which in turn lead to an expansion for the Jacobi field $\eta$ through the Raychaudhuri equation \eqref{classEC_Raychaudhuri}. This expansion must be perturbative, since $\eta$ characterises the distance between neighbouring geodesics, and the geodesics are expanded perturbatively by assumption.

Having defined perturbative expansions for the relevant quantities, let us now be more specific about the class of states we will consider. Since we are attempting to prove the AANEC, we restrict to a background geometry (described by $g_{ab(0)}$ and called a background because it is sourced by the classical contribution $\mean{T_{ab}}_{(0)}$) which allows for a complete achronal null geodesic $\gamma_{(0)}$. This geodesic is part of a null congruence $K_0\subseteq M$ with $\sigma^2_{(0)} = \omega^2_{(0)} = \d\eta_{(0)}/\d\lambda = 0$ \cite{Galloway_2000}; we assume that $\gamma_{(0)} + \gamma_{(1)}\subset K_0$, although it will typically not be one of the geodesics with respect to $g_{ab(0)}$ that form $K_0$. By \eqref{H1_SemiclassEFE} and \eqref{classEC_Raychaudhuri}, the vanishing of $\sigma^2_{(0)}$, $\omega^2_{(0)}$, and $\d\eta_{(0)}/\d\lambda$ implies that $\mean{T_{ab}}_{(0)}l^a_{(0)}l^b_{(0)} = 0$ on $K_0$. Please note the similarity between this background and the one described in section \ref{sec_AANEC_Background}. Having defined our background, we expand $I$ as:
\begin{align}
    I \alis \int_{\lambda_-}^{\lambda_+}\eta^d_{(0)}\roha{16\pi G_N\mean{T_{ab}}_{(0)}l^a_{(0)}l^b_{(1)} + 8\pi G_N\mean{T_{ab}}_{(1)}l^a_{(0)}l^b_{(0)} + \sigma^2_{(1)}}\d\lambda + \order(\hbar^2)\ , \label{NewCANEC_scCANEC_Iexp}
\end{align}
where the integral is taken along $\gamma_{(0)}$, since the zeroth order contribution to the integrand vanishes on $\gamma_{(0)} + \gamma_{(1)}$. \eqref{NewCANEC_scCANEC_Iexp} can be simplified by distinguishing matter fields from a cosmological constant $\Lambda$, as $\mean{T_{ab}}_\psi = \smallmean{T^\mathrm{m}_{ab}}_\psi - \Lambda g_{ab}/(8\pi G_N)$. Then $\mean{T_{ab}}_{(0)}l^a_{(0)}l^b_{(0)} = 0$ implies that $\mean{T^\mathrm{m}_{ab}}_{(0)} = V_0g_{ab(0)}$ (since $l^a_{(0)}$ is null with regard to $g_{ab(0)}$), where $V_0$ is constant due to the covariant conservation of $\mean{T_{ab}}_{(0)}$ with respect to $g_{ab(0)}$. $V_0$ can be combined with $\Lambda$ by defining $\smallmean{T^\mathrm{m}_{ab}}'_\psi = \smallmean{T^\mathrm{m}_{ab}}_\psi - V_0g_{ab}$ and $\Lambda' = \Lambda - 8\pi G_NV_0$, so that $\smallmean{T^\mathrm{m}_{ab}}'_\psi - \Lambda'g_{ab}/(8\pi G_N) = \smallmean{T^\mathrm{m}_{ab}}_\psi - \Lambda g_{ab}/(8\pi G_N)$. It follows that $\smallmean{T^\mathrm{m}_{ab}}'_{(0)} = 0$, and since the difference between $\smallmean{T^\mathrm{m}_{ab}}_\psi$ and $\smallmean{T^\mathrm{m}_{ab}}'_\psi$ is only in the way the matter fields are split from the cosmological constant, we can drop the primes and write $\smallmean{T_{ab}}_{(0)} = -\Lambda g_{ab(0)}/(8\pi G_N)$. Substituting this into \eqref{NewCANEC_scCANEC_Iexp} leads to
\begin{align}
    I \alis \int_{\lambda_-}^{\lambda_+}\eta_{(0)}^d\roha{8\pi G_N\smallmean{T^\mathrm{m}_{ab}}_{(1)}l^a_{(0)}l^b_{(0)} - \Lambda\viha{2g_{ab(0)}l^a_{(0)}l^b_{(1)} + g_{ab(1)}l^a_{(0)}l^b_{(0)}} + \sigma^2_{(1)}}\d\lambda + \order(\hbar^2)\ ,
\end{align}
where we have also split $\mean{T_{ab}}_{(1)}$ into a contribution from matter fields and from $\Lambda$. The combination in square brackets can be recognised as the first order term in the expansion of $g_{ab}l^al^b = 0$ and therefore it vanishes, leaving us with the following expansion for $I$:
\begin{align}
    I \alis \int_{\lambda_-}^{\lambda_+}\eta_{(0)}^d\roha{8\pi G_N\smallmean{T^\mathrm{m}_{ab}}_{(1)}l^a_{(0)}l^b_{(0)} + \sigma^2_{(1)}}\d\lambda + \order(\hbar^2)\ .
\end{align}
According to \eqref{NewCANEC_scCANEC_exact}, $I$ is bounded below by boundary terms involving $\eta^{d-1}\d\eta/\d\lambda$. Due to the presence of the derivative, we already know that there will be no zeroth order contribution to this quantity, meaning that the leading order is evaluated on $\gamma_{(0)}$; in fact, we find 
\begin{align}
    -(d-2)\viha{\eta^{d-1}\diff{\eta}{\lambda}}^{\lambda_+}_{\lambda_-} = -(d-2)\eta_{(0)}^{d-1}\viha{\diff{\eta_{(1)}}{\lambda}}^{\lambda_+}_{\lambda_-} + \order(\hbar^2)\ , \label{NewCANEC_scCANEC_bound}
\end{align}
where the evaluation of the term in square brackets occurs at the indicated points on $\gamma_{(0)}$. According to \eqref{NewCANEC_scCANEC_exact}, this serves as a lower bound to $I$; for that to be true, the leading order of $I$ must be bounded from below by the leading order in \eqref{NewCANEC_scCANEC_bound}. Hence, we find that
\begin{align}
    \int_{\lambda_-}^{\lambda_+}\roha{8\pi G_N\smallmean{T^\mathrm{m}_{ab}}_{(1)}l^a_{(0)}l^b_{(0)} + \sigma^2_{(1)}}\d\lambda \geq -(d-2)\eta_{(0)}^{-1}\viha{\diff{\eta_{(1)}}{\lambda}}^{\lambda_+}_{\lambda_-}\ ,
\end{align}
in which we were allowed to move $\eta_{(0)}^d$ out of the integral since it is a positive constant. Since $\gamma_{(0)}$ is complete, we can take the limit $\lambda_\pm\rightarrow\pm\infty$. Our final assumption, then, is that in this limit $\eta_{(1)}$ tends to a constant. This can be justified by once again appealing to our assumption that the expansion of $\gamma$ is perturbative, combined with the assumption that any physical system is confined to some finite region of spacetime, so that its influence may be disregarded altogether in the asymptotic past and future. We thus arrive at \eqref{curveANEC_ShearInclusive}, the perturbative statement of the shear-inclusive AANEC \cite{Wall}:
\begin{align}
    \int_{\gamma_{(0)}}\roha{8\pi G_N\smallmean{T^\mathrm{m}_{ab}}_{(1)}l^a_{(0)}l^b_{(0)} + \sigma^2_{(1)}}\d\lambda \geq 0\ . \label{NewCANEC_scCANEC_eind}
\end{align}
The derivation as presented above is very remarkable indeed: based on pure geometry and semi-classical gravity, we seem to be able to formulate a general property of quantum fields, a claim which warrants considerable scrutiny. We will discuss two potential loopholes in our argument, namely the restrictiveness of the assumed background, and the validity of the assumptions required to define our perturbative framework.

Our first concern is with the restrictions imposed on the background geometry: we demanded that it should allow for a complete achronal null geodesic. This is a very restrictive property, but it is only relevant to the background; there are still many states allowed on top of this background, as argued in \cite{Wall}. We should also note that the maximally symmetric spacetimes are all allowed as background geometries, which are of great practical concern since many realistic systems can be described as quantum fields on a maximally symmetric spacetime (e.g. Minkowski spacetime for particle physics, or dS spacetime for cosmology).

Our second concern is that the assumptions which we require to formulate the perturbative expansions are too restrictive. The relevant assumptions are that the $\hbar$-expansion of $\mean{T_{ab}}_\psi$ is perturbative, and that this leads to a perturbative expansion of $\gamma$. Other assumptions, such as the existence of a perturbative expansion for the Jacobi field and the asymptotic vanishing of $\d\eta_{(1)}/\d\lambda$ along $\gamma_{(0)}$, can be justified based on these assumptions. Unfortunately, the second assumption (that $\gamma$ can be expanded perturbatively around $\gamma_{(0)}$) excludes the possibility that quantum effects completely defocus a null congruence, since in that case one of the `corrections' $\gamma_{(n\geq1)}$ must diverge at some point. Quantitatively, we can see that the perturbative expansion breaks down for a defocused congruence when $\eta_{(n\geq1)}\sim\eta_{(0)}$.

Excluding quantum effects that defocus null congruences is problematic because it defeats the purpose of energy conditions. Normally, the idea is that we use an energy condition to impose restrictions on the geometry, whereas the argument we have presented here goes backwards: given a condition on the geometry, we derive an energy condition. We are therefore unable to say whether states that result in the defocusing of a null congruence are physically acceptable or not, which is the entire purpose of an energy condition. Interestingly, the assumption that quantum effects shift the background geodesic perturbatively also plays an important role in the calculation of \cite{Wall} (or section \ref{sec_AANEC}). However, it enters into that calculation somewhat differently: it is assumed that the null geodesic $H_\mathrm{past}$ deviates perturbatively from the background geodesic $H$ to the past of some point $p$, and that another null geodesic $H_\mathrm{fut}$ deviates perturbatively from $H$ to the future of $p$. What $H_\mathrm{past}$ ($H_\mathrm{fut}$) does to the future (past) of $p$ remains unspecified, but it is also irrelevant, because we are ultimately interested in quantities that are integrated along $H_\mathrm{past}$ to the past of $p$ and along $H_\mathrm{fut}$ to the future of $p$, as can be seen in \eqref{curveANEC_AANECintegraal}. 

In summary, this chapter has argued that the CANEC is more broadly applicable than its original holographic derivation \cite{CANEC_odd,CANEC_even} may suggest. In particular, we have found that it (or a related condition) holds for holographic CFTs on spacetimes for which the method of section \ref{sec_CANEC} yields a trivial constraint and that it holds for general CFTs when considering the Weyl transformed null plane in conformally flat spacetimes. Finally, this section showed that a self-consistent, shear-inclusive CANEC integral \eqref{NewCANEC_scCANEC_tussen0} seems to lead to the perturbative shear-inclusive AANEC \eqref{curveANEC_ShearInclusive}, although we argued that this was due to the initial assumptions. To further cement the CANEC as a viable generalisation of the ANEC, the next chapter will try to formulate it in Minkowski spacetime for a null congruence other than the null plane.
\chapter{Light-ray operators}\label{chap_LO}

In the preceding chapter, we have demonstrated that the Weyl transformation of the Minkowski spacetime averaged null energy condition (ANEC) leads to the conformally invariant ANEC (CANEC) for a general CFT in a conformally flat spacetime. This required identifying the ANEC in Minkowski spacetime as the CANEC applied to the null plane. However, there are many null congruences in Minkowski spacetime for which the Jacobi field is not constant, and the question naturally arises whether the CANEC also holds on these congruences.

To get an idea of what the CANEC would look like on these other congruences, let us consider an important null congruence with non-constant $\eta$ in Minkowski spacetime: the lightcone. The future lightcone $\partial J^+(p)$ of any point $p$ can be used to define a congruence $\partial J^+(\Gamma)$, the union of the future lightcones of points on a timelike curve $\Gamma$. To describe this congruence more explicitly, we equip Minkowski spacetime with Cartesian coordinates and define the congruence as the union of the future lightcones of the curve $\ve{x} = \ve{0}$; the congruence thus consists of future-directed null geodesics emanating radially from the spatial origin. Any geodesic $\gamma$ from this congruence can be described by picking a point $p$ on $\gamma$ with coordinates $x^\mu = (t_0,\ve{x}_0)$ and constructing a null geodesic which passes through both $p$ and the spatial origin. If we choose the affine parameter $\lambda$ such that $\abs{\ve{x}(\lambda)} = \lambda$ (recall that we consider null geodesics starting at $\ve{x} = \ve{0}$, so they are incomplete), the tangent vector of $\gamma$ has components
\begin{align}
    l^\mu = \roha{1 , \ve{x}_0/\abs{\ve{x}_0}}\ . \label{LO_NullConeTangent}
\end{align}
We can extend this vector to the tangent vector field of the entire congruence by dropping the subscript 0. With this vector field, the expansion of the congruence follows from \eqref{AppA_Exp}:
\begin{align}
    \frac{d-2}{\eta}\diff{\eta}{\lambda} = \theta = \partial_\mu l^\mu = \frac{d-2}{\abs{\ve{x}}}\ , \label{LO_NullConeExpansion}
\end{align}
where we also recalled the definition of $\eta$ from \eqref{AppA_JacobiDef}. \eqref{LO_NullConeTangent} is based on choosing an affine parameter such that $\abs{\ve{x}(\lambda)} = \lambda$ along an arbitrary geodesic. Substituting this into \eqref{LO_NullConeExpansion} and solving the resulting differential equation for $\eta$ yields $\eta(\lambda) = \lambda$.

However, this future lightcone congruence consists of incomplete geodesics. To complete them, we consider the past lightcone congruence, constructed analogously to the congruence of future lightcones but using future-directed null geodesics which converge towards the line $\ve{x} = \ve{0}$ instead of emanating from it (i.e. it is $\partial J^-(\Gamma)$ for the same choice of $\Gamma$ as before). As a consequence, the affine parameter can be taken to obey $\abs{\ve{x}} = -\lambda$, switching the sign of the spatial components in \eqref{LO_NullConeTangent}, reversing the sign of $\theta$, and thereby implying $\eta(\lambda) = -\lambda$. By considering a complete null geodesic to be composed of a geodesic from the past lightcone and one from the future lightcone, joined at the spatial origin, we find $\eta(\lambda) = \abs{\lambda}$ for null geodesics embedded in a lightcone. The CANEC, formulated on a lightcone, would therefore claim that
\begin{align}
    \int_{-\infty}^\infty\abs{\lambda}^{d}\mean{T_{ab}}_\psi l^al^b\,\d\lambda\geq0\ ,\label{LO_NullConeCANEC}
\end{align}
for any state $\ket{\psi}$. The operator that is allegedly bounded in \eqref{LO_NullConeCANEC} bears a remarkable similarity to the light-ray operators that were introduced in \cite{LO_original} and further studied in e.g. \cite{LO_Mathys}:
\begin{align}
    L_s = \int_{-\infty}^\infty\lambda^{s+2}T_{ab}l^al^b\,\d\lambda\ , \label{LO_def}
\end{align}
with $s\in\mathbb{Z}$. In particular, the CANEC on a lightcone would indicate that for even $d$, $\mean{L_{d-2}}_\psi\geq0$. Inspired by this similarity, we study the even light-ray operators $L_{2n}$ with $n\geq-1$, using the non-minimally coupled but otherwise free real scalar field $\phi$ as a toy model. We use this particular theory because, as discussed in section \ref{sec_ConformalFields}, it is a CFT for $m=0$ and conformal coupling $\xi = \frac{1}{4}(d-2)/(d-1)$ and a regular quantum field theory otherwise.

In section \ref{sec_NonMinimalScalar}, we review some important aspects of (the quantisation of) the non-minimally coupled free real scalar field. We use this background in section \ref{sec_NonMinimalLO} to prove that the light-ray operators obeying \eqref{LO_NonMinConstraint} are positive-semidefinite (in the sense that $\mean{L_{2n}}_\psi\geq0$ for any state $\ket{\psi}$), including $L_{d-2}$ for even $d$ and sufficiently small non-minimal coupling. However, \ref{sec_OtherCongruence} demonstrates via an explicit example that the CANEC cannot apply to all congruences. 

\section{Quantising the non-minimally coupled scalar field}\label{sec_NonMinimalScalar}
In this section, we discuss the quantisation of the non-minimally coupled free real scalar field in $d$ dimensions. Although the quantisation of the field operators themselves is also discussed in section \ref{sec_CanonicalQuantisation}, this section goes slightly further by also considering the quantisation of the null-null component of the energy-momentum tensor that is involved in the light-ray operators.

In $d$ dimensions, the classical action of the non-minimally coupled but otherwise free real scalar field $\phi$ with non-minimal coupling $\xi$ is
\begin{align} 
    S = -\frac{1}{2}\int\d^dx\,\sqrt{-g}\roha{\nabla_a\phi\nabla^a\phi + (m^2 + \xi R)\phi^2}\ , \label{LO_NonMinAction}
\end{align}
where $m$ is the mass of the field. The classical energy-momentum tensor of this theory can be found by varying $S$ with respect to the metric, leading to \cite{NonMinimalScalar}
\begin{align}
    T_{ab} = \nabla_a\phi\nabla_b\phi - \frac{1}{2} \roha{\nabla_c\phi\nabla^c\phi + m^2\phi^2}g_{ab} + \xi\roha{R_{ab} - \frac{1}{2}g_{ab}R + g_{ab}\nabla^c\nabla_c - \nabla_a\nabla_b}\phi^2\ .
\end{align}
Now we choose an arbitrary null geodesic and parameterise it with affine parameter $\lambda$ such that it has tangent vector $l^a$. Using that $l^a\nabla_a = \DIFF{}{\lambda}$, which reduces to the usual directional derivative when acting on scalars, we find that
\begin{align}
    T_{ab}l^al^b = \roha{\diff{\phi}{\lambda}}^2 - \xi\diff{^2\phi^2}{\lambda^2} + \xi\phi^2R_{ab}l^al^b\ . \label{LO_NullNullEMT}
\end{align}
The equations of motion for $\phi$ can be found by varying $S$ with respect to $\phi$, leading to
\begin{align}
    \roha{\nabla_a\nabla^a - m^2 - \xi R}\phi = 0\ . \label{LO_EoMscalar}
\end{align}
In Minkowski spacetime, $R_{ab} = 0 = R$, simplifying both \eqref{LO_NullNullEMT} and \eqref{LO_EoMscalar}: in the former, the last term drops out, while in the latter, it becomes clear that a solution is provided by plane waves $e^{\pm ip\cdot x}$ if $p\cdot p = -m^2$. Using these plane wave modes, the field can be quantised as
\begin{align}
    \hat{\phi}(x) = \int\frac{\d^{d-1}\ve{p}}{(2\pi)^{d-1}}\frac{1}{\sqrt{2\omega_\ve{p}}}\roha{\hat{a}_\ve{p}e^{ip\cdot x} + \hat{a}^\dagger_\ve{p} e^{-ip\cdot x}}\ ,
\end{align}
where we have chosen a Cartesian coordinate system $(t,\ve{x})$ in which the components of the momentum are $p^\mu = (\omega_\ve{p},\ve{p})$ and we normalised the plane wave modes with respect to the hypersurface $t=0$ using \eqref{AppB_KGinprod}. We will be interested in the operator $T_{ab}l^al^b$, which by inspection of \eqref{LO_NullNullEMT} means that the (vacuum subtracted) two-point function is an important quantity. It can be evaluated to be
\begin{align}
    \normal{\hat{\phi}(x)\hat{\phi}(y)}\ \alis \hat{\phi}(x)\hat{\phi}(y) - \bra{0}\hat{\phi}(x)\hat{\phi}(y)\ket{0} \non
    \alis \int\frac{\d^{d-1}\ve{p}}{(2\pi)^{d-1}}\frac{\d^{d-1}\ve{k}}{(2\pi)^{d-1}}\frac{1}{\sqrt{4\omega_\ve{p}\omega_\ve{k}}}\roha{e^{-i(p\cdot x - k\cdot y)}\hat{a}^\dagger_\ve{p}\hat{a}_\ve{k} + e^{i(p\cdot x + k\cdot y)}\hat{a}_\ve{p}\hat{a}_\ve{k}} + \mathrm{h.c.}\ ,
\end{align}
where `h.c.' stands for hermitian conjugate and $\ket{0}$ is the Minkowski vacuum state. Let us consider this quantity on a null geodesic, parametrised (without loss of generality) as $x^+ = \lambda$, $x^- = 0$, and $\ve{x}^\perp = \ve{0}^\perp$; along this null geodesic, we have $p\cdot x(\lambda) = p_+\lambda$. We can then change variables from $p^1$ and $k^1$ to $p_+$ and $k_+$. Since $p_+ = -\frac{1}{2}(\omega_\ve{p} - p^1)$ we see that $p_+,k_+\leq0$ and that $\d p_+/p_+ = -\d p^1/\omega_\ve{p}$; the minus signs of the Jacobians cancel, and we find
\begin{align}
    \normal{\hat{\phi}(\lambda)\hat{\phi}(\lambda')}\ \alis \int_{-\infty}^0\frac{\d p_+}{2\pi}\frac{\d k_+}{2\pi}\int\frac{\d^{d-2}\ve{p}^\perp}{(2\pi)^{d-2}}\frac{\d^{d-2}\ve{k}^\perp}{(2\pi)^{d-2}}\frac{\sqrt{\omega_\ve{p}\omega_\ve{k}}}{2p_+k_+}\left(e^{-i(p_+\lambda - k_+\lambda')}\hat{a}^\dagger_\ve{p}\hat{a}_\ve{k} \right. \non
    &\qquad\ \qquad\ \qquad\ \qquad\ \qquad\ \qquad\ \qquad\ \qquad\ \qquad\ \qquad\ \left. + e^{i(p_+\lambda + k_+\lambda')}\hat{a}_\ve{p}\hat{a}_\ve{k}\right) + \mathrm{h.c.} \non
    \alis \frac{1}{2}\int_{-\infty}^0\frac{\d p_+}{2\pi}\frac{\d k_+}{2\pi}\roha{\hat{A}^\dagger_p\hat{A}_ke^{-i(p_+\lambda - k_+\lambda')} + \hat{A}_p\hat{A}_ke^{i(p_+\lambda + k_+\lambda')}} + \mathrm{h.c.}\ , \label{LO_TwoPointLightcone}
\end{align}
where we consider $\ve{p}$ and $\ve{k}$ to be functions of $p_+$ and $k_+$ respectively, and we have defined the shorthands $\hat{A}_p$ and $\hat{A}_k$ through
\begin{align}
    \hat{A}_p = \frac{1}{p_+}\int\frac{\d^{d-2}\ve{p}^\perp}{(2\pi)^{d-2}}\sqrt{\omega_\ve{p}}\,\hat{a}_\ve{p}\ . \label{LO_PencilOperator}
\end{align}
We construct the operator $T_{ab}l^al^b$ from \eqref{LO_TwoPointLightcone} by applying the appropriate differential operator, read off from \eqref{LO_NullNullEMT}, and taking the coincidence limit $\lambda'\rightarrow\lambda$; this leads to
\begin{align}
    T_{ab}(\lambda)l^al^b \alis \lim_{\lambda'\rightarrow\lambda}\diff{}{\lambda}\diff{}{\lambda'}\normal{\hat{\phi}(\lambda)\hat{\phi}(\lambda')} - \xi \diff{^2}{\lambda^2}\normal{\hat{\phi}(\lambda)\hat{\phi}(\lambda)} \non
    \alis \frac{1}{2}\int_{-\infty}^0\frac{\d p_+}{2\pi}\frac{\d k_+}{2\pi}\left[\roha{p_+k_+ + (p_+ - k_+)^2\xi}e^{-i(p_+ - k_+)\lambda}\hat{A}^\dagger_p\hat{A}_k \right. \non
    &\qquad\ \qquad\ \qquad\ \qquad\ \quad \left. - \roha{p_+k_+ - (p_+ + k_+)^2\xi}e^{i(p_+ + k_+)\lambda}\hat{A}_p\hat{A}_k \right] + \mathrm{h.c.}\ , \label{LO_NullNullOperatorEMT}
\end{align}
where we have kept the $\lambda$-dependence of $T_{ab}$ explicit to emphasise that this expression is only valid on this particular null geodesic.

\section{Light-ray operators for the free scalar}\label{sec_NonMinimalLO}
Having constructed the operator $T_{ab}l^al^b$ along the null geodesic, we turn our attention to the light-ray operators themselves in this section. If we substitute \eqref{LO_NullNullOperatorEMT} into \eqref{LO_def} for $s = 2n$, the integrals over $\lambda$ are of the form
\begin{align}
    \int_{-\infty}^\infty\d\lambda\,\lambda^{2n+2}e^{-i\alpha\lambda} \alis \frac{1}{(-i)^{2n+2}}\partial_\alpha^{2n+2}\int_{-\infty}^\infty\d\lambda\,e^{-i\alpha\lambda} = 2\pi(-1)^{n+1}\delta^{(2n+2)}(\alpha)\ . \label{LO_DeltaFunction}
\end{align}
The superscript on the delta function denotes the $(2n+2)^\text{th}$ derivative with respect to the argument of the delta function. Thus, substituting \eqref{LO_NullNullOperatorEMT} into \eqref{LO_def} for $s=2n$ and applying \eqref{LO_DeltaFunction} for $\alpha = \pm p_+ - k_+$, we obtain
\begin{align}
    L_{2n} \alis \frac{(-1)^{n+1}}{4\pi}\int_{-\infty}^0\d p_+\d k_+\,\left[\roha{p_+k_+ + (p_+ - k_+)^2\xi}\hat{A}^\dagger_p\hat{A}_k\delta^{(2n+2)}(p_+ - k_+) \right. \non
    &\qquad\ \qquad\ \qquad\ \qquad\ \quad\ \left. - \roha{p_+k_+ - (p_+ + k_+)^2\xi}\hat{A}_p\hat{A}_k\delta^{(2n+2)}(p_+ + k_+) \right] + \mathrm{h.c.} \label{LO_Volledig}
\end{align}
Now we consider the fact that integrating a function multiplied by a (derivative of a) delta function will always produce quantities constructed from (derivatives of) the integrand, evaluated at points where the argument of the delta function vanishes. However, for a massive field ($m\neq0$) we have $p_+,k_+ < 0$ for all finite momenta, meaning that the second delta function in \eqref{LO_Volledig} has no support at finite momenta. If we also assume that no physical state excites modes with infinite momentum, the second term in \eqref{LO_Volledig} has no support at all and does not contribute to the expectation value of $L_{2n}$.

For $m=0$, the situation is more complex, although the same conclusion should hold. To see why, consider the expression for $p_+$ in terms of $\ve{p}$:
\begin{align}
    p_+ = -\frac{1}{2}\roha{\omega_\ve{p} - p^1} = -\frac{1}{2}\roha{\sqrt{m^2 + (\ve{p}^\perp)^2 + (p^1)^2} - p^1}\ .
\end{align}
It can be seen from this expression that $\ve{p}^\perp \neq \ve{0}^\perp$ fulfils the same role as a non-zero mass in ensuring that $p_+ < 0$ for finite $p^1$ (and we continue to assume that physically relevant states do not excite modes with infinite momentum). One may still have $p_+ = 0$ for $\ve{p}^\perp = \ve{0}^\perp$, but this is a single point in the $d-2$ dimensional space over which the integral in \eqref{LO_PencilOperator} is taken. Therefore, we do not expect this to lead to a contribution to $\mean{L_{2n}}_\psi$ of the second term in \eqref{LO_Volledig} for any reasonable state (for which $\mean{\hat{a}_\ve{p}\hat{a}_\ve{k}}_\psi$ is a sufficiently smooth function of $\ve{p}$ and $\ve{k}$ which does not diverge at $\ve{p}^\perp = \ve{0}^\perp$ or $\ve{k}^\perp = \ve{0}^\perp$, among other conditions). Importantly, these arguments don't work for $d=2$, where there is no $\ve{p}^\perp$. 

Put together, we assume that we are working in $d\geq3$ or with a field of non-vanishing mass, so that the second term in \eqref{LO_Volledig} can be ignored and we can focus on evaluating the contribution of the first term. This term naturally splits into a term which is independent of $\xi$ and one which isn't; we refer to these terms as the minimal and non-minimal coupling terms respectively. Let us consider them individually, and define $u = p_+ + k_+$ and $v = p_+ - k_+$. Suppressing the hermitian conjugate, the minimal coupling term becomes
\begin{align}
    L_{2n}^\text{min} \alis \frac{(-1)^{n+1}}{8\pi}\int_{-\infty}^0\d u\int_{-u}^u\d v\, p_+k_+\hat{A}^\dagger_p\hat{A}_k\partial_v^{2n+2}\delta(v) \non
    \alis \frac{(-1)^n}{8\pi}\int_{-\infty}^0\d u \int_{-u}^{u}\d v\,\partial_v\roha{p_+k_+\hat{A}^\dagger_p\hat{A}_k}\partial_v^{2n+1}\delta(v) \non
    &\qquad\ \qquad\ \qquad\ \qquad\ \qquad\ \qquad\ + \frac{(-1)^{n+1}}{8\pi}\int_{-\infty}^0\d u\, \viha{p_+k_+\hat{A}^\dagger_p\hat{A}_k\partial_v^{2n+2}\delta(v)}^u_{-u}\ . \label{LO_FirstStepMin}
\end{align}
The step from the first to the second line is an integration by parts, for which we need to take the derivative of creation and annihilation operators. This is uncommon, but we interpret it as a formal operation which only really acquires its meaning inside an expectation value. To be explicit, consider the following state:
\begin{align}
    \ket{h} = \int\prod_{j\in\N}\roha{\frac{\d^{d-1}\ve{k}_j}{(2\pi)^{d-1}}\frac{\hat{a}^\dagger_{\ve{k}_j}}{\sqrt{2\omega_{\ve{k}_j}}}}h\roha{\ve{k}_1,\ldots,\ve{k}_N}\ket{0}\ , \label{LO_GeneralState}
\end{align}
with $\N = \cuha{1,\ldots,N}$, the set of positive integers up to and including $N$, and $h$ some sufficiently smooth function which vanishes sufficiently fast at large momenta. This state includes the possibility of exciting a single mode multiple times, since we are allowed to make $h$ sharply peaked when $\ve{k}_i - \ve{k}_j$ approaches $\ve{0}$ for any pair $i,j$. Very general states can therefore be constructed as superpositions of the vacuum and states like $\ket{h}$, with different choices of $h$ (and accordingly different $N$). We then define $\partial_{p_+}^n\hat{A}_p$ as the operator which acts on $\ket{h}$ as
\begin{align}
    \partial_{p_+}^n\hat{A}_p\ket{h} = \int\frac{\d^{d-2}\ve{p}^\perp}{(2\pi)^{d-2}}\sum_{i=1}^N\int\prod_{j\in\N\setminus i}\roha{\frac{\d^{d-1}\ve{k}_j}{(2\pi)^{d-1}}\frac{\hat{a}^\dagger_{\ve{k}_j}}{\sqrt{2\omega_{\ve{k}_j}}}}\partial_{p_+}^n\roha{\frac{h\roha{\ve{k}_1,\ldots,\ve{k}_i=\ve{p},\ldots,\ve{k}_N}}{\sqrt{2}\,p_+}}\ket{0}\ .
\end{align}
The rationale behind this definition is that this is exactly what one would get by acting on $\ket{h}$ with $\hat{A}_p$ and then differentiating $n$ times. Importantly, with this construction there are no derivatives of operators any more, only well-defined derivatives of functions. To further ensure that these derivatives are well-behaved, we define
\begin{align}
    \roha{\partial_{p_+}^{n_1} + \partial_{k_+}^{n_2}}\roha{\alpha\hat{A}_p + \beta\hat{A}_k} \alis \alpha \partial_{p_+}^{n_1}\hat{A}_p + \beta\partial_{k_+}^{n_2}\hat{A}_k \ ,\\
    \partial_{p_+}^{n_1}\partial_{k_+}^{n_2}\roha{\hat{A}_p\hat{A}_k} \alis \partial_{p_+}^{n_1}\hat{A}_p\partial_{k_+}^{n_2}\hat{A}_k\ , \\
    \partial_{p_+}\roha{f(p_+)\hat{A}_p} \alis (\partial_{p_+}f)\hat{A}_p + f\partial_{p_+}\hat{A}_p\ ,
\end{align}
for constant $\alpha,\beta\inC$ and a differentiable function $f$.

Having explained what the derivative of an operator entails, we now consider the boundary terms in \eqref{LO_FirstStepMin}. The same arguments as before apply: the derivative of the delta function will only be non-zero when its argument (here $\pm u$) vanishes, which at finite momenta it never does for the massive field. For the massless field, a non-zero $\ve{p}^\perp$ takes on the role of the mass while the integral in \eqref{LO_PencilOperator} is expected to prevent the modes with $\ve{p}^\perp = \ve{0}^\perp$ and finite $p^1$ from making any contribution to the expectation value. Modes with infinite momentum are deemed unphysical and are therefore never excited, so that in any reasonable state the expectation value of $\hat{A}^\dagger_p\hat{A}_k$ vanishes sufficiently quickly when $p_+,k_+\rightarrow0$. Putting all this together and recalling that we are assuming that either $m\neq0$ or $d\geq3$, we recognise that the boundary term in \eqref{LO_FirstStepMin} can be discarded. Thus, we are allowed to integrate by parts $2n+2$ times in total:
\begin{align}
    L_{2n}^\mathrm{min} \alis \frac{(-1)^{n+1}}{8\pi}\int_{-\infty}^0\d u\int_{-u}^u\d v\,\delta(v)\partial_v^{2n+2}\roha{p_+k_+\hat{A}^\dagger_p\hat{A}_k} \non
    \alis \frac{(-1)^{n+1}}{4\pi}\int_{-\infty}^0\d p_+\d k_+\,\frac{\delta(p_+ - k_+)}{2^{2n+2}}\sum_{j=0}^{2n+2}\roha{\begin{array}{c}
        2n+2 \\
        j
    \end{array}}(-1)^{j}\partial_{p_+}^{2n+2-j}\partial_{k_+}^{j}\roha{p_+k_+\hat{A}^\dagger_p\hat{A}_k} \non
    \alis \frac{(-1)^{n+1}}{4\pi\times2^{2n+2}}\sum_{j=0}^{2n+2}\roha{\begin{array}{c}
        2n+2 \\
        j
    \end{array}}(-1)^j\int_{-\infty}^0\d p_+\,\partial_{p_+}^{2n+2-j}(p_+\hat{A}^\dagger_p)\partial_{p_+}^j(p_+\hat{A}_p)\ ,
\end{align}
where the step from the first to the second line used that $\partial_v = \frac{1}{2}(\partial_{p_+} - \partial_{k_+})$. Because we have restricted to states and models in which the modes with either $p_+ \rightarrow -\infty$ or $p_+ \rightarrow 0$ are not excited, we can integrate by parts in this expression while discarding the boundary terms until the $2n+2$ derivatives are distributed evenly between $p_+\hat{A}^\dagger_p$ and $p_+\hat{A}_p$. For this, we split the sum into three parts: $j < n+1$, $j = n+1$, and $j > n+1$, and integrate by parts until there are $n+1$ derivatives acting on both $p_+\hat{A}^\dagger_p$ and $p_+\hat{A}_p$, leading to
\begin{align}
    L^\mathrm{min}_{2n} \alis \frac{(-1)^{n+1}}{4\pi\times2^{2n+2}}\sum_{j=0}^n\roha{\begin{array}{c}
        2n+2 \\
        j
    \end{array}}(-1)^{n+1}\int_{-\infty}^0\d p_+\,\partial_{p_+}^{n+1}(p_+\hat{A}^\dagger_p)\partial_{p_+}^{n+1}(p_+\hat{A}_p) \non
    &\ \ \ + \frac{(-1)^{n+1}}{4\pi\times2^{2n+2}}\roha{\begin{array}{c}
        2n+2 \\
        n+1
    \end{array}}(-1)^{n+1}\int_{-\infty}^0\d p_+\,\partial_{p_+}^{n+1}(p_+\hat{A}^\dagger_p)\partial_{p_+}^{n+1}(p_+\hat{A}_p) \non
    &\ \ \ + \frac{(-1)^{n+1}}{4\pi\times2^{2n+2}}\sum_{j=n+2}^{2n+2}\roha{\begin{array}{c}
        2n+2 \\
        j
    \end{array}}(-1)^{-n-1}\int_{-\infty}^0\d p_+\,\partial_{p_+}^{n+1}(p_+\hat{A}^\dagger_p)\partial_{p_+}^{n+1}(p_+\hat{A}_p)\ . \label{LO_SplitSum}
\end{align}
The combined sum of binomial coefficients can be evaluated using the following identity:
\begin{align}
    \sum_{j=0}^{2n+2}\roha{\begin{array}{c}
        2n+2 \\
        j
    \end{array}} = 2^{2n+2}\ . \label{LO_SumIdentity}
\end{align}
We therefore arrive at the following expression:
\begin{align}
    L^\mathrm{min}_{2n} \alis \frac{1}{4\pi}\int_{-\infty}^0\d p_+\,\roha{p_+\partial_{p_+}^{n+1}\hat{A}^\dagger_p + (n+1)\partial_{p_+}^n\hat{A}^\dagger_p}\roha{p_+\partial_{p_+}^{n+1}\hat{A}_p + (n+1)\partial_{p_+}^n\hat{A}_p} \non
    \alis \frac{1}{4\pi}\int_{-\infty}^0\d p_+\,\left[(p_+)^2\partial_{p_+}^{n+1}\hat{A}^\dagger_p\partial_{p_+}^{n+1}\hat{A}_p + (n+1)^2\partial_{p_+}^n\hat{A}^\dagger_p\partial_{p_+}^n\hat{A}_p \right. \non
    &\qquad\ \qquad\ \qquad\ \qquad\ \qquad\ \qquad\  \left. + (n+1)\partial_{p_+}\roha{\partial_{p_+}^n\hat{A}^\dagger_p\partial_{p_+}^n\hat{A}_p}p_+ \right] \non
    \alis \frac{1}{4\pi}\int_{-\infty}^0\d p_+\,\viha{(p_+)^2\partial_{p_+}^{n+1}\hat{A}^\dagger_p\partial_{p_+}^{n+1}\hat{A}_p + n(n+1)\partial_{p_+}^n\hat{A}^\dagger_p\partial_{p_+}^n\hat{A}_p}\ , \label{LO_FinalMin}
\end{align}
where in the step to the second line we combined the cross terms into a single derivative, and we integrated by parts in the step to the third line (again discarding boundary terms). Note that the derivatives in the second term of \eqref{LO_FinalMin} don't make sense for $n=-1$; since the $n$-dependent prefactor of this term vanishes in this case, we don't consider this a problem.

Next, we consider the non-minimal coupling term. Using the same $u$ and $v$ as in \eqref{LO_FirstStepMin} and using the same arguments to discard the boundary terms that arise in the integration by parts, we find that we can express this term (again suppressing the hermitian conjugate) as
\begin{align}
    L^\mathrm{non-min}_{2n} \alis \frac{(-1)^{n+1}\xi}{8\pi}\int_{-\infty}^0 \d u\int_{-u}^u\d v\,v^2\hat{A}^\dagger_p\hat{A}_k\partial_v^{2n+2}\delta(v) \non
    \alis \frac{(-1)^{n+1}(2n+2)(2n+1)\xi}{8\pi}\int_{-\infty}^0 \d u\int_{-u}^u\d v\,\delta(v)\partial_v^{2n}\roha{\hat{A}^\dagger_p\hat{A}_k}\ , \label{LO_FirstStepNonMin}
\end{align}
where we kept only those terms in the expansion of $\partial_v^{2n+2}(v^2\hat{A}^\dagger_p\hat{A}_k)$ that have precisely two derivatives acting on $v^2$; after all, more derivatives would annihilate $v^2$, while fewer would lead to surviving factors of $v$ that vanish upon evaluation of the integral over $v$.

By using that $\partial_v = \frac{1}{2}(\partial_{p_+} - \partial_{k_+})$, \eqref{LO_FirstStepNonMin} can be written in terms of $p_+$ and $k_+$ again, after which the delta function can be used to evaluate the integral over $k_+$:
\begin{align}
    L^\mathrm{non-min}_{2n} \alis \frac{(-1)^{n+1}(2n+1)(2n+2)\xi}{4\pi\times2^{2n}}\sum_{j=0}^{2n}\roha{\begin{array}{c}
        2n \\
        j
    \end{array}}(-1)^j\int_{-\infty}^0 \d p_+ \,\partial_{p_+}^{2n-j}\hat{A}_p^\dagger\partial_{p_+}^j\hat{A}_p \ .
\end{align}
For the non-minimal coupling term, the last step is to apply the procedure that led to \eqref{LO_SplitSum}, i.e. split the sum into terms with $j < n$, $j = n$, and $j > n$ such that we can integrate by parts to distribute the $2n$ derivatives equally between $\hat{A}^\dagger_p$ and $\hat{A}$, upon which we can apply the identity \eqref{LO_SumIdentity} to evaluate the remaining sum. The outcome is that
\begin{align}
    L^\mathrm{non-min}_{2n} \alis -\frac{(2n+1)(2n+2)\xi}{4\pi}\int_{-\infty}^0\d p_+\,\partial_{p_+}^n\hat{A}^\dagger_p\partial_{p_+}^n\hat{A}_p \ . \label{LO_FinalNonMin}
\end{align}
The overall minus sign arises due to the fact that we now have $2n$ derivatives to distribute, instead of the $2(n+1)$ derivatives in the minimal coupling term. Combining \eqref{LO_FinalMin} and \eqref{LO_FinalNonMin} and noting that our expressions for $\hat{L}^\mathrm{min}_{2n}$ and $\hat{L}^\mathrm{non-min}_{2n}$ are hermitian, we obtain the following (formal) expression for the light-ray operators:
\begin{align}
    L_{2n} \alis \int_{-\infty}^0\frac{\d p_+}{2\pi}\viha{(p_+)^2\partial_{p_+}^{n+1}\hat{A}^\dagger_p\partial_{p_+}^{n+1}\hat{A}_p + (n+1)\roha{n-2\xi(2n+1)}\partial_{p_+}^n\hat{A}^\dagger_p\partial_{p_+}^n\hat{A}_p}\ .
\end{align}
To make further progress, we eliminate the explicit dependence on $p_+$ by defining
\begin{align}
    \hat{B}(p) = \sqrt{-\frac{p_+}{\mu}}\partial_{p_+}^n\hat{A}_p\ ,
\end{align}
with $\mu > 0$ some arbitrary and fixed scale with units of energy. If we then make the change of variables $p_+ = -\mu e^{-q}$ and denote $\hat{B}_q = \hat{B}\roha{p(q)}$, we find that
\begin{align}
    L_{2n} \alis \int_{-\infty}^\infty\frac{\mu e^{-q}\d q}{2\pi}\viha{\partial_q (e^{q/2}\hat{B}_q^\dagger)\partial_q (e^{q/2}\hat{B}_q) + (n+1)\roha{n-2\xi(2n+1)}e^q\hat{B}_q^\dagger\hat{B}_q} \non
    \alis \int_{-\infty}^\infty\frac{\mu\d q}{2\pi}\viha{\partial_q\hat{B}^\dagger_q\partial_q\hat{B}_q + \frac{1}{2}\partial_q(\hat{B}^\dagger_q\hat{B}_q) + \cuha{\frac{1}{4} + (n+1)(n - 2\xi(2n+1))}\hat{B}^\dagger_q\hat{B}_q}\ , \label{LO_FinalForm}
\end{align}
where we expanded $\partial_q (e^{q/2}\hat{B}_q^\dagger)\partial_q(e^{q/2}\hat{B}_q)$ and combined the cross terms into the full derivative $\partial_q(\hat{B}^\dagger_q\hat{B}_q)$; we can ignore this term since it becomes a boundary term which vanishes due to our assumption that no states with infinite momentum have been excited. 

We then note that the remaining operators in \eqref{LO_FinalForm} form a linear combination of operators multiplied by their respective hermitian conjugates, meaning that any expectation value would be a sum of (positive definite) norms of states. We can choose states in which the expectation value of $\hat{B}_q^\dagger\hat{B}_q$ is either oscillating rapidly or almost constant, leading to the dominance of respectively the first or the last term in \eqref{LO_FinalForm}; hence, $L_{2n}$ will be positive-semidefinite if and only if the term in curly brackets is non-negative. This condition is trivially satisfied for $n=-1$, reproducing the result of \cite{Klinkhammer}; for $n\geq0$, we can formulate it in two equivalent ways, namely as a bound on $\xi$ given $n$ or as a bound on $n$ given $\xi$:
\begin{align}
    \xi \leq \frac{2n+1}{4(2n+2)}& &2n+2\geq\frac{1}{1 - 4\xi}\ , \label{LO_NonMinConstraint}
\end{align}
where the second form has assumed that $\xi < \frac{1}{4}$; inspection of the first form reveals that for $\xi\geq\frac{1}{4}$, there is no $L_{2n}$ which only has non-negative expectation values for $n\geq0$.

\eqref{LO_NonMinConstraint} has immediate consequences for the CANEC on the lightcone. To see them, consider two particular values of the non-minimal coupling. For minimal coupling, $\xi = 0$, \eqref{LO_NonMinConstraint} indicates that $L_{2n}$ is positive-semidefinite for any integer $n\geq-1$. On the other hand, a conformally coupled scalar field, with $\xi = \frac{1}{4}(d-2)/(d-1)$, will by \eqref{LO_NonMinConstraint} only have positive-semidefinite $L_{2n}$ for $n=-1$ or $2n+2\geq d-1$. For values of $\xi$ between minimal and conformal coupling, $L_{2n}$ is positive-semidefinite for $n=-1$ or $n$ greater than some value between zero and $(d-3)/2$. Importantly, in all these cases the light-ray operator corresponding to the CANEC operator on the lightcone for even $d$ (i.e. $L_{d-2}$) is positive-semidefinite. Thus, the free scalar field with non-minimal coupling $0\leq\xi\leq\frac{1}{4}(d-2)/(d-1)$ obeys the CANEC on the lightcone for even $d$.

For odd $d$, the situation is much less nice. We can consider the odd-numbered light-ray operators $L_{2n+1}$; by the same methods as before, this operator can be expressed as
\begin{align}
    L_{2n+1} \alis -\frac{i}{2}\int_{-\infty}^0\frac{\d p_+}{2\pi}\left[(p_+)^2\partial_{p_+}^{n+2}\hat{A}_p^\dagger\partial_{p_+}^{n+1}\hat{A}_p + (n+1)(n + 1 - 2\xi(2n+3))\partial_{p_+}^{n+1}\hat{A}^\dagger_p\partial_{p_+}^n\hat{A}_p \right. \non
    &\qquad\ \qquad\ \qquad\ \qquad\ \qquad\ \qquad\ \qquad\ \qquad\ \qquad\ \left. + p_+\partial_{p_+}^{n+1}\hat{A}_p^\dagger\partial_{p_+}^{n+1}\hat{A}_p \right] + \mathrm{h.c.}
\end{align}
To simplify this, we can introduce the operator $\hat{B}(p) = \partial_{p_+}^n\hat{A}_p$, parameterise $p_+$ as $p_+ = -\mu e^{-q}$ again, and denote $\hat{B}_q = \hat{B}\roha{p(q)}$; together, this leads to
\begin{align}
    L_{2n+1} = -\frac{i}{2}\int_{-\infty}^\infty\frac{\d q}{2\pi}\viha{\partial_q^2\hat{B}_q^\dagger\partial_q\hat{B}_q + (n+1)(n+1 - 2\xi(2n+3))(\partial_q\hat{B}_q^\dagger)\hat{B}_q}\ .
\end{align}
The expectation value of this operator is not a linear combination of norms, and it does not have a definite sign. However, $L_{2n+1}$ is not quite of the right operator for the CANEC on the lightcone; that would be an operator like
\begin{align}
    \tilde{L}_{2n+1} = \int_{-\infty}^\infty \abs{\lambda}^{2n+1}T_{ab}l^al^b\,\d\lambda\ . \label{LO_LoAsCANEC}
\end{align}
After substituting \eqref{LO_NullNullOperatorEMT} into this operator, the $\lambda$-integrals are of the form
\begin{align}
    \int_{-\infty}^\infty\abs{\lambda}^{2n+1}e^{-i\alpha\lambda}\d\lambda \alis -\int_{-\infty}^0\lambda^{2n+1}e^{-i\alpha\lambda}\d\lambda + \int_0^\infty\lambda^{2n+1}e^{-i\alpha\lambda}\d\lambda \non
    \alis (-1)^n\partial_\alpha^{2n+1}\int_0^\infty2\sin(\alpha\lambda)\d\lambda\ . \label{LO_Integraal}
\end{align}
The final integral is not convergent, but it can be regularised as
\begin{align}
    \int_0^\infty\sin(\alpha\lambda)\d\lambda \alis \lim_{\varepsilon\downarrow0}\int_0^\infty\sin(\alpha\lambda)e^{-\varepsilon\lambda}\d\lambda = \frac{1}{2}\lim_{\varepsilon\downarrow0}\roha{\frac{1}{\alpha + i\varepsilon} + \frac{1}{\alpha - i\varepsilon}} = \frac{1}{\alpha}\ .
\end{align}
One can verify that the limit commutes with the derivatives in \eqref{LO_Integraal} by taking derivatives of the penultimate form before sending $\varepsilon\rightarrow0$. Thus, the $\lambda$-integrals in \eqref{LO_LoAsCANEC} take the form
\begin{align}
    \int_{-\infty}^\infty\abs{\lambda}^{2n+1}e^{-i\alpha\lambda}\d\lambda = \frac{(-1)^{n+1}(2n+1)!}{\alpha^{2n+2}}\ .
\end{align}
Importantly, this quantity has support when $\alpha\neq0$, meaning that the term in \eqref{LO_NullNullOperatorEMT} containing $\hat{A}_p\hat{A}_k$ will contribute to the expectation value of $\tilde{L}_{2n+1}$. This happens, for example, in a state like $\ket{\psi} = a\ket{0} + b\ket{2}$, with $\abs{a}^2 + \abs{b}^2 = 1$ and $\ket{2}$ a normalised state created by acting on the vacuum with two creation operators. If we choose $\abs{a}\gg\abs{b}$, the term $\smallmean{\hat{A}_p\hat{A}_k}_\psi = \bar{a}b\bra{0}\hat{A}_p\hat{A}_k\ket{2}$ and its complex conjugate will dominate $\smallmean{\tilde{L}_{2n+1}}_\psi$; since its sign is indeterminate, we conclude that expectation values of $\tilde{L}_{2n+1}$ can take on any sign.

\section{The CANEC on a cylindrical congruence}\label{sec_OtherCongruence}
The previous section demonstrated that for the free, non-minimally coupled real scalar field there is an important range of values of $\xi$ for which $L_{d-2}$ is positive-semidefinite for even $d$. However, this does not in general suggest that the CANEC holds on the lightcone, since neither $\mean{L_{d-2}}_\psi$ nor $\smallmean{\tilde{L}_{d-2}}_\psi$ are all of the same sign for odd $d$. To further emphasise that the CANEC cannot hold on all null congruences in Minkowski spacetime, not even when $d$ is even, this section constructs the CANEC operator for a specific congruence in $d=4$. We then argue that this operator must be of indefinite sign for a class of non-minimally coupled free scalar fields as a consequence of \eqref{LO_NonMinConstraint}.

To construct this congruence, we consider $d=4$ Minkowski spacetime equipped with cylindrical coordinates. The metric then reads
\begin{align}
    \d s^2 = -\d t^2 + \d r^2 + r^2\d\varphi^2 + \d z^2\ .
\end{align}
By considering the equations of motion in these coordinates, it follows that $t = \lambda$, $r = \abs{\lambda}$ describes a complete null geodesic if $\varphi$ and $z$ are kept constant, save a shift by $\pi$ in $\varphi$ as the geodesic passes through $r=0$. The components of the tangent vector to any such geodesic are
\begin{align}
    l^\mu = (\dot{t},\dot{r},\dot{\varphi},\dot{z}) = (1,\mathrm{sgn}(\lambda),0,0)\ . \label{LO_AnotherTangent}
\end{align}
Since this does not depend on the choice of geodesic, \eqref{LO_AnotherTangent} can be taken to give the tangent vector field for an entire congruence, with geodesics labelled by their values of $\varphi$ and $z$. We refer to this congruence as a `cylindrical' congruence, as it is constructed from geodesics that propagate radially towards/from the $z$-axis. The expansion of this congruence follows from \eqref{AppA_Exp}, combined with \eqref{AppA_Christoffel}:
\begin{align}
    \theta \alis \partial_\mu l^\mu + \Gamma^\mu_{\mu\nu}l^\nu = \frac{\mathrm{sgn}(\lambda)}{r}\ . \label{LO_AnotherExpansion}
\end{align}
For all geodesics in the cylindrical congruence, we have $r = \abs{\lambda}$; thus, \eqref{LO_AnotherExpansion} leads to $\theta = \lambda^{-1}$. Using the definition of the Jacobi field \eqref{AppA_JacobiDef} for $d=4$, we find that the Jacobi field in this congruence is
\begin{equation}
    \eta(\lambda) = \eta_0\sqrt{\frac{\abs{\lambda}}{\lambda_0}}\ ,
\end{equation}
where we introduced arbitrary length scales $\eta_0$ and $\lambda_0$. The CANEC on a cylindrical congruence is therefore concerned with
\begin{align}
    \int_{-\infty}^\infty\eta^4 T_{ab}l^al^b\,\d\lambda = \frac{\eta_0^4}{\lambda_0^2}\int_{-\infty}^\infty\lambda^2 T_{ab}l^al^b\,\d\lambda = \frac{\eta_0^4}{\lambda_0^2}L_0\ .
\end{align}
By \eqref{LO_NonMinConstraint}, a non-minimally coupled but otherwise free scalar field only has $\mean{L_0}_\psi\geq0$ for all $\ket{\psi}$ if $\xi\leq\frac{1}{8}$. This upper limit excludes the important case of $\xi = \frac{1}{6}$, conformal coupling for $d = 4$; the conformally coupled scalar does not obey the CANEC on a cylindrical congruence.

In summary, this chapter has demonstrated that there are many positive-semidefinite light-ray operators for a non-minimally coupled but otherwise free scalar field. Among these operators are $L_{-2}$, reproducing the result of \cite{Klinkhammer}, and $L_{d-2}$ for even $d$ and $0\leq\xi\leq\frac{1}{4}(d-2)/(d-1)$, which corresponds to the CANEC operator on the lightcone. However, this does not suggest that the CANEC holds for general null congruences in Minkowski spacetime: the CANEC does not hold on the lightcone for odd $d$, and it also does not hold on a cylindrical congruence for the free, conformally coupled scalar field when $d=4$. Altogether, this suggests that there may be some deeper principle which determines the congruences to which the CANEC applies.

\chapter{Conclusions and outlook}\label{chap_Conc}

In the previous chapters, we have studied various energy conditions with the goal of motivating the conformally invariant averaged null energy condition (CANEC) as an appropriate energy condition for QFTs in curved spacetime by extending the context in which the CANEC applies; we have also argued that many light-ray operators are positive-semidefinite for the non-minimally coupled real scalar. This chapter summarises the steps we have taken to address these aims and discusses our results, before pointing out possible lines of future research.

The CANEC was motivated over the course of chapters \ref{chap_ClassEC}, \ref{chap_ANEC}, and \ref{chap_curveANEC}. First, chapter \ref{chap_ClassEC} justified our interest in constraints on the null-null component of the energy-momentum tensor by demonstrating that several classical energy conditions imply the null energy condition (NEC), which also underlies Penrose's singularity theorem. Then, chapter \ref{chap_ANEC} showed that for QFTs in Minkowski spacetime, the NEC only holds on average along a complete null geodesic, a condition known as the averaged NEC (ANEC). This chapter also used several examples to argue that the ANEC must fail for QFTs in curved spacetimes. This failure was partially addressed in chapter \ref{chap_curveANEC}, which introduced the CANEC as an energy condition for holographic CFTs which arises naturally from the no-bulk-shortcut property. 

Extensions of the CANEC were mainly found in \ref{chap_BeyondCANEC}, and suggest that the CANEC holds more generally than just for holographic CFTs in certain fixed spacetimes. First, we proved the ANEC for holographic CFTs in maximally symmetric spacetimes by extending the results from \cite{holographic_ANEC,Rosso_2020} to AdS spacetime. Since maximally symmetric spacetimes allow one to choose a null congruence with a constant Jacobi field, this proof doubles as a proof of CANEC for holographic CFTs in maximally symmetric spacetimes. In turn, the conformal flatness of maximally symmetric spacetimes means that this proof corroborates the result of section \ref{sec_CFTinConFlat}, where we found that the CANEC holds for general CFTs on a conformally flat background. This result was actually obtained earlier, in \cite{Rosso_2020}, although it was not recognised explicitly as the CANEC.

Based on the result of section \ref{sec_CFTinConFlat}, one may be tempted to suspect that the general form \eqref{H5_CANECform} of the CANEC is correct for any CFT in any spacetime. However, this temptation should be resisted based on the calculation in section \ref{sec_HolographicCFTinSchwarz}, which demonstrated that a holographic CFT in Minkowski spacetime perturbed by static Newtonian sources will obey an energy condition that is related to the CANEC but uses a different weighting factor than the one in \eqref{H5_CANECform}. The difference may be related to the fact that the congruence considered in section \ref{sec_HolographicCFTinSchwarz} is not shear-free, in contrast to the one considered in section \ref{sec_CFTinConFlat}. 

Furthermore, it is interesting to note that \cite{ANECviolation_Schwarzschild} found that the NEC is violated everywhere outside the event horizon of a Schwarzschild black hole in the Boulware vacuum, making the identification of a positive-semidefinite operator constructed from an average of the null energy in section \ref{sec_HolographicCFTinSchwarz} quite remarkable. One possible resolution of this tension is that the Boulware vacuum is simply a bad state, and conclusions drawn about it should not be taken to apply in general. However, it is mostly considered a bad state due to its behaviour near the event horizon, a region explicitly avoided by the calculation in section \ref{sec_HolographicCFTinSchwarz}; one may therefore imagine creating a state which approximates the Boulware vacuum at large distances and avoids its singular behaviour close to the event horizon. An alternative resolution of the tension between \cite{ANECviolation_Schwarzschild} and section \ref{sec_HolographicCFTinSchwarz} could be to ascribe it to the difference in the CFTs that are being considered. After all, section \ref{sec_HolographicCFTinSchwarz} considered holographic CFTs, while \cite{ANECviolation_Schwarzschild} considered the free scalar CFT. These two (types of) theories are characterised by vastly different central charges \cite{Kiritsis}, which suggests that the lower bound of the operator identified in section \ref{sec_HolographicCFTinSchwarz} does not vanish for general CFTs, but approaches zero as the central charge increases. 

The final result from chapter \ref{chap_BeyondCANEC} was concerned with a perturbative derivation of the self-consistent achronal ANEC (AANEC) from the shear-inclusive CANEC-integral \eqref{NewCANEC_scCANEC_tussen0}. Since this derivation used nothing but geometry and the semi-classical Einstein equations \eqref{H1_SemiclassEFE}, its validity would be highly surprising as it would essentially trivialise the AANEC. Closer inspection revealed that the calculation had indeed made a much stronger assumption, namely that quantum effects neither focus nor defocus the null congruence containing a complete achronal null geodesic. As discussed in section \ref{sec_FailedProof}, we believe this subverts the typical reasoning behind proofs of an energy condition, meaning that \eqref{NewCANEC_scCANEC_eind} identifies a quantity that is non-negative if a congruence is neither focused nor defocused. Since a proof of an energy condition should identify a quantity which is non-negative regardless of its effect on congruences (and then draw conclusions about this effect), we do not consider section \ref{sec_FailedProof} to prove the AANEC. The proof of the AANEC from \cite{Wall} (as described in section \ref{sec_AANEC}) is not subject to this criticism because it considers two null geodesics and only cares about the regions in which they are truly perturbatively different from the unperturbed null geodesic.

Finally, in chapter \ref{chap_LO} we attempted to address a limitation of our proof of the CANEC for CFTs on conformally flat spacetimes, namely that it is only formulated for geodesics in the Weyl transformed Minkowski null plane. One might hope that the CANEC could be formulated for more general congruences, and to investigate this possibility we considered the even-numbered light-ray operators $L_{2n}$ as defined in \eqref{LO_def}. Working with the non-minimally coupled but otherwise free real scalar field in Minkowski spacetime to adapt the calculation in \cite{Klinkhammer}, it was shown that $L_{2n}$ is positive-semidefinite if \eqref{LO_NonMinConstraint} is satisfied. 

An important consequence of this result is that massive scalar fields with non-minimal coupling parameter $\xi\in[0,\frac{1}{4}(d-2)/(d-1)]$ obey the CANEC on the lightcone for even $d$ (as well as many other constraints for any $d$), since $L_{d-2}$ is positive-semidefinite for $d=2n$. The same is likely to hold for the massless case in even $d\geq4$. This does not, however, imply that the CANEC holds on any congruence in (even-dimensional) Minkowski spacetime. A counterexample to this idea was provided by constructing a congruence in $d=4$ in such a way that the corresponding CANEC-operator for the conformally coupled scalar field is one of the light-ray operators that do not satisfy \eqref{LO_NonMinConstraint}. This means that CANEC would be violated on this congruence.

Chapter \ref{chap_LO} also briefly considered the odd-numbered light-ray operators $L_{2n+1}$, arguing that these objects have an indefinite sign. This is in tension with a result from \cite{LO_Mathys}, which showed that the ANEC-operator $L_{-2}$ can be mapped to $L_{d-2}$ via a conformal coordinate transformation. For odd $d$, we would therefore argue that $L_{d-2}$ can have negative expectation values, whereas the result from \cite{LO_Mathys} would suggest that this should be impossible due to the ANEC. The cause of this discrepancy remains unknown, although it might be related to the breakdown of the algebra of light-ray operators studied in \cite{LO_Mathys}.

Overall, we have shown that the CANEC can be applied in contexts beyond the setting of \cite{CANEC_odd,CANEC_even}. It (or a related condition) can be formulated for holographic theories in spacetimes for which the argument in section \ref{sec_CANEC} yields a trivial condition; it can be formulated for general CFTs in a conformally flat spacetime; and it might be possible to formulate it in Minkowski spacetime for certain congruences besides the null plane. We believe this warrants further study of the CANEC and mention several approaches to this future research in the next section.

\section{Outlook}
Several lines of future research suggest themselves at this point. For one, the difference between the proposed general form of the CANEC \eqref{H5_CANECform} and the condition derived in section \ref{sec_HolographicCFTinSchwarz} invites us to investigate which constraints are imposed by the no-bulk-shortcut on holographic CFTs in more general curved spacetimes. It may, for example, be interesting to consider holographic CFTs in the actual Schwarzschild spacetime, or in spacetimes with a different topology (such as a torus). This way, it might be possible to discern the dependence of the CANEC on the geometry and topology of the spacetime in which the CFTs lives.

Another possible avenue of future research would be to address the discrepancy between \cite{ANECviolation_Schwarzschild} and the result from section \ref{sec_HolographicCFTinSchwarz}. We have suggested that general CFTs may obey a bound similar to the one derived in section \ref{sec_HolographicCFTinSchwarz}, but with a lower bound that approaches zero as the central charge increases. Testing this hypothesis would be a way of discerning more clearly the way in which the CANEC depends on the theory it is applied to, and perhaps even pave the way towards a CANEC for non-conformally invariant QFTs.

As a third direction for future research, the light-ray operators studied in chapter \ref{chap_LO} warrant further scrutiny. For example, it may be valuable to consider massless non-minimally coupled scalar fields in more detail, in particular for $d=2$, since only massless fields can be CFTs. Other ways of generalising our work could be to consider these operators for different theories, or to investigate the disagreement between chapter \ref{chap_LO} and the result from \cite{LO_Mathys} more closely. Pinpointing exactly where the two arguments are incompatible may help with delineating the class of congruences to which CANEC applies. Another way to make progress in the same direction might be finding a physical interpretation of the observation that certain light-ray operators only have non-negative expectation values. If we understand what these operators (collectively) represent, it may be possible to reformulate the CANEC in a way that unambiguously selects a class of congruences.

Finally, this thesis has focused almost exclusively on formulating an energy condition, without considering the physical consequences (for e.g. cosmology or gravitational physics). We believe this approach to be justifiable, as it would be inappropriate to consider the consequences of an energy condition we cannot yet state in general. Nevertheless, if the CANEC really does hold more generally, it becomes interesting to consider its wider implications for singularity theorems or exotic spacetimes. Perhaps it will one day be a condition related to the CANEC that summarises those features of general relativity that are physically realistic, and not mere mathematical artefacts.

\section*{Acknowledgements}

I would like to extend my gratitude to dr. Ben Freivogel for the fascinating discussions, helpful remarks, and endless enthusiasm. My thanks also go out to dr. Diego Hofman for his sharp comments and interesting insights. Finally, my gratitude goes out to my fellow students; to Amélie, Diogo, Duarte, Emanuel, Helena, Justin, Karel, Lihan, Merlijne, Sam, Shayan, Thomas, and Tiago. Thank you for your questions, your ideas, and your endlessly enjoyable distractions; thank you for the collaboration in creating our own miniature master's office; and thank you for all the inspiration and the motivation.

\appendix

\chapter{Geometry in general relativity}\label{app_GR}
In this appendix, we remind the reader of some geometrical aspects of general relativity. First, section \ref{sec_BasicDefs} reviews the definitions of basic objects in general relativity and some aspects of tensor calculus. This is followed by a discussion of geodesics and curvature in section \ref{sec_GeodCurv}, upon which section \ref{sec_CongRay} builds by considering congruences and the Raychaudhuri equation. Finally, we summarise a few results regarding Weyl transformations in section \ref{sec_Weyl}.

For a more pedagogical introduction to general relativity, see e.g. \cite{HawkingEllis}, \cite{Carroll_2019}, or \cite{Hartle_2014}, in order of decreasing mathematical rigour; we mostly follow \cite{Carroll_2019}.

\section{Basic definitions and tensor calculus}\label{sec_BasicDefs}
The central object in the geometry of curved spacetimes is the manifold. To describe manifolds, we define charts, atlases, manifolds, and submanifolds as follows:
\begin{definition}[Charts and atlases]
    A chart $(\phi,U)$ is the combination of a subset $U\subseteq M$ of some set $M$ and an injective map $\phi: U\rightarrow\Rd$ such that the image $\phi(U)$ is an open subset of \Rd. An atlas is an indexed set of charts $\cuha{(\phi_\alpha,U_\alpha)}$ such that $M = \bigcup_\alpha U_\alpha$ (i.e. the $U_\alpha$ cover $M$) and if $U_\alpha\cap U_\beta \neq \emptyset$, then the coordinate transformation $\phi_\alpha\circ\phi_\beta^{-1}:\phi_\beta(U_\alpha\cap U_\beta) \rightarrow \phi_\alpha(U_\alpha\cap U_\beta)$ is a smooth bijective map.
\end{definition}
\begin{definition}[Manifolds]\label{def_Manifold}
    A smooth manifold $M$ of dimension $d$ is a set equipped with a maximal atlas, i.e. an atlas containing all charts mapping subsets of $M$ to open subsets of $\Rd$.
\end{definition}
\begin{definition}[Submanifolds]
    Let $M$ be a manifold. The subset $N\subseteq M$ is a submanifold if $N$ is a manifold as well.
\end{definition}
Note that definition \ref{def_Manifold} does not allow $M$ to have a boundary, since points on the boundary could not be mapped to an open subset of \Rd. We then define curves and vectors:
\begin{definition}[Curves]
    Let $M$ be a manifold and $I\subseteq\mathbb{R}$ an open interval. A curve in $M$ is then a map $\gamma:I\rightarrow M$, which labels any point $p\in M$ along $\gamma$ by a value of the parameter $\lambda\in I$.
\end{definition}
\begin{definition}[Vectors]
    Let $M$ be a manifold and $\gamma$ a curve in $M$ parametrised by $\lambda$; a vector at $p\in M$ on $\gamma$ is the directional derivative $\diff{}{\lambda}$. The set of directional derivatives at $p$ for all curves through $p$ forms a vector space called the tangent space $T_p$.
\end{definition}
A natural basis for $T_p$ follows from the chain rule: let $(\phi,U)$ be a chart with $p\in U$, $f:M\rightarrow\mathbb{R}$ a smooth function, and $\gamma:I\rightarrow M$ a curve parametrised by $\lambda$ with $\gamma(I)\subset U$. Choose coordinates $x^\mu$ on $\phi(U)$ and denote $f\circ\phi^{-1}:\Rd\rightarrow \mathbb{R}$ as $f(x)$ and $\phi\circ\gamma:I\rightarrow\Rd$ as $x^\mu(\lambda)$. Then the action of a vector $V = \diff{}{\lambda}$ on $f$ is given by
\begin{align}
    V(f) = \diff{f\circ\gamma}{\lambda} = \diff{x^\mu(\lambda)}{\lambda}\partial_\mu f(x) \ , \label{H1_VectorOpScalar}
\end{align}
which introduces the summation convention: repeated indices are implicitly summed over. Since $f$ was arbitrary we identify $V = \diff{x^\mu}{\lambda}\partial_\mu$, and because $\diff{x^\mu}{\lambda}$ depends on the choice of $V$, we use $\partial_\mu$ as a basis known as the coordinate basis for $T_p$. The components of $V$ are then $V^\mu = \diff{x^\mu}{\lambda}$.
\par
The cotangent space $T_p^*$ consists of the linear maps $\omega:T_p\rightarrow\mathbb{R}$. Assuming a basis $\{e_{(\mu)}\}$ (the index now labels different vectors) of $T_p$, we define a dual basis $\{e^{(\mu)}\}$ of $T_p^*$ as the operators such that $e^{(\mu)}(e_{(\nu)}) = \delta^\mu_\nu$, the Kronecker delta; the dual basis of the coordinate basis is denoted as $\d x^\mu$. We define the action of a vector $V\in T_p$ on a covector $\omega\in T_p^*$ as $V(\omega) = \omega(V)$, hence
\begin{equation}
    V(\omega) = \omega(V) = \omega_\mu e^{(\mu)} (V^\nu e_{(\nu)}) = \omega_\mu V^\mu\ , \label{H1_CoVector}
\end{equation}
Finally, a tensor of rank $(m,n)$ is a multilinear map from $m$ covectors and $n$ vectors to a scalar:
\begin{align}
    T = T^{\mu_1\ldots\mu_m}_{\ \ \ \ \ \ \ \ \nu_1\ldots\nu_n}e_{(\mu_1)}\otimes\ldots\otimes e_{(\mu_m)}\otimes e^{(\nu_1)}\otimes\ldots\otimes e^{(\nu_n)}\ . \label{H1_TensorDef}
\end{align}
$\otimes$ is the tensor product, defined such that for a $(k,l)$ tensor $T$ and an $(m,n)$ tensor $S$, the action of $T\otimes S$ on the covectors $\omega_{(1)},\ldots,\omega_{(k+m)}$ and the vectors $V_{(1)},\ldots,V_{(l+n)}$ is
\begin{align}
    T\otimes S\roha{\omega_{(1)},\ldots,\omega_{(k+m)},V_{(1)},\ldots,V_{(l+n)}} \alis \non
    T\left(\omega_{(1)},\ldots,\omega_{(k)},V_{(1)},\ldots,V_{(l)}\right)&S\left(\omega_{(k+1)},\ldots,\omega_{(k+m)},V_{(l+1)},\ldots,V_{(l+n)}\right)\ .
\end{align}
With \eqref{H1_TensorDef} and \eqref{H1_CoVector}, acting on $m$ covectors and $n$ vectors with an $(m,n)$ tensor $T$ yields
\begin{align}
    T\roha{\omega_{(1)},\ldots,\omega_{(m)},V_{(1)},\ldots,V_{(n)}} \alis T^{\mu_1\ldots\mu_m}_{\ \ \ \ \ \ \ \ \nu_1\ldots\nu_n}\omega_{(1)\mu_1}\ldots\omega_{(m)\mu_m}V^{\nu_1}_{(1)}\ldots V^{\nu_n}_{(n)}\ .
\end{align}
Thus, we often only need the components of a (co-)vector or tensor with respect to some basis. Nevertheless, it will be useful to make coordinate-independent statements. This can be done with the abstract index notation from \cite{abstracte_indices}, where Latin indices from the beginning of the alphabet indicate how many vectors or covectors a tensor should act on to give a scalar, with the positioning differentiating between the two, e.g. $T_{ab}$ is a $(0,2)$ tensor with components $T_{\mu\nu}$. Tensor operations are denoted as they would be in normal indices; e.g. for a covector $\omega$ and a vector $V$, $\omega(V) = \omega_a V^a$, and if $S$ has components $S_{\mu\nu} = T^\lambda_{\,\ \mu\lambda\nu}$, we write $S_{ab} = T^c_{\,\ acb}$.

We can equip a manifold with a metric, a symmetric $(0,2)$ tensor $g_{ab}$. This metric defines an inner product $\mean{\cdot,\cdot}:T_p\times T_p \rightarrow\mathbb{R}$ with $\mean{X,Y} = g_{ab}X^aY^b$ for $X^a,Y^a\in T_p$; we call two vectors orthogonal if $g_{ab}X^aY^b = 0$, and the norm of $X^a$ is $g_{ab}X^aX^b$. We assume that the matrix of metric components, denoted $(g_{\mu\nu})$, is invertible, and define the inverse metric as the $(2,0)$ tensor $g^{ab}$ with $g^{ac}g_{cb} = g_{bc}g^{ca} = \delta^a_b$ (the Kronecker delta). The metric and its inverse establish an isomorphism between $T_p$ and $T_p^*$ by assigning a unique covector to any vector as $V_a = g_{ab}V^b$ and vice versa as $\omega^a = g^{ab}\omega_b$. In the context of general relativity, a manifold $M$ is called a spacetime if it is connected (there exists a curve between any two points in $M$), time-orientable (all points in $M$ have a consistent sense of `past' and `future'), and equipped with a smooth metric of Lorentzian signature ($(g_{\mu\nu})$ has one negative eigenvalue). Due to the Lorentzian signature, $g_{ab}X^aX^b$ can be positive, zero, or negative, classifying $X^a$ as spacelike, null, or timelike respectively.

The partial derivative of a tensor is often not a tensor, and hence doesn't define a coordinate-independent notion of a derivative. This notion is provided by the covariant derivative:
\begin{definition}[Covariant derivative]
    Let $X$ and $Y$ be smooth vector fields, and let $S$ and $T$ be smooth tensor fields of arbitrary rank. The covariant derivative $\nabla$ is then a map which sends $X$ and $T$ to the smooth tensor field $\nabla_X T$ of the same rank as $T$, such that
    \begin{enumerate}[(1)]
        \item for arbitrary functions $f,g$, we have $\nabla_{fX+gY}T = f\nabla_XT + g\nabla_YT$;
        \item $\nabla_XT$ is linear in $T$, i.e. if $T$ and $S$ have the same rank, $\nabla_X(\alpha T + \beta S) = \alpha\nabla_XT + \beta\nabla_XS$ for $\alpha,\beta\inR$;
        \item $\nabla$ obeys the Leibniz rule, i.e. $\nabla_X\roha{T\otimes S} = (\nabla_X T) \otimes S + T\otimes(\nabla_XS)$;
        \item $\nabla$ reduces to the usual directional derivative for functions $f:M\rightarrow\mathbb{R}$, i.e. $\nabla_Xf = X(f)$.
    \end{enumerate}
\end{definition}
By property (1), the map $\nabla T$ which acts on a vector $X^a$ as $X\rightarrow\nabla_XT$ is linear and hence defines an $(m,n+1)$ tensor (if $T$ is an $(m,n)$ tensor) which we denote as $\nabla_c T^{a_1\ldots a_m}_{\ \ \ \ \ \ \ \ b_1\ldots b_n}$; we demand that covariant differentiation commutes with contraction. We can find the components of $\nabla T$ with respect to a basis $\{e_{(\mu)}\}$ of $T_p$ if we define the connection coefficients $\Gamma^\rho_{\mu\nu}$ as
\begin{align}
    \nabla_\mu e_{(\nu)} \equiv \nabla_{e_{(\mu)}}e_{(\nu)} = \Gamma^\rho_{\mu\nu}e_{(\rho)}\ .
\end{align}
By combining properties (3) and (4) of the covariant derivative, one can derive its action on the basis covectors $e^{(\mu)}$, after which the components of $\nabla T$ are straightforwardly found. For example, the components of $\nabla_a V^b$ and $\nabla_a V_b$ in the coordinate basis are
\begin{align}
    \nabla_\mu V^\nu = \partial_\mu V^\nu + \Gamma^\nu_{\mu\rho}V^\rho& , &\nabla_\mu V_\nu = \partial_\mu V_\nu - \Gamma^\rho_{\mu\nu}V_\rho\ ,
\end{align}
The Levi-Civita connection is the unique torsion-free metric compatible connection, i.e. with $\Gamma^\alpha_{\mu\nu} = \Gamma^\alpha_{\nu\mu}$ and $\nabla_a g_{bc} = 0$. In the coordinate basis, it is given by
\begin{equation}
    \Gamma^\rho_{\mu\nu} = \frac{1}{2}g^{\rho\lambda}\roha{\partial_\mu g_{\lambda\nu} + \partial_\nu g_{\lambda\mu} - \partial_\lambda g_{\mu\nu}}\ . \label{AppA_Christoffel}
\end{equation}

\section{Geodesics and curvature}\label{sec_GeodCurv}
For a manifold $M$, the covariant derivative provides a notion of parallel transport along a curve $\gamma:I\rightarrow M$, parametrised by $\lambda\in I$. In flat space, a tensor would be parallel transported along $\gamma$ if the directional derivative of its components vanishes. However, a directional derivative is generally not metric compatible. This means that for two vector fields with vanishing directional derivative the directional derivative of their inner product does not vanish, so the angle between these vectors is changing, contradicting our intuitive notion of parallel transport. This is resolved by defining the covariant directional derivative as
\begin{equation}
    \DIFF{}{\lambda} \equiv \diff{x^\mu}{\lambda}\nabla_\mu = \nabla_X\ ,
\end{equation}
where $X^a$ is the tangent vector to $\gamma$. Thus, $V^a$ is parallel transported along $\gamma$ if $X^a\nabla_a V^b = 0$, or in the coordinate basis
\begin{align}
    \DIFF{V^\mu}{\lambda} = X^\nu\nabla_\nu V^\mu = \diff{V^\mu}{\lambda} + X^\nu\Gamma^\mu_{\nu\rho}V^\rho = 0\ . \label{AppA_geodeet}
\end{align}
If $\gamma$ parallel transports its own tangent vector it is called an affinely parametrised geodesic, and its defining property $X^a\nabla_a X^b = 0$ is called the geodesic equation. The qualification `affinely parametrised' is required because, if $\lambda$ is an affine parameter, we could choose another parameter $\sigma = \sigma(\lambda)$, assumed to be strictly increasing or decreasing with $\lambda$. This leads to a new tangent vector $Y^a = \diff{\lambda}{\sigma}X^a \equiv X^a/f$ for which $Y^a\nabla_a Y^b = -Y(f)Y^b/f$; an affine parameter is therefore one for which the right-hand side of this equality vanishes. $\sigma$ would also be an affine parameter if $Y(f) = 0$, implying that $f = \d\sigma/\d\lambda$ is constant. Affine parameters $\lambda$ and $\sigma$ are therefore related as $\sigma = a\lambda+b$ for constant $a,b\inR$. 

Note that since geodesics parallel transport their own tangent vectors, the norm of this tangent vector does not change. Hence, if the metric has a Lorentzian signature, a geodesic which has a timelike tangent vector anywhere will have a timelike tangent vector everywhere and is therefore called timelike itself. The same holds for the null and spacelike cases.

If the manifold $M$ is a spacetime, an alternative definition of a timelike geodesic is possible. Consider two points $p,q\in M$ that can be connected by a curve with everywhere timelike tangent vector $X^a$ and parameter $\lambda$ such that $\lambda(p) = 0$ and $\lambda(q) = 1$. The length of this curve is
\begin{equation}
    l = \int_0^1L(\lambda)\d\lambda \equiv \int_0^1\sqrt{-g_{ab}X^aX^b}\,\d\lambda = \int_0^1\sqrt{-g_{\mu\nu}\diff{x^\mu}{\lambda}\diff{x^\nu}{\lambda}}\,\d\lambda\ . \label{H1_KrommeLengte}
\end{equation}
We may now find the curve for which $l$ has a critical point, i.e. for which a variation $\delta x^\mu$ leads to $\delta l = 0$. The Euler-Lagrange equations show that this is achieved by a curve with
\begin{equation}
    X^a\nabla_aX^b = \frac{1}{L}\diff{L}{\lambda}X^b\ . \label{H1_GeodeetDef2}
\end{equation}
We can make the right-hand side vanish through a judicious choice of parameterisation; a straightforward choice is the proper time $\tau$, which along a curve $\gamma$ with tangent vector $X^a$ is defined through $\diff{\tau}{\lambda} = \sqrt{-g_{ab}X^aX^b}$; one can check that this implies $L = 1$. Therefore, timelike geodesics are critical points of the path length, and the proper time is an affine parameter. If we parameterise a timelike geodesic with tangent vector $P^a$ with $\tau$, the geodesic equation becomes
\begin{align}
    \DIFF{P^a}{\tau} = 0\ . \label{AppA_NewtonCurved}
\end{align}
This can be seen as a curved spacetime generalisation of the Newtonian equation of motion for a free particle, so we postulate that free massive particles move on timelike geodesics while free massless particles move along null geodesics.

We can also use geodesics to construct a natural coordinate system. We begin by defining the exponential map:
\begin{definition}[The exponential map]
    Let $M$ be a smooth manifold and let $p\in M$ with tangent space $T_p$. Let $\gamma_X(\lambda)$ be an affinely parametrised geodesic with tangent vector $X^a\in T_p$ at $p$ and $\gamma_X(0) = p$. The exponential map $\mathrm{exp}_p:T_p\rightarrow M$ is then defined as the map $\exp_p(X^a) = \gamma_X(1)$. 
\end{definition}
Note that $\exp_p(\lambda X^a) = \gamma_X(\lambda)$; hence, the exponential map is only well-defined for all $X^a\in T_p$ if the affine parameter of all $\gamma_X$ is unbounded. We call a geodesic with unbounded affine parameter complete, which leads to the notion of a geodesically complete manifold:
\begin{definition}[Geodesic completeness]
    A smooth manifold $M$ in which the exponential map is well-defined on all of $T_p$ at all $p\in M$, i.e. in which the affine parameter of any geodesic is unbounded, is called geodesically complete.
\end{definition}
Even if the exponential map is not defined on the entire manifold, it can be used to define so-called Riemann normal coordinates in a neighbourhood $\N_p$ of $p$. If $\N_p$ is sufficiently small, the geodesic $\gamma_X$ connecting $p$ and $q = \exp_p(X^a)\in \N_p$ will be unique, and if we then choose basis vectors $\{e_{(\mu)}^a\}$ for $T_p$ such that $X^a = x^\mu e_{(\mu)}^a$, we can assign the coefficients $x^\mu$ to $q$ as its coordinates; because $\gamma_X$ is unique, this procedure labels all points in $\N_p$ uniquely. The region $\N_p$ in which this procedure is viable is called a normal neighbourhood of $p$; if we choose $\N_p$ such that it is a normal neighbourhood of all $q\in\N_p$ (i.e. the geodesic connecting any two points in $\N_p$ is unique and contained in $\N_p$), then $\N_p$ is a normal convex neighbourhood.

Finally, a central goal in the geometry of general relativity is to quantify the amount of curvature at each point of the manifold we use to model spacetime. One way to do this is to parallel transport a vector around an infinitesimal loop and measure the curvature as the difference between the initial and final vector. If we span the loop with two vectors $X^a$ and $Y^a$, this procedure is equivalent to the action of the commutator of covariant derivatives on a vector $V^a$: the commutator calculates the difference between the vector obtained by transporting first along $X^a$ and then $Y^b$, and the opposite order. We thus measure curvature with the Riemann tensor $R^a_{\,\ bcd}$, which for the Levi-Civita connection is defined as
\begin{align}
    \viha{\nabla_c,\nabla_d}V^a = R^a_{\,\ bcd}V^b\ . \label{AppA_RiemannDef}
\end{align}
The components of $R^a_{\,\ bcd}$ with respect to the coordinate basis are given in \eqref{H1_RiemannComponenten}. Much of the information in the Riemann tensor is also contained in its various traces, known as the Ricci tensor $R_{ab} = R^c_{\ \,acb}$ and the Ricci scalar $R = g^{ab}R_{ab}$.

\section{Congruences and the Raychaudhuri equation}\label{sec_CongRay}
A useful way to study the curvature of a spacetime $M$ is to consider the behaviour of a set of geodesics that are neighbouring in the sense that there exists an open region $N\subseteq M$, so that every point in $N$ lies on exactly one geodesic. Such a set is referred to as a congruence. We parameterise the geodesics in the congruence with an affine parameter $\lambda$ such, that the tangent vector field is $l^a$ and $\lambda=0$ defines a smooth manifold.

We wish to analyse the behaviour of neighbouring geodesics in terms of their separation. To define this, pick a point $p\in N$ with $\lambda=0$ and find the subspace $H_p\subset T_p$ of the tangent space $T_p$ at $p$ containing all vectors that are orthogonal to and independent from $l^a$. Since $M$ is a spacetime, the dimension of $H_p$ depends on the type of congruence. After all, we can use $l^a$ as a basis vector for $T_p$, leaving $d-1$ basis vectors that are linearly independent of $l^a$. If $l^a$ is time- or spacelike, these vectors can all be chosen orthogonal to $l^a$, so $H_p$ is $d-1$ dimensional; if $l^a$ is null, it is orthogonal to itself, so there will only be $d-2$ basis vectors of $T_p$ that are linearly independent of $l^a$ and can be chosen to be orthogonal to it. In this case, $H_p$ is $d-2$ dimensional. In general, if $H_p$ is $d-n$ dimensional, we say that $H_p$ has codimension $n$.

Next, we choose coordinates $x^\mu$ on $N$ (or at least some open region of $N$ which contains $p$) such, that the coordinate basis of $H_p$ is $\cuha{\partial_j}$. Points around $p$ that only have different $x^j$ must lie on different geodesics than $p$, since curves in the direction of $\partial_j$ must be orthogonal to the geodesic through $p$; we can, however, assign these points $\lambda=0$ as well. Hence, we characterise the geodesics in a neighbourhood of $p$ by their $x^j$ at $\lambda=0$; we denote the geodesics by $\gamma(\lambda;x^j)$, and the coordinate curves they describe as $x^\mu(\lambda;x^j)$. A measure of the orthogonal separation of neighbouring geodesics is then given by the deviation vectors $S^a_{(j)}$, with components
\begin{align}
    S^\mu_{(j)} = \parti{x^\mu(\lambda;x^j)}{x^j}\ .
\end{align}
It is then natural to ask how the deviation vectors change along the congruence, and consider
\begin{align}
    \DIFF{S^a_{(j)}}{\lambda} = l^b\nabla_bS^a_{(j)} = [l,S_{(j)}]^a + (\nabla_b l^a)S^b_{(j)}\ , \label{AppA_Deviation}
\end{align}
where we have introduced the commutator $[X,Y]^a = X^b\nabla_bY^a - Y^b\nabla_bX^a$. The components of $[l,S_{(j)}]^a$ in the coordinate basis can be checked to be
\begin{align}
    [l,S_{(j)}]^\mu \alis l^\nu\partial_\nu S_{(j)}^\mu - S^\nu_{(j)}\partial_\nu l^\mu \non
    \alis \partial_\lambda\partial_jx^\mu - \partial_j\partial_\lambda x^\mu = 0\ , \label{AppA_Commutator}
\end{align}
where we have used that $\lambda$ can be seen as one of the coordinates $x^\mu$ to rewrite the directional derivative to $\partial_\lambda$, and that partial derivatives commute. Since \eqref{AppA_Commutator} is manifestly tensorial, we conclude that $[l,S_{(j)}]^a=0$ and hence the quantity $\nabla_b l^a$ describes, according to \eqref{AppA_Deviation}, the failure of $S^a_{(j)}$ to be parallel transported, summarising the behaviour we are interested in. To proceed, we decompose it into its trace, traceless symmetric part, and anti-symmetric part as
\begin{align}
    \nabla_al_b = \frac{\theta}{d-n}P_{ab} + \sigma_{ab} + \omega_{ab}\ . \label{AppA_Decompositie}
\end{align}
Here, $\theta$ is called the expansion, $P_{ab} = P_{ba}$ is the projection tensor onto $H_p$ at any $p\in N$ (or equivalently the metric of $H_p$), $\sigma_{ab} = \sigma_{ba}$ is the traceless shear tensor, and $\omega_{ab} = -\omega_{ba}$ is the vorticity. The appearance of $P_{ab}$ follows from the observation that $l^a\nabla_bl_a = 0 = l^b\nabla_bl_a$, the first by virtue of the chain rule and the second by virtue of $l^a$ being the tangent vector to an affinely parametrised geodesic. For a time- or spacelike congruence, it is straightforward to see that a satisfactory expression for $P_{ab}$ is
\begin{align}
    P_{ab} = g_{ab} \pm l_al_b\ , \label{AppA_ProjectieNietNull}
\end{align}
where the plus holds for timelike and the minus for spacelike congruences. This means that for any $V^a,W^b\in H_p$, $P_{ab}$ acts as the metric since $P_{ab}V^aW^b = g_{ab}V^aW^b$, while any vector parallel to $l^a$ is annihilated. For a null congruence, the choice of $P_{ab}$ is less natural since $H_p$ is of codimension two. We therefore choose another null vector $k^a$ which has $k_al^a = -1$ and is parallel transported, and define
\begin{align}
    P_{ab} = g_{ab} + l_ak_b + k_al_b\ ; \label{AppA_ProjectieNull}
\end{align}
again, we notice that for $V^a,W^b\in H_p$, $P_{ab}V^aW^b = V_aW^a$ while vectors that are parallel to either $l^a$ or $k^a$ are annihilated. With \eqref{AppA_ProjectieNietNull} and \eqref{AppA_ProjectieNull}, we note that $g^{ab}P_{ab} = d-n$, which we use to express the expansion, shear, and vorticity in terms of $\nabla_al_b$:
\begin{align}
    \theta \alis \nabla_al^a\ , \label{AppA_Exp}\\
    \sigma_{ab} \alis \frac{1}{2}\roha{\nabla_al_b + \nabla_bl_a} - \frac{\nabla_cl^c}{d-n}P_{ab}\ , \label{AppA_Shear} \\
    \omega_{ab} \alis \frac{1}{2}\roha{\nabla_al_b - \nabla_bl_a}\ . \label{AppA_Vort}
\end{align}
Note that $\sigma_{ab}l^a = \omega_{ab}l^a = 0$, and that for a null congruence $\sigma^{ab}k_b$ and $\omega^{ab}k_b$ are perpendicular to both $l^a$ and $k^a$. Hence, $\sigma_{ab}$ and $\omega_{ab}$ are `spatial' in the sense that there exists a coordinate system in which they have no temporal components; it follows that $\sigma_{ab}\sigma^{ab}\geq0$ and $\omega_{ab}\omega^{ab}\geq0$. Much of the content of $\nabla_al_b$ is summarised by $\theta$, which measures the change in the volume of a $d-n$ dimensional sphere of test particles travelling along the geodesic (besides \cite{HawkingEllis}, \cite{Witten_causality} discusses this interpretation). Its evolution is described by the Raychaudhuri equation, which we derive by considering the derivative of $\theta$ along the congruence:
\begin{align}
    \diff{\theta}{\lambda} = l^a\nabla_a(\nabla_bl^b) \alis l^a\nabla_b\nabla_al^b + l^aR^b_{\ \,cab}l^c \non
    \alis \nabla_b(l^a\nabla_al^b) - \roha{\nabla_bl_a}(\nabla^al^b) - R^b_{\ \,cba}l^al^c\ . \label{AppA_Raychaudhuri1}
\end{align}
In the first line, we used \eqref{AppA_RiemannDef}. Since the geodesics in the congruence are affinely parametrised, the first term in \eqref{AppA_Raychaudhuri1} vanishes, and we can substitute \eqref{AppA_Decompositie} and $R_{ab} = R^c_{\ \,acb}$ to find
\begin{align}
    \diff{\theta}{\lambda} \alis -\frac{\theta^2}{(d-n)^2}P_{ba}P^{ab} - \sigma_{ba}\sigma^{ab} - \omega_{ba}\omega^{ab} - \frac{\theta}{d-n}(P_{ba}\sigma^{ab} + \sigma_{ba}P^{ab}) - R_{ab}l^al^b\ .
\end{align}
Using \eqref{AppA_ProjectieNietNull} and \eqref{AppA_ProjectieNull}, we find $P_{ab}P^{ab} = d-n$, and combining them with \eqref{AppA_Shear} shows that $P_{ab}\sigma^{ab} = 0$, so that we obtain the final form of the Raychaudhuri equation:
\begin{align}
    \diff{\theta}{\lambda} \alis -\frac{\theta^2}{d-n} - \sigma_{ab}\sigma^{ab} + \omega_{ab}\omega^{ab} - R_{ab}l^al^b\ . \label{AppA_RaychaudhuriExpansion}
\end{align}
A slight simplification of this equation can be found by realising that if $\theta$ measures changes in $d-n$ dimensional volumes, there should be a length scale $\eta$, also known as a Jacobi field, associated with these volumes. In general, we define the Jacobi field as
\begin{align}
    \theta = \frac{1}{\eta^{d-n}}\diff{\eta^{d-n}}{\lambda} = \frac{d-n}{\eta}\diff{\eta}{\lambda}\ , \label{AppA_JacobiDef}
\end{align}
which allows for a straightforward interpretation of $\theta$ as the relative change in a volume $\eta^{d-n}$. Furthermore, points where neighbouring geodesics intersect are now characterised by $\eta\rightarrow0$; we refer to such points as caustics. In terms of $\eta$, the Raychaudhuri equation reads
\begin{align}
    \frac{d-n}{\eta}\diff{^2\eta}{\lambda^2} = - \sigma_{ab}\sigma^{ab} + \omega_{ab}\omega^{ab} - R_{ab}l^al^b\ . \label{AppA_Raychaudhuri2}
\end{align}
Physically, one can interpret $\sigma_{ab}$ as describing the influence of distant inhomogeneities on a geodesic (describing e.g. gravitational lensing), so $\sigma_{ab} = 0$ only in rather special spacetimes \cite{shearfree}. On the other hand, it is quite straightforward to find an (initially) irrotational congruence, in which $\omega_{ab} = 0$. This can be seen by constructing Gaussian normal coordinates, in which one can explicitly calculate $\omega_{ab}=0$ \cite{Senovilla_1998,Witten_causality}; we will discuss this construction for null congruences (following \cite{Witten_causality}), since timelike congruences demonstrate similar behaviour.

Consider a spacetime $M$, and a codimension 2 spacelike submanifold $W\subset M$ (i.e. a $d-2$ dimensional submanifold with a timelike normal vector). Since $W$ is spacelike, one of its normal vectors is timelike; the other one must be spacelike, because $g_{ab}$ has Lorentzian signature. If $W$ is orientable, the normal vectors define four orientations for null congruences emanating orthogonally from $W$: past- and future-directed, and propagating along or opposite to the spacelike normal vector\footnote{As \cite{Senovilla_1998} points out, the null geodesics emanating orthogonally from $W$ do not form congruences, as they don't cover an open region of spacetime. Instead they form null hypersurfaces (codimension 1 submanifolds with a null normal vector), each of which can be made into a congruence by deforming $W$ in a direction not included in the respective hypersurface and emanating null geodesics from the deformed $W$ with the same spatio-temporal orientation as the geodesics covering the hypersurface. We proceed assuming this has been done.}. Without loss of generality, we consider the congruence of future-directed null geodesics propagating along the spacelike normal vector of $W$ and define $N\subset M$ as the codimension 1 submanifold covered by this congruence. We can equip (a subset of) $W$ with coordinates $x^I$ (with $I\in\cuha{2,\ldots,d-1}$), and choose an affine parameter $u$ for the null geodesics covering $N$ such that $u$ vanishes at $W$ and increases towards the future of $W$; this uniquely defines $u$ up to multiplication by a positive function of $x^I$. By declaring $x^I$ to be constant along the geodesics, we define a coordinate system on $N$ which is non-degenerate until the geodesics forming $N$ intersect, i.e. until they encounter a caustic.

To extend these coordinates to a $d$ dimensional region in $M$, we follow \cite{Witten_causality} by choosing a codimension 1 spacelike submanifold $S\subset M$ with $W\subset S$. We then define a function $v:S\rightarrow\mathbb{R}$ such that $S$ is the union of codimension 2 spacelike submanifolds $W_\alpha = \cuha{p\in S:v(p)=\alpha,\alpha\in K\subseteq\mathbb{R}}$, with $W = W_0$. If the orientation of each $W_\alpha$ is smoothly connected to that of $W$, the procedure from the previous paragraph can be repeated: extend the $x^I$ to functions on $S$ to equip every $W_\alpha$ with coordinates and choose an affine parameter $u$ for the future-directed null geodesics emanating orthogonally from $W_\alpha$ along the spacelike normal vector, with $u = 0$ on $W_\alpha$. By declaring $v$ and $x^I$ to be constant along these geodesics, we have constructed a coordinate system $(u,v,x^I)$ for at least the region close to $W$ and $N$. 

Let us now examine the metric components in these coordinates. One can set $g_{uu} = 0$, since $u$ is the affine parameter of a null geodesic with constant $v$ and $x^I$. Every such null geodesic is orthogonal to some $W_\alpha$, so $g_{uI} = 0$ on all $W_\alpha$; it follows that $g_{uI}$ only depends on $u$. However, a curve with constant $v$ and $x^I$ is by definition a null geodesic with affine parameter $u$, so according to \eqref{AppA_geodeet} $\Gamma^\mu_{uu} = 0$. By \eqref{AppA_Christoffel} this means that $\partial_ug_{u\mu} = 0$; hence $g_{uI} = 0$ identically. $g_{uv}$ on the other hand cannot vanish (otherwise $g_{ab}$ would be degenerate), so we define $g_{uv} = -e^q$, where $q = q(v,x^I)$ and the sign is chosen such that $u - \beta v = x^I = 0$ defines a timelike curve for sufficiently big $\beta>0$. Thus, in Gaussian normal coordinates the metric is
\begin{align}
    \d s^2 = -e^q\roha{\d u\d v + \d v\d u} + g_{vv}\d v^2 + g_{IJ}\d x^I\d x^J + g_{vI}\roha{\d v\d x^I + \d x^I\d v}\ , \label{AppA_GaussianNormal}
\end{align}
where $g_{vv}$, $g_{vI}$, and $g_{IJ}$ can all be functions of all coordinates. To calculate the vorticity in these coordinates, we recall that every null geodesic covering $N$ can be described by fixing $v$ and $x^I$ and choosing $u=\lambda$. Every such geodesic has a tangent vector with components $l^\mu = \delta^\mu_u$, which means that the components of the vorticity tensor are
\begin{align}
    \omega_{\mu\nu} = \frac{1}{2}\roha{\nabla_\mu l_\nu - \nabla_\nu l_\mu} = \frac{1}{2}\roha{\nabla_\mu g_{\nu u} - \nabla_\nu g_{\mu u}} = 0\ , \label{AppA_NoVort}
\end{align}
where we made use of the metric compatibility of the covariant derivative. Since \eqref{AppA_NoVort} is manifestly tensorial, we conclude that any null congruence emanating orthogonally from a codimension 2 submanifold has $\omega_{ab} = 0$, at least up to a caustic.

\section{Weyl transformations}\label{sec_Weyl}
To finish this appendix, we consider the effect of a Weyl transformation on the geometry and geodesics of a manifold. A Weyl transformation is a local rescaling of the metric tensor, so if a spacetime $\tilde{M}$ has a metric $\tilde{g}_{ab}$, a Weyl transformation assigns it a new metric $g_{ab}$ by mapping
\begin{align}
    \tilde{g}_{ab}(x) \rightarrow g_{ab}(x) = \Omega^2(x)\tilde{g}_{ab}(x)\ , \label{AppA_Weyl}
\end{align}
where $\Omega:\tilde{M}\rightarrow\mathbb{R}$ is a smooth, strictly positive function (since we assume that $g^{ab}$ exists). Alternatively, one may consider two manifolds $\tilde{M}$ and $M$ which can (partially) be identified through a diffeomorphism. If, whenever this identification can be made, the metrics $\tilde{g}_{ab}$ and $g_{ab}$ of $\tilde{M}$ and $M$ respectively are related as in \eqref{AppA_Weyl}, then a Weyl transformation is a map from $\tilde{M}$ to $M$. This allows us to interpret two types of points which warrant special attention, namely those for which $\Omega\rightarrow0$ and those for which $\Omega\rightarrow\infty$. The latter case corresponds to an edge of $M$ because continuing $g_{ab}$ beyond a divergence would mean that $g_{ab}$ is no longer a smooth tensor field. Assuming that the Weyl transformation is invertible we see that $\Omega\rightarrow0$ corresponds to the edge of $\tilde{M}$ by the same argument.

Given that $g_{ab}(x) = \Omega^2(x)\tilde{g}_{ab}(x)$, it is straightforward to calculate the transformed Levi-Civita connection in the coordinate basis:
\begin{align}
    \Gamma^\rho_{\mu\nu} = \frac{1}{2}g^{\rho\lambda}\roha{\partial_\mu g_{\lambda\nu} + \partial_\nu g_{\lambda\mu} - \partial_\lambda\ g_{\mu\nu}} = \tilde{\Gamma}^\rho_{\mu\nu} + \Omega^{-1}\roha{\delta^\rho_\mu\partial_\nu\Omega + \delta^\rho_\nu\partial_\mu\Omega - \tilde{g}_{\mu\nu}\tilde{g}^{\rho\lambda}\partial_\lambda\Omega}\ .
\end{align}
Following \cite{conforme_rand}, we recognise that the partial derivatives in the brackets can be replaced by covariant derivatives, allowing us to give the effect of a Weyl transformation on the covariant derivative of an arbitrary vector $X^a$ in a manifestly coordinate-independent way:
\begin{align}
    \nabla_bX^a = \tilde{\nabla}_bX^a + \Omega^{-1}\roha{\delta^a_b\tilde{\nabla}_c\Omega + \delta^a_c\tilde{\nabla}_b\Omega - \tilde{g}_{bc}\tilde{g}^{ad}\tilde{\nabla}_d\Omega}X^c \ .\label{AppA_WeylCovDer}
\end{align}
Consider now a geodesic $\gamma$ with tangent vector $\tilde{X}^a$ and affine parameter $\tilde{\lambda}$ in $\tilde{M}$. Due to the diffeomorphism between $\tilde{M}$ and $M$, $\gamma$ corresponds to a curve in $M$ with the same tangent vector, but let us choose a different parameter $\lambda$, so that the tangent vector of $\gamma$ in $M$ is $X^a = \diff{\tilde{\lambda}}{\lambda}\tilde{X}^a$. It is then again a matter of computation to find
\begin{align}
    X^a\nabla_aX^b \alis \roha{\diff{\tilde{\lambda}}{\lambda}}^2\tilde{X}^a\tilde{\nabla}_a\tilde{X}^b + \diff{}{\lambda}\roha{\Omega^2\diff{\tilde{\lambda}}{\lambda}}\Omega^{-2}\tilde{X}^b - \roha{\diff{\tilde{\lambda}}{\lambda}}^2\tilde{g}_{cd}\tilde{X}^c\tilde{X}^d\Omega^{-1}\tilde{g}^{ab}\tilde{\nabla}_a\Omega\ .
\end{align}
The first term on the right-hand side will vanish due to the affine parameterisation of $\gamma$ in $\tilde{M}$. The final term will, in general, not vanish if $\gamma$ is time- or spacelike in $\tilde{M}$, meaning that $\gamma$ will not be a geodesic in $M$ in these cases. If however $\gamma$ is a null geodesic in $\tilde{M}$ then it will be a null geodesic in $M$ as well, and $\lambda$ an affine parameter if we choose
\begin{align}
    \diff{\tilde{\lambda}}{\lambda} = \Omega^{-2}\ . \label{AppA_WeylAffine}
\end{align}
It may also be useful to know how the quantities describing a congruence transform. Since time- and spacelike geodesics are not mapped to geodesics, we consider a null congruence, with tangent vector field $l^a = \Omega^{-2}\tilde{l}^a$; if we choose the auxiliary vector fields $k^a$ and $\tilde{k}^a$ from the definition \eqref{AppA_ProjectieNull} of the projection tensor such that $k_al^a = \tilde{k}_a\tilde{l}^a = -1$, it follows that $k^a = \tilde{k}^a$ and $P_{ab} = \Omega^2\tilde{P}_{ab}$. By then applying \eqref{AppA_WeylCovDer} repeatedly to \eqref{AppA_Exp}-\eqref{AppA_Vort}, we obtain
\begin{align}
    \theta \alis \Omega^{-2}\tilde{\theta} + (d-2)\Omega^{-2}\diff{\ln\Omega}{\tilde{\lambda}}\ , \label{AppA_ExpTrans} \\
    \sigma_{ab} \alis \tilde{\sigma}_{ab} - \tilde{l}_a\tilde{\nabla}_b\ln\Omega - \tilde{l}_b\tilde{\nabla}_a\ln\Omega - \roha{\tilde{l}_a\tilde{k}_b + \tilde{l}_b\tilde{k}_a}\diff{\ln\Omega}{\tilde{\lambda}}\ , \\
    \omega_{ab} \alis \tilde{\omega}_{ab}\ ,
\end{align}
where of course the indices of quantities with and without tildes are raised and lowered by $\tilde{g}_{ab}$ and $g_{ab}$ respectively. We conclude that while a Weyl transformation cannot introduce vorticity, it seemingly introduces a shear; however, note that $\sigma_{ab}\sigma^{ab} = \Omega^{-4}\tilde{\sigma}_{ab}\tilde{\sigma}^{ab}$. The transformation of the Jacobi field can be found by combining \eqref{AppA_ExpTrans} with \eqref{AppA_JacobiDef}, leading to
\begin{align}
    \eta = \Omega\tilde{\eta}\ . \label{AppA_WeylJacobi}
\end{align}
Finally, we can analyse the effect of a Weyl transformation on the curvature of the manifold. This can be done by using \eqref{H1_RiemannComponenten} to compute the components of the transformed Riemann tensor explicitly and write them in a coordinate independent way; since we do not need the full Riemann tensor, we only quote the result for the Ricci tensor and scalar here:
\begin{align}
    R_{ab} \alis \tilde{R}_{ab} - (d-2)\roha{\tilde{\nabla}_a\tilde{\nabla}_b\ln\Omega - \tilde{\nabla}_a\ln\Omega\tilde{\nabla}_b\ln\Omega} \non
    &\qquad\ \qquad\ \qquad\ \qquad\ \qquad\ - \tilde{g}^{cd}\roha{\tilde{\nabla}_c\tilde{\nabla}_d\ln\Omega + (d-2)\tilde{\nabla}_c\ln\Omega\tilde{\nabla}_d\ln\Omega}\tilde{g}_{ab}\ , \label{AppA_ConformalRicciT}\\
    R \alis \Omega^{-2}\tilde{R} - (d-1)\Omega^{-2}\tilde{g}^{ab}\roha{2\tilde{\nabla}_a\tilde{\nabla}_b\ln\Omega + (d-2)\tilde{\nabla}_a\ln\Omega\tilde{\nabla}_b\ln\Omega}\ . \label{AppA_ConformalRicciS}
\end{align}
\chapter{Field theoretical background}\label{app_QFT}

This appendix will remind the reader of some basic aspects of quantum field theory (QFT). First, section \ref{sec_ClassicalFields} discusses the formalism of classical field theory. Section \ref{sec_CanonicalQuantisation} then quantises these classical fields using canonical quantisation; section \ref{sec_DensityMatrices} presents an alternative way of representing calculations in QFT. This appendix concludes with a discussion of conformally invariant field theories (CFTs) in section \ref{sec_ConformalFields}.

There is much literature on quantum field theory. Two texts that make use of canonical quantisation are \cite{PeskinSchroeder} and \cite{Schwartz}; our treatment is mostly based on the former.

\section{Classical field theory}\label{sec_ClassicalFields}

As explained in section \ref{sec_Causality}, physics is believed to obey the postulate of local causality: a causal influence can only propagate between two points $p$ and $q$ if there exists a curve with an everywhere timelike or null tangent vector between them \cite{HawkingEllis}. Said differently, if a signal can affect the state of a physical system at some point $p$, then it must have travelled between its source and $p$ at a speed not exceeding the speed of light. A natural way to encode this principle into natural laws is to use local fields. These act as the medium through which interactions propagate, and limit the speed of propagation by demanding that at every point, the field can only influence its immediate (causally connected) surroundings. Examples of fields in classical physics include the electromagnetic field and, in general relativity, the metric tensor.

To formulate the dynamics of a classical field in a diffeomorphism invariant way, we turn to the principle of least action, for which a quantity called the action $S$ is required. In a non-relativistic context, $S$ would be defined as $S = \int L\,\d t$, the time integral of a Lagrangian $L$; a diffeomorphism invariant generalisation defines $S$ with an integral over spacetime, namely
\begin{align}
    S = \int\L\roha{\Phi,\nabla_a\Phi,g_{ab}}\sqrt{-g}\,\d^dx\ . \label{AppB_CurveActie}
\end{align}
Here, \L\ is the Lagrangian density (which we also call the Lagrangian), assumed to be a functional of the field $\Phi$ and its first derivative $\nabla_a\Phi\,$, and $g$ is the determinant of $(g_{\mu\nu})$; the factor $\sqrt{-g}$ ensures that the volume element $\sqrt{-g}\,\d^dx$ is diffeomorphism invariant.

According to the principle of least action, the configuration of $\Phi$ will now be such that $S$ is minimised. Such a minimum occurs at a stationary point of $S$, which is a field configuration such, that an infinitesimal variation in $\Phi$ does not change $S$ at first order in the variation (this is analogous to the way the first derivative of a function vanishes at an extremum). In general, a variation $\Phi\rightarrow\Phi+\delta\Phi$ leads to a change in $S$ of
\begin{align}
    \delta S \alis \int\roha{\parti{\L}{\Phi}\delta\Phi + \parti{\L}{\nabla_a\Phi}\delta\nabla_a\Phi}\sqrt{-g}\,\d^dx + \order(\delta\Phi^2) \non
    \alis \int\roha{\parti{\L}{\Phi} - \nabla_a\roha{\parti{\L}{\nabla_a\Phi}}}\delta\Phi\sqrt{-g}\,\d^dx + \int\nabla_a\roha{\parti{\L}{\nabla_a\Phi}\delta\Phi}\sqrt{-g}\,\d^dx + \order(\delta\Phi^2)\ , \label{AppB_VarActie}
\end{align}
where we have assumed that $\delta\nabla_a\Phi = \nabla_a\delta\Phi$, integrated by parts, and used that $\nabla_a\sqrt{-g} = 0$ due to the metric compatibility of the covariant derivative. The second integral in \eqref{AppB_VarActie} is a boundary term, which vanishes if we assume that $\delta\Phi\rightarrow0$ at the edge of spacetime. At a stationary point of $S$, we should have $\delta S = \order(\delta\Phi^2)$ for all $\delta\Phi$, which by \eqref{AppB_VarActie} implies that
\begin{align}
    \parti{\L}{\Phi} - \nabla_a\roha{\parti{\L}{\nabla_a\Phi}} = 0 \label{AppB_EulerLagrange}
\end{align}
at a stationary point of $S$. We therefore conclude that these are the equations of motion (or Euler-Lagrange equations) which a classical field obeys; if there are multiple fields, \eqref{AppB_EulerLagrange} holds for every one of them. Finally, we may define a conjugate momentum $\Pi(x)$ to the field $\Phi(x)$; analogously to the non-relativistic definition $p = \parti{L}{\dot{x}}$, we do this by defining 
\begin{align}
    \Pi(x) = \parti{\sqrt{-g}\L}{(\nabla_t\Phi)}\ ,
\end{align}
where $t$ is some timelike coordinate.

So far, we have not specified the field $\Phi$; it may be a scalar field, but the same equations apply if it is a vector or tensor field (a treatment of spinors in curved spacetime is beyond the scope of this thesis). Let us therefore consider two important examples of classical field theories: the non-minimally coupled but otherwise free real scalar field $\phi$ and general relativity. The Lagrangian of the former is given by
\begin{align}
    \L = -\frac{1}{2}\nabla_a\phi\nabla^a\phi - \frac{1}{2}\roha{m^2 + \xi R}\phi^2\ , \label{AppB_FreeScalar}
\end{align}
where $m$ is the mass of the field, $R$ the Ricci scalar, and $\xi$ the non-minimal coupling parameter; the latter is named for the fact that $\xi\neq0$ connects $\phi$ to the geometry of spacetime in a way besides the factor of $\sqrt{-g}$ in \eqref{AppB_CurveActie}. By \eqref{AppB_EulerLagrange}, the equations of motion for this field are
\begin{align}
    \roha{\nabla_a\nabla^a - (m^2 + \xi R)}\phi = 0\ . \label{AppB_KleinGordon}
\end{align}
This equation is rather difficult to solve in general. However, in many circumstances the geometry is adequately modelled by Minkowski spacetime, which has $R = 0$ and can be equipped with Cartesian coordinates $x^\mu = (t,\ve{x})$. \eqref{AppB_KleinGordon} then reduces to a familiar form of the Klein-Gordon equation, namely $\roha{\partial_\mu\partial^\mu - m^2}\phi = 0$. To solve this equation, we may expand the dependence of $\phi$ on $\ve{x}$ in Fourier modes:
\begin{align}
    \phi(t,\ve{x}) = \int\frac{\d^{d-1}\ve{p}}{(2\pi)^{d-1}}\phi(t,\ve{p})e^{i\ve{p}\cdot\ve{x}}\ ,
\end{align}
where $\ve{p}\cdot\ve{x} = \delta_{ij}p^ix^j$ with $\delta_{ij}$ the Kronecker delta, and $\phi(t,\ve{p}) = \phi^*(t,-\ve{p})$ to ensure $\phi(t,\ve{x})\inR$. We find that the Klein-Gordon equation is satisfied if $\partial_t^2\phi(t,\ve{p}) + (m^2 + \ve{p}^2)\phi(t,\ve{p}) = 0$, which can be recognised as the equation of motion for a harmonic oscillator with frequency $\omega_\ve{p} = \sqrt{m^2 + \ve{p}^2}$. Its solutions are $\phi(t,\ve{p})\propto\exp(\pm i\omega_\ve{p}t)$, and so any solution to \eqref{AppB_KleinGordon} in Minkowski spacetime can, for some choice of coefficients $\phi_\ve{p}$, be expanded as
\begin{align}
    \phi(x) = \int\frac{\d^{d-1}\ve{p}}{(2\pi)^{d-1}}\roha{\phi_\ve{p}e^{ip\cdot x} + \phi_\ve{p}^*e^{-ip\cdot x}}\ , \label{AppB_KleinGordonExpansion1}
\end{align}
where $p\cdot x = p_\mu x^\mu$ and $p^0 = \omega_{\ve{p}}$. As a matter of aesthetics, one would like this expansion to be in terms of functions that are orthonormal in some appropriate sense. This appropriate sense is provided by the Klein-Gordon inner product, which for two functions $f_1(x)$ and $f_2(x)$ is \cite{Carroll_2019,Fewster_2012}
\begin{align}
    (f_1,f_2) = -i\int_\Sigma\roha{f_1\nabla_\mu f_2^* - f_2^*\nabla_\mu f_1}n^\mu\sqrt{g_\Sigma}\,\d^{d-1}x\ , \label{AppB_KGinprod}
\end{align}
where $\sqrt{g_\Sigma}\,\d^{d-1}x$ is the area element of a spacelike hypersurface $\Sigma$ with unit normal vector $n^\mu$ (i.e. $g_\Sigma$ is the determinant of the restriction of the metric to $\Sigma$). From \eqref{AppB_KleinGordonExpansion1}, we see that the basis functions in which $\phi(x)$ is expanded are $f_\ve{p}(x)\propto e^{ip\cdot x}$; we fix the normalisation by demanding that $(f_{\ve{p}},f_{\ve{q}}) = (2\pi)^{d-1}\delta(\ve{p} - \ve{q})$ if $\Sigma$ is the hypersurface defined by $t=0$. With this normalisation, the expansion \eqref{AppB_KleinGordonExpansion1} can be rewritten as
\begin{align}
    \phi(x) = \int\frac{\d^{d-1}\ve{p}}{(2\pi)^{d-1}}\frac{1}{\sqrt{2\omega_\ve{p}}}\roha{a_\ve{p}e^{ip\cdot x} + a^*_\ve{p}e^{-ip\cdot x}} \label{AppB_KleinGordonExpansion2}
\end{align}
for some coefficients $a_\ve{p}$. Finally, the conjugate momentum of $\phi$ in Minkowski spacetime is $\pi(x) = \partial_t\phi(x)$.

Another important classical field theory is general relativity. In this theory, the metric $g_{ab}$ is promoted to a dynamical field and assigned the Einstein-Hilbert action; writing $S_\mathrm{mat}$ for the action of the relevant matter fields, the total action for general relativity is \cite{HawkingEllis}
\begin{align}
    S_\mathrm{gr} = \frac{1}{16\pi G_N}\int R\,\sqrt{-g}\,\d^dx + S_\mathrm{mat}\ ,
\end{align}
where the first term on the right-hand side is the Einstein-Hilbert action. Applying the steps from \eqref{AppB_CurveActie}-\eqref{AppB_EulerLagrange} to $S_\mathrm{gr}$ to obtain equations of motion for the metric is complicated by two factors: the Ricci scalar $R$ is not constructed from covariant derivatives of $g_{ab}$ and since we want to know the equations of motion for the metric, there should be a contribution from the variation of the volume element in \eqref{AppB_VarActie}. Nevertheless, one can vary the metric as $g_{ab} \rightarrow g_{ab} + \delta g_{ab}$; the resulting change in $S_\mathrm{gr}$ is \cite{Carroll_2019}
\begin{align}
    \delta S_\mathrm{gr} = \frac{1}{16\pi G_N}\int\roha{R_{ab} - \frac{1}{2}g_{ab}R}\delta g^{ab}\,\sqrt{-g}\,\d^dx + \delta S_\mathrm{mat} + \order\roha{(\delta g^{ab})^2}\ . \label{AppB_VarGrav1}
\end{align}
To simplify the notation, we introduce the functional derivative of $S$ with respect to $g^{ab}$ via
\begin{align}
    \delta S = \int\funci{S}{g^{ab}}\delta g^{ab}\,\d^dx + \order\roha{(\delta g^{ab})^2}\ .
\end{align}
Here, $\delta S$ is the change in $S$ due to a variation in the metric and $\funci{S}{g^{ab}}$ is the functional derivative of $S$ with respect to $g^{ab}$. With this definition, \eqref{AppB_VarGrav1} can be written as
\begin{align}
    \delta S_\mathrm{gr} = \frac{1}{16\pi G_N}\int\roha{R_{ab} - \frac{1}{2}g_{ab}R + \frac{16\pi G_N}{\sqrt{-g}}\funci{S_\mathrm{mat}}{g^{ab}}}\delta g^{ab}\,\sqrt{-g}\,\d^dx + \order\roha{(\delta g^{ab})^2}\ . \label{AppB_VarGrav2}
\end{align}
Because of the principle of least action, the configuration of the metric will be such that $\delta S_\mathrm{gr} = \order\roha{(\delta g^{ab})^2}$ for any $\delta g^{ab}$; by \eqref{AppB_VarGrav2}, this implies that the metric should satisfy
\begin{align}
    R_{ab} - \frac{1}{2}g_{ab}R = -\frac{16\pi G_N}{\sqrt{-g}}\funci{S_\mathrm{mat}}{g^{ab}}\ .
\end{align}
By comparing this to the field equations of general relativity in \eqref{H1_EFE}, the energy-momentum tensor of the matter fields must be identified as
\begin{align}
    T_{ab} = -\frac{2}{\sqrt{-g}}\funci{S_\mathrm{mat}}{g^{ab}}\ . \label{AppB_EMT}
\end{align}
This can be applied to the non-minimally coupled but otherwise free real scalar field from \eqref{AppB_FreeScalar}, leading to an energy-momentum tensor \cite{NonMinimalScalar}
\begin{align}
    T_{ab} = \nabla_a\phi\nabla_b\phi - \frac{1}{2} \roha{\nabla_c\phi\nabla^c\phi + m^2\phi^2}g_{ab} + \xi\roha{R_{ab} - \frac{1}{2}g_{ab}R + g_{ab}\nabla^c\nabla_c - \nabla_a\nabla_b}\phi^2\ . \label{AppB_NonMinimalEMT}
\end{align}

\section{Canonical quantisation}\label{sec_CanonicalQuantisation}
The previous section described several aspects of classical field theories; such classical theories can be developed into QFTs by a process known as `quantisation'. In this section, we discuss canonical quantisation, a rather simple way of going about this process, by quantising the free, real scalar field $\phi$ (which we considered in the previous section) in Minkowski spacetime.

The essential idea of canonical quantisation is that we reinterpret the dynamical variables of a classical field theory as operators acting on a Hilbert space and obeying the canonical commutation relations. For example, non-relativistic quantum mechanics is usually concerned with systems that are classically described by a set of particles with generalised coordinates $q_j$ and conjugate momenta $p_j$; these variables are promoted to operators $\hat{q}_j$ and $\hat{p}_j$, which obey the canonical commutation relations $[\hat{q}_j,\hat{p}_k] = i\delta_{jk}$ and $[\hat{q}_j,\hat{q}_k] = [\hat{p}_j,\hat{p}_k] = 0$. The analogous quantities for field theories in flat spacetime are the field $\Phi(t,\ve{x})$ and its conjugate momentum $\Pi(t,\ve{y})$ evaluated at the same time coordinate, which are promoted to operators $\hat{\Phi}(t,\ve{x})$ and $\hat{\Pi}(t,\ve{y})$. If we take $\Phi$ to be the real scalar field $\phi$, which was discussed in the previous section, then the generalisation of the non-relativistic canonical commutation relations is straightforward:
\begin{align}
    [\hat{\phi}(t,\ve{x}),\hat{\pi}(t,\ve{y})] = i\delta\roha{\ve{x} - \ve{y}}\ , & &[\hat{\phi}(t,\ve{x}),\hat{\phi}(t,\ve{y})] = [\hat{\pi}(t,\ve{x}),\hat{\pi}(t,\ve{y})] = 0\ .\label{AppB_CanoniekKwant1}
\end{align}
Of course, promoting $\phi$ and $\pi$ to operators implies that the coefficients $a_\ve{p}$ and $a_\ve{p}^*$ from \eqref{AppB_KleinGordonExpansion2} are also promoted to operators, respectively $\hat{a}_\ve{p}$ and $\hat{a}^\dagger_\ve{p}$. Using \eqref{AppB_KleinGordonExpansion2} and the fact that classically, $\pi(t,\ve{x}) = \partial_t\phi(t,\ve{x})$, one can express $\hat{a}_\ve{p}$ in terms of $\hat{\phi}$ and $\hat{\pi}$:
\begin{align}
    \hat{a}_\ve{p} = \frac{1}{\sqrt{2\omega_\ve{p}}}\int\roha{i\hat{\pi}(x) + \omega_\ve{p}\hat{\phi}(x)}e^{-ip\cdot x}\d^{d-1}\ve{x}\ . \label{AppB_Annihilator}
\end{align}
Then, using \eqref{AppB_Annihilator} in \eqref{AppB_CanoniekKwant1}, the commutation relations for $\hat{a}_\ve{p}$ and $\hat{a}^\dagger_\ve{p}$ can be found:
\begin{align}
    [\hat{a}_\ve{p},\hat{a}^\dagger_\ve{q}] = (2\pi)^{d-1}\delta(\ve{p} -\ve{q})& &[\hat{a}_\ve{p},\hat{a}_\ve{q}] = [\hat{a}^\dagger_\ve{p},\hat{a}^\dagger_\ve{q}] = 0\ .\label{AppB_CanoniekKwant2}
\end{align}
To interpret the effect of acting with $\hat{a}_\ve{p}$ and $\hat{a}^\dagger_\ve{p}$, note from \eqref{AppB_Annihilator} and \eqref{AppB_CanoniekKwant2} that they are the QFT analogues of the annihilation and creation operators of the quantum mechanical harmonic oscillator \cite{Griffiths_QM}. Thus, acting with $\hat{a}^\dagger_\ve{p}$ on a state is interpreted as adding an excitation with momentum $\ve{p}$ to the field, while acting with $\hat{a}_\ve{p}$ is interpreted as removing that same excitation. This way, $\hat{a}^\dagger_\ve{p}$ and $\hat{a}_\ve{p}$ define particles: a particle with momentum $\ve{p}$ is an excitation created by $\hat{a}^\dagger_\ve{p}$ and annihilated by $\hat{a}_\ve{p}$. $\hat{a}^\dagger_\ve{p}$ and $\hat{a}_\ve{p}$ also define a vacuum state $\ket{0}$ as the state without particles, i.e. such that $\hat{a}_\ve{p}\ket{0} = 0$ for all $\ve{p}$; by repeatedly acting on $\ket{0}$ with $\hat{a}^\dagger_{\ve{p}}$ for various $\ve{p}$, one obtains the Hilbert space \calH\ of the theory. Physical observables are obtained as expectation values; for example, the expectation value of the field in a state $\ket{\psi}\in\calH$ is $\bra{\psi}\hat{\phi}(x)\ket{\psi} = \smallmean{\hat{\phi}(x)}_\psi$.

Besides the field operators themselves, all other dynamical quantities from the classical theory are promoted to operators in the QFT. The usual strategy for this is to take the classical expression for some quantity in terms of $\phi(x)$ and $\pi(x)$, and then replace $\phi\rightarrow\hat{\phi}$ and $\pi\rightarrow\hat{\pi}$. However, if one would do this with e.g. the energy-momentum tensor in \eqref{AppB_NonMinimalEMT}, one needs to understand how to handle products of operators like $\hat{\phi}(x)\hat{\phi}(x)$. The naive expression, obtained by simply multiplying the operator version of \eqref{AppB_KleinGordonExpansion2}, is problematic:
\begin{align}
    \hat{\phi}(x)\hat{\phi}(x) \alis \int\frac{\d^{d-1}\ve{p}}{(2\pi)^{d-1}}\frac{\d^{d-1}\ve{q}}{(2\pi)^{d-1}}\frac{1}{\sqrt{4\omega_\ve{p}\omega_\ve{q}}}\left(\hat{a}_\ve{p}\hat{a}_\ve{q}e^{i(p+q)\cdot x} + \hat{a}^\dagger_\ve{p}\hat{a}^\dagger_\ve{q}e^{-i(p+q)\cdot x} \right. \non
    &\qquad\ \qquad\ \qquad\ \qquad\ \left. +\ \hat{a}^\dagger_\ve{p}\hat{a}_\ve{q}e^{-i(p-q)\cdot x} + \hat{a}^\dagger_\ve{q}\hat{a}_\ve{p}e^{i(p-q)\cdot x}\right) + \int\frac{\d^{d-1}\ve{p}}{(2\pi)^{d-1}}\frac{1}{2\omega_\ve{p}}\ .
\end{align}
The ordering of the creation and annihilation operators, with every creation operator to the left of the annihilation operators, is chosen so that the operator part of this expression has a vanishing expectation value in the vacuum state. However, this required commuting $\hat{a}_\ve{p}$ and $\hat{a}^\dagger_\ve{q}$, giving rise to the final term. This term is divergent, meaning that as an operator, $\hat{\phi}(x)\hat{\phi}(x)$ does not make much sense: it gives an infinite expectation value in every state.

A resolution to this problem is provided by a procedure known as vacuum subtraction \cite{Wald_1994}. For this procedure, we first point-split the quantity we are interested in; in the present context, this would mean that we consider $\hat{\phi}(x)\hat{\phi}(y)$ for $x\neq y$. Then, we subtract the expectation value of the resulting operator in the vacuum state to remove the offending term(s). Finally, we take the coincidence limit $y\rightarrow x$ to obtain an operator that comes as close to the divergent original operator as possible. For $\hat{\phi}(x)\hat{\phi}(x)$, this procedure is carried out as follows:
\begin{align}
    \normal{\hat{\phi}^2(x)}\ \alis \lim_{y\rightarrow x}\roha{\hat{\phi}(x)\hat{\phi}(y) - \bra{0}\hat{\phi}(x)\hat{\phi}(y)\ket{0}} \non
    \alis  \int\frac{\d^{d-1}\ve{p}}{(2\pi)^{d-1}}\frac{\d^{d-1}\ve{q}}{(2\pi)^{d-1}}\frac{1}{\sqrt{4\omega_\ve{p}\omega_\ve{q}}}\roha{\hat{a}_\ve{p}\hat{a}_\ve{q}e^{i(p+q)\cdot x} + \hat{a}^\dagger_\ve{p}\hat{a}_\ve{q}e^{-i(p-q)\cdot x}} + \mathrm{h.c.}\ ,
\end{align}
where the double colon indicates a vacuum subtracted operators and the abbreviation `h.c.' denotes the Hermitian conjugate of the terms to its left. The point-splitting ensures that we are dealing with sensible (divergence-free) operators at every step; it is therefore a way of regulating the divergence in $\hat{\phi}(x)\hat{\phi}(x)$. Vacuum subtraction as a whole is an example of a renormalisation procedure, wherein one systematically removes (regulated) divergences.

\section{Density matrices}\label{sec_DensityMatrices}
The previous section demonstrated how a classical field theory can be used to construct a QFT via canonical quantisation. The resulting theory revolved around operators acting on states from the Hilbert space $\calH$, and extracted observables by taking expectation values. In this section, we consider an alternative way of obtaining observables: the density matrix \cite{Griffiths_QM}.

Consider a state $\ket{\psi}\in\calH$; its density matrix is defined as $\rho^\psi = \ket{\psi}\bra{\psi}$. Assuming that there exists a complete basis\footnote{For simplicity, this basis will be assumed to be discrete. In practice one would have to take a continuum limit, for example using a path integral.} $\cuha{\ket{\psi_i}}$ of $\calH$, the expectation value of an operator $\hat{\order}$ is
\begin{align}
    \bra{\psi}\hat{\order}\ket{\psi} = \sum_i\bra{\psi}\hat{\order}\ket{\psi_i}\inprod{\psi_i}{\psi} = \sum_i\bra{\psi_i}\rho\hat{\order}\ket{\psi_i} \equiv \Tr(\rho\hat{\order})\ , \label{AppB_DensityExpectation}
\end{align}
which also defined the trace operation. This can be generalised to mixed states, field configurations that cannot be described by an element of \calH, by defining the density matrix as
\begin{align}
    \rho^\psi = \sum_\alpha c^\psi_\alpha\ket{\psi_\alpha}\bra{\psi_\alpha}\ .
\end{align}
In this definition, $\ket{\psi_\alpha}\in\calH$ are unit vectors and the $c^\psi_\alpha\in[0,1]$ are normalised such that $\sum_\alpha c^\psi_\alpha = 1$. The expectation value of an operator is then defined analogously to \eqref{AppB_DensityExpectation}, as
\begin{align}
    \smallmean{\hat{\order}}_\psi = \Tr(\rho^\psi\hat{\order})\ . \label{AppB_DensityExpMixed}
\end{align}
We may also consider a system for which the Hilbert space factorises, as $\calH = \calH_1\otimes\calH_2$. In that case, any state $\ket{\psi}\in\calH$ may be written as $\ket{\psi} = \smallket{\psi^1}\smallket{\psi^2}$, and the density matrix for subsystem 1 can be defined by taking a trace over subsystem 2 as follows:
\begin{align}
    \Tr_2(\rho^\psi) \alis \sum_i\smallbra{\psi^2_i}\roha{\sum_\alpha c^\psi_\alpha\smallket{\psi^1_\alpha}\smallket{\psi^2_\alpha}\smallbra{\psi^2_\alpha}\smallbra{\psi^1_\alpha}}\smallket{\psi^2_i} \non
    \alis \sum_{\alpha}\sum_i\abs{\smallinprod{\psi^2_i}{\psi^2_\alpha}}^2c^\psi_\alpha\smallket{\psi^1_\alpha}\smallbra{\psi^1_\alpha} \non
    \alis \sum_\alpha c^\psi_\alpha\smallket{\psi^1_\alpha}\smallbra{\psi^1_\alpha} \equiv \rho^\psi_1\ , \label{AppB_ReducedDensity}
\end{align}
where we have chosen a basis $\cuha{\ket{\psi^2_i}}$ for $\calH_2$ and assumed that $\ket{\psi^1_\alpha}\in\calH_1$ and $\ket{\psi^2_\alpha}\in\calH_2$ are unit vectors. To confirm that $\rho^\psi_1$ really acts like a density matrix, we can consider the expectation value of an operator $\hat{\order}_1$ which only acts on subsystem 1, which can be written as
\begin{align}
    \smallmean{\hat{\order}_1}_\psi \alis \Tr(\rho^\psi\hat{\order}_1) \non
    \alis \sum_{i,j}\smallbra{\psi^2_i}\smallbra{\psi^1_j} \rho^\psi\hat{\order}_1\smallket{\psi^1_j}\smallket{\psi^2_i} \non
    \alis \sum_j\smallbra{\psi^1_j}\rho^\psi_1\hat{\order}_1 \smallket{\psi^1_j} \non
    \alis \Tr_1(\rho^\psi_1\hat{\order}_1)\ .
\end{align}
With the necessary replacements, this is indeed analogous to \eqref{AppB_DensityExpMixed} when the trace and the density matrix are restricted to subsystem 1; $\rho^\psi_1$ is therefore called the reduced density matrix. 

\section{Conformal field theory}\label{sec_ConformalFields}
The action as defined in section \ref{sec_ClassicalFields} is, by construction, invariant under general diffeomorphisms. However, inspired by section \ref{sec_Weyl} one may ask how a Weyl transformation \eqref{AppA_Weyl} would affect fields other than the metric, or equivalently, how one would relate a field theory in a spacetime $\tilde{M}$ with metric $\tilde{g}_{ab}$ to the same theory in a spacetime $M$ with metric $g_{ab} = \Omega^2\tilde{g}_{ab}$. In this section, we briefly discuss two instances in which this question can be answered to some degree: a Weyl invariant classical field and conformally invariant quantum fields (CFTs). There are many much more pedagogical treatments of CFTs available, e.g. \cite{CFT_Modave,CFT_Rychkov}.

\subsection{Weyl invariance}
To discuss Weyl invariance at the classical level, we must first understand how Weyl transformations affect classical fields. We start by realising that the action is dimensionless (we set $\hbar = c = 1$). Since the volume element has mass dimension $-d$ (i.e. $\sqrt{-g}\,\d^dx$ has dimensions $(\text{mass})^{-d}$), the Lagrangian must have mass dimension $d$. From this restriction, the mass dimension $\Delta_\Phi$ of any field $\Phi$ can be found. If one then rescales all distances by a factor $\lambda$, the fields are naturally rescaled as
\begin{align}
    \Phi(x) \rightarrow \Phi'(\lambda x) = \lambda^{-\Delta_\Phi}\Phi(x)\ .
\end{align}
A Weyl transformation $\tilde{g}_{ab} \rightarrow g_{ab} = \Omega^2\tilde{g}_{ab}$ can be interpreted as a local rescaling of distances by a factor $\Omega(x)$. For sufficiently local fields, we therefore guess that they transform as
\begin{align}
    \tilde{\Phi}(x) \rightarrow \Phi(x) = \Omega^{-\Delta_{\tilde{\Phi}}}(x)\tilde{\Phi}(x)\ . \label{AppB_WeylVeld}
\end{align}
Fields which transform like this are called primary fields; a non-primary field is e.g. $\nabla_a\Phi$.

With the transformation \eqref{AppB_WeylVeld}, we can begin to understand how the dynamics of $\Phi$ differ from those of $\tilde{\Phi}$. As an example, consider the non-minimally coupled real scalar $\phi$. From its Lagrangian \eqref{AppB_FreeScalar} one finds that $\Delta_\phi = \frac{1}{2}d - 1$, so a Weyl transformation $\tilde{g}_{ab} \rightarrow g_{ab} = \Omega^2\tilde{g}_{ab}$ transforms the Lagrangian as
\begin{align}
    -\frac{1}{2}\tilde{g}^{ab}\tilde{\nabla}_a\tilde{\phi}\tilde{\nabla}_b\tilde{\phi} - \frac{1}{2}\roha{m^2 + \xi\tilde{R}}\tilde{\phi}^2 \longrightarrow&\ -\frac{1}{2}g^{ab}\nabla_a\phi\nabla_b\phi - \frac{1}{2}\roha{m^2 + \xi R}\phi^2 \non
    \alis -\frac{1}{2}\Omega^{-d}\viha{\tilde{g}^{ab}\tilde{\nabla}_a\tilde{\phi}\tilde{\nabla}_b\tilde{\phi} + \roha{m^2\Omega^2 + \xi\tilde{R}}\tilde{\phi}^2} \non
    &\ - \frac{d-1}{2\Omega^d}\roha{\frac{d-2}{4(d-1)} - \xi}\tilde{\phi}^2\tilde{g}^{ab}\left(2\tilde{\nabla}_a\tilde{\nabla}_b\ln\Omega \right. \non
    &\qquad\ \qquad\ \qquad\ \quad\ \left. +\ (d-2)\tilde{\nabla}_a\ln\Omega\tilde{\nabla}_b\ln\Omega\right)\ , \label{AppB_WeylLagrangeGeneral}
\end{align}
where we have used \eqref{AppA_ConformalRicciS} and \eqref{AppB_WeylVeld}. From \eqref{AppB_WeylLagrangeGeneral}, one can observe that for the particular choice of parameters $m=0$ and $\xi = \frac{1}{4}(d-2)/(d-1)$, the Lagrangian transforms as $\L\rightarrow\Omega^{-d}\L$. By the definition \eqref{AppB_CurveActie}, this in turn implies that for $m=0$ and $\xi = \frac{1}{4}(d-2)/(d-1)$, we have
\begin{align}
    S[\phi,g_{ab}] = S[\tilde{\phi},\tilde{g}_{ab}] \label{AppB_WeylInvariant}
\end{align}
with $\phi = \Omega^{1-d/2}\tilde{\phi}$ and $g_{ab} = \Omega^2\tilde{g}_{ab}$; any field theory which obeys \eqref{AppB_WeylInvariant} is called Weyl invariant. A useful property of Weyl invariant theories is that the energy-momentum tensors $\tilde{T}_{ab}$ and $T_{ab}$ of the theory in $\tilde{M}$ and $M$ can be related, which can be seen by realising that a variation $\delta\tilde{g}^{ab}$ induces a variation $\delta g^{ab} = \Omega^{-2}\delta\tilde{g}^{ab}$. Then, by \eqref{AppB_EMT}, we have
\begin{align}
    T_{ab} = -\frac{2}{\sqrt{-g}}\funci{S[\Phi,g_{ab}]}{g^{ab}} = -\frac{2}{\Omega^d\sqrt{-\tilde{g}}}\frac{\delta S[\tilde{\Phi},\tilde{g}_{ab}]}{\Omega^{-2}\delta\tilde{g}^{ab}} = \Omega^{2-d}\tilde{T}_{ab}\ , \label{AppB_ClassicalEMTtrans}
\end{align}
where we generalised to any Weyl invariant field theory. Another useful property is that the energy-momentum tensor is traceless, which follows from performing an infinitesimal Weyl transformation. In effect, this varies the metric by $\delta g^{ab} = \sigma(x)g^{ab}$, which in turn changes $S$ by
\begin{align}
    \delta S = \int\funci{S}{g^{ab}}\delta g^{ab}\d^dx + \order\roha{(\delta g^{ab})^2} = -\frac{1}{2}\int\sigma(x)T_{ab}g^{ab}\,\sqrt{-g}\,\d^dx + \order(\sigma^2)\ . \label{AppB_EMTtrace}
\end{align}
If the theory is Weyl invariant, then any infinitesimal Weyl transformation should lead to $\delta S = 0$; by \eqref{AppB_EMTtrace}, this implies that $g^{ab}T_{ab} = 0$.

For QFTs, the effect of Weyl transformations is more involved. For one, quantising a Weyl invariant classical field theory does not necessarily lead to a Weyl invariant QFT; in fact, the invariance is usually broken by quantum effects in interactions. Even in the case that one has a Weyl invariant QFT, not all results from the classical theory carry over to the QFT. Although the transformation \eqref{AppB_WeylVeld} can be reinterpreted as a statement about primary operators, the transformation \eqref{AppB_ClassicalEMTtrans} of the energy-momentum tensor must in general be modified to
\begin{align}
    U\tilde{T}_{ab}U^\dagger = \Omega^{d-2}\roha{T_{ab} - X_{ab}}\ , \label{AppB_CFTEMT}
\end{align}
where $U:\tilde{\calH}\rightarrow\calH$ is a unitary map from the Hilbert space $\tilde{\calH}$ of the QFT in a geometry described by $\tilde{g}_{ab}$ to the Hilbert space $\calH$ of the same QFT in a geometry described by $g_{ab} = \Omega^2\tilde{g}_{ab}$; $X_{ab}$ is called the conformal or Weyl anomaly. It is completely determined by the geometry, vanishes for odd $d$, and arises due to the need to introduce geometric counterterms that render the action of a QFT finite \cite{Conforme_anomalie,holographic_EMT,duff_anomaly}; from the point of view of vacuum subtraction, one might say that the vacuum expectation value which should be subtracted from the energy-momentum tensor depends on the geometry in a way that is captured by $X_{ab}$ \cite{Wald_1994,Wald_TraceAnomaly}. Similarly, renormalisation in curved spacetime leads to the violation of the classical identity $g^{ab}T_{ab} = 0$; instead, the trace of the energy-momentum tensor becomes equal to a local geometric quantity which vanishes in Minkowski spacetime \cite{Wald_1994}. 

\subsection{Conformal invariance}
Closely related to Weyl invariance is the concept of conformal invariance. A conformally invariant field theory is not demanded to be invariant under general Weyl transformations but only under those Weyl transformations that leave the metric invariant up to a diffeomorphism, i.e. such that
\begin{align}
    g_{\mu\nu}(x) = \Omega^2(x)\tilde{g}_{\mu\nu}(x) = \parti{\tilde{x}^\rho}{x^\mu}\parti{\tilde{x}^\sigma}{x^\nu}\tilde{g}_{\rho\sigma}(\tilde{x})\ .
\end{align}
The combination of such a Weyl transformation with a diffeomorphism that removes the factor $\Omega^2$ is called a conformal transformation\footnote{Some authors reserve the term `conformal transformation' for the Weyl transformation \cite{WeylVsConformal}, while others apply it only to the diffeomorphism \cite{CFT_Modave}.}; many spacetimes do not allow conformal transformations, because there are no diffeomorphisms that only rescale their metric. However, they are allowed in Minkowski spacetime. Specifically, Lorentz boosts and translations are conformal transformations when combined with a trivial Weyl transformation ($\Omega=1$), as are rigid scale transformations (dilatations) when they follow a uniform Weyl transformation, with $\Omega = \lambda$ a constant. Finally, one may do a Weyl transformation with $\Omega^2 = (1-2b\cdot x + b^2x^2)^{-2}$ for some fixed vector $b^a$ and $x^2 = \eta_{\mu\nu}x^\mu x^\nu$; one obtains a conformal transformation by following this Weyl transformation with a diffeomorphism which sends
\begin{align}
    x^\mu \rightarrow x'^\mu = \frac{x^\mu - b^\mu x^2}{1 - b_\nu x^\nu + b^2x^2}\ .
\end{align}
These transformations (boosts, translations, dilatations and special conformal transformations) leave the Minkowski metric invariant, making them highly relevant to QFTs in flat spacetime. If the dynamics of a QFT are invariant under conformal transformations, the QFT is called a conformal field theory (CFT); an explicit example of a CFT is the real scalar described by \eqref{AppB_FreeScalar} for $m = 0$ and $\xi = \frac{1}{4}(d-2)/(d-1)$. This is the reason that this value of $\xi$ is also called conformal coupling.

CFTs are closely related to Weyl invariant theories since conformal transformations are based on a subset of all possible Weyl transformations. This means that our discussion of Weyl invariant theories also applies to CFTs; in particular, \eqref{AppB_WeylVeld} and \eqref{AppB_CFTEMT} continue to describe the transformation of respectively primary operators and the energy-momentum tensor under the Weyl transformation involved in a conformal transformation (the diffeomorphism may introduce an additional transformation). In fact, the connection between CFTs and Weyl invariant QFTs goes even deeper: it can be shown that field theories that are conformally invariant in Minkowski spacetime are also Weyl invariant in general spacetimes \cite{WeylVsConformal}.

\bibliographystyle{unsrt}
\bibliography{sources}

\begin{thebibliography}{10}

\bibitem{Einstein}
A.~{Einstein}.
\newblock {Die Feldgleichungen der Gravitation}.
\newblock {\em Sitzungsberichte der Königlich Preußischen Akademie der Wissenschaften}, pages 844--847, January 1915.

\bibitem{HawkingEllis}
S.~W. Hawking and G.~F.~R. Ellis.
\newblock {\em {The large scale structure of space-time}}.
\newblock {Cambridge University Press}, Cambridge, May 1973.

\bibitem{Carroll_2019}
S.~M. Carroll.
\newblock {\em {Spacetime and geometry}}.
\newblock {Cambridge University Press}, 2019.

\bibitem{Oppenheimer_1939}
J.~R. Oppenheimer and H.~Snyder.
\newblock {On Continued Gravitational Contraction}.
\newblock {\em {Phys. Rev.}}, 56:455--459, September 1939.

\bibitem{Kontou_review}
E.~A. Kontou and K.~Sanders.
\newblock Energy conditions in general relativity and quantum field theory.
\newblock {\em {Classical and Quantum Gravity}}, 37(19):193001, September 2020.

\bibitem{Olum_1998}
K.~D. Olum.
\newblock {Superluminal Travel Requires Negative Energies}.
\newblock {\em {Phys. Rev. Lett.}}, 81:3567--3570, October 1998.

\bibitem{AANEC}
N.~Graham and K.~D. Olum.
\newblock Achronal averaged null energy condition.
\newblock {\em {Phys. Rev. D}}, 76:064001, September 2007.

\bibitem{wormholes_ANEC}
M.~Visser, S.~Kar, and N.~Dadhich.
\newblock {Traversable Wormholes with Arbitrarily Small Energy Condition Violations}.
\newblock {\em {Phys. Rev. Lett.}}, 90:201102, May 2003.

\bibitem{Senovilla_1998}
J.~M.~M. Senovilla.
\newblock Singularity theorems and their consequences.
\newblock {\em {General Relativity and Gravitation}}, 30(5):701–848, May 1998.

\bibitem{Penrose_1965}
R.~Penrose.
\newblock {Gravitational Collapse and Space-Time Singularities}.
\newblock {\em Phys. Rev. Lett.}, 14:57--59, January 1965.

\bibitem{EGJ_NietPuntsgewijs}
H.~Epstein, V.~Glaser, and A.~Jaffe.
\newblock Nonpositivity of the energy density in quantized field theories.
\newblock {\em {Il Nuovo Cimento}}, 36(3):1016–1022, April 1965.

\bibitem{ANEC_Entropy}
T.~Faulkner, R.~G. Leigh, O.~Parrikar, and H.~Wang.
\newblock {Modular Hamiltonians for deformed half-spaces and the averaged null energy condition}.
\newblock {\em {Journal of High Energy Physics}}, 2016(9), September 2016.

\bibitem{ANEC_causaliteit}
T.~Hartman, S.~Kundu, and A.~Tajdini.
\newblock {Averaged null energy condition from causality}.
\newblock {\em {Journal of High Energy Physics}}, 2017(7), July 2017.

\bibitem{ANECviolation_Schwarzschild}
M.~Visser.
\newblock {Gravitational vacuum polarization. II. Energy conditions in the Boulware vacuum}.
\newblock {\em {Phys. Rev. D}}, 54:5116--5122, October 1996.

\bibitem{ANEC_violation_old}
D.~Urban and K.~D. Olum.
\newblock Averaged null energy condition violation in a conformally flat spacetime.
\newblock {\em {Phys. Rev. D}}, 81:024039, January 2010.

\bibitem{ANEC_violation_new}
A.~Ishibashi, K.~Maeda, and E.~Mefford.
\newblock {Achronal averaged null energy condition, weak cosmic censorship, and AdS/CFT duality}.
\newblock {\em {Phys. Rev. D}}, 100:066008, September 2019.

\bibitem{Rosenfeld_1963}
L.~Rosenfeld.
\newblock {On quantization of fields}.
\newblock {\em {Nuclear Physics}}, 40:353–356, January 1963.

\bibitem{Fewster_2012}
C.~J. Fewster.
\newblock Lectures on quantum energy inequalities, August 2012.

\bibitem{QEI_2D}
C.~J. Fewster and S.~Hollands.
\newblock Quantum energy inequalities in two-dimensional conformal field theory.
\newblock {\em {Reviews in Mathematical Physics}}, 17(05):577–612, June 2005.

\bibitem{DSNEC_scalar}
J.~R. Fliss, B.~W. Freivogel, and E.~A. Kontou.
\newblock {The double smeared null energy condition}.
\newblock {\em {SciPost Phys.}}, 14:024, 2023.

\bibitem{DSNEC_fermion}
D.~Fragoso and L.~Guo.
\newblock The fermionic double smeared null energy condition, May 2024.

\bibitem{CANEC_odd}
N.~Iizuka, A.~Ishibashi, and K.~Maeda.
\newblock {Conformally invariant averaged null energy condition from AdS/CFT}.
\newblock {\em {Journal of High Energy Physics}}, 2020(3), March 2020.

\bibitem{CANEC_even}
N.~Iizuka, A.~Ishibashi, and K.~Maeda.
\newblock {The averaged null energy conditions in even dimensional curved spacetimes from AdS/CFT duality}.
\newblock {\em {Journal of High Energy Physics}}, 2020(10), October 2020.

\bibitem{LO_original}
H.~Casini, E.~Testé, and G.~Torroba.
\newblock {Modular Hamiltonians on the null plane and the Markov property of the vacuum state}.
\newblock {\em {Journal Of Physics A: Mathematical And Theoretical}}, 50(36):364001, August 2017.

\bibitem{LO_Mathys}
A.~Belin, D.~M. Hofman, G.~Mathys, and M.~T. Walters.
\newblock On the stress tensor light-ray operator algebra.
\newblock {\em {Journal of High Energy Physics}}, 2021(5), May 2021.

\bibitem{holographic_ANEC}
W.~R. Kelly and A.~C. Wall.
\newblock Holographic proof of the averaged null energy condition.
\newblock {\em {Phys. Rev. D}}, 90:106003, Nov 2014.

\bibitem{Rosso_2020}
F.~Rosso.
\newblock {Global Aspects of Conformal Symmetry and the ANEC in dS and AdS}.
\newblock {\em {Journal of High Energy Physics}}, 2020(3), March 2020.

\bibitem{Wall}
A.~C. Wall.
\newblock Proving the achronal averaged null energy condition from the generalized second law.
\newblock {\em {Phys. Rev. D}}, 81:024038, January 2010.

\bibitem{abstracte_indices}
R.~Penrose and W.~Rindler.
\newblock {\em {Spinors and Space-Time: Volume 1, Two-Spinor Calculus and Relativistic Fields}}.
\newblock {Spinors and Space-time}. {Cambridge University Press}, May 1987.

\bibitem{Witten_causality}
E.~Witten.
\newblock {Light rays, singularities, and all that}.
\newblock {\em Rev. Mod. Phys.}, 92, November 2020.

\bibitem{Curiel2017}
E.~Curiel.
\newblock {\em A Primer on Energy Conditions}, pages 43--104.
\newblock Springer New York, New York, NY, 2017.

\bibitem{Earman_2014}
J.~Earman.
\newblock {No superluminal propagation for classical relativistic and relativistic quantum fields}.
\newblock {\em {Studies in History and Philosophy of Science Part B: Studies in History and Philosophy of Modern Physics}}, 48:102--108, November 2014.

\bibitem{EED_Kontou}
P.~Brown, C.~J. Fewster, and E.~A. Kontou.
\newblock {A singularity theorem for Einstein–Klein–Gordon theory}.
\newblock {\em {General Relativity and Gravitation}}, 50(10), September 2018.

\bibitem{classNEC_stab1}
R.~V. Buniy, S.~D.~H. Hsu, and B.~M. Murray.
\newblock The null energy condition and instability.
\newblock {\em {Phys. Rev. D}}, 74, September 2006.

\bibitem{classNEC_stab2}
S.~Dubovsky, T.~Grégoire, A.~Nicolis, and R.~Rattazzi.
\newblock Null energy condition and superluminal propagation.
\newblock {\em {Journal of High Energy Physics}}, 2006(03):025, March 2006.

\bibitem{classNEC_stab3}
S.~M. Carroll, M.~Hoffman, and M.~Trodden.
\newblock Can the dark energy equation-of-state parameter $\ensuremath{w}$ be less than $\ensuremath{-}1?$.
\newblock {\em {Phys. Rev. D}}, 68:023509, July 2003.

\bibitem{HawkingA}
S.~A. Hayward.
\newblock General laws of black-hole dynamics.
\newblock {\em {Phys. Rev. D}}, 49:6467--6474, June 1994.

\bibitem{classGSL}
E.~E. Flanagan, D.~Marolf, and R.~M. Wald.
\newblock {Proof of classical versions of the Bousso entropy bound and of the generalized second law}.
\newblock {\em {Phys. Rev. D}}, 62:084035, September 2000.

\bibitem{weak_thms1}
A.~Borde.
\newblock Geodesic focusing, energy conditions and singularities.
\newblock {\em Classical and Quantum Gravity}, 4(2):343, March 1987.

\bibitem{weak_thms2}
T.~A. Roman.
\newblock {On the ``averaged weak energy condition'' and Penrose's singularity theorem}.
\newblock {\em {Phys. Rev. D}}, 37:546--548, January 1988.

\bibitem{weak_thms3}
C.~J. Fewster and G.~J. Galloway.
\newblock Singularity theorems from weakened energy conditions.
\newblock {\em {Classical and Quantum Gravity}}, 28(12):125009, May 2011.

\bibitem{CasimirEffect}
H.~B.~G. Casimir.
\newblock {On the attraction between two perfectly conducting plates}.
\newblock {\em Indag. Math.}, 10(4):261--263, 1948.

\bibitem{Haag_1996}
R.~Haag.
\newblock {\em {Local Quantum Physics: Fields, Particles, Algebras}}.
\newblock Springer, New York and Heidelberg and Berlin, 1996.

\bibitem{Witten_ReehSchlieder}
E.~Witten.
\newblock {APS Medal for Exceptional Achievement in Research: Invited article on entanglement properties of quantum field theory}.
\newblock {\em {Rev. Mod. Phys.}}, 90:045003, October 2018.

\bibitem{Reeh_Schlieder_1961}
H.~Reeh and S.~Schlieder.
\newblock {Bemerkungen zur Unitäräquivalenz von Lorentzinvarianten Feldern}.
\newblock {\em {Il Nuovo Cimento}}, 22(5):1051–1068, December 1961.

\bibitem{BlancoCasini_2013}
D.~D. Blanco and H.~Casini.
\newblock {Localization of Negative Energy and the Bekenstein Bound}.
\newblock {\em {Phys. Rev. Lett.}}, 111:221601, November 2013.

\bibitem{Griffiths_QM}
D.~J. Griffiths and D.~F. Schroeter.
\newblock {\em {Introduction to Quantum Mechanics}}.
\newblock {Cambridge University Press}, August 2018.

\bibitem{Kiritsis}
E.~Kiritsis.
\newblock {\em {String Theory In A Nutshell Ed. 2}}.
\newblock {Princeton University Press}, January 2019.

\bibitem{Klinkhammer}
G.~Klinkhammer.
\newblock Averaged energy conditions for free scalar fields in flat spacetime.
\newblock {\em {Phys. Rev. D}}, 43:2542--2548, April 1991.

\bibitem{WeylVsConformal}
K.~Farnsworth, M.~A. Luty, and V.~Prilepina.
\newblock {Weyl versus conformal invariance in quantum field theory}.
\newblock {\em {The Journal of High Energy Physics}}, 2017(10), October 2017.

\bibitem{Conforme_anomalie}
S.~Deser and A.~Schwimmer.
\newblock Geometric classification of conformal anomalies in arbitrary dimensions.
\newblock {\em {Physics Letters B}}, 309(3):279--284, July 1993.

\bibitem{ANEC_violation_VeryOld}
M.~Visser.
\newblock Scale anomalies imply violation of the averaged null energy condition.
\newblock {\em {Physics Letters B}}, 349(4):443--447, April 1995.

\bibitem{Maldacena}
J.~M. Maldacena.
\newblock {The large N limit of superconformal field theories and supergravity}.
\newblock {\em {International Journal of Theoretical Physics}}, 38(4):1113–1133, January 1999.

\bibitem{Polchinski_AdS/CFT}
J.~Polchinski.
\newblock {Introduction to Gauge/Gravity Duality}, October 2010.

\bibitem{FeffermanGraham}
C.~Fefferman and C.~R. Graham.
\newblock {Conformal invariants}.
\newblock In {\em {Élie Cartan et les mathématiques d'aujourd'hui - Lyon, 25-29 juin 1984}}, number S131 in {Astérisque}, pages 95--116. {Société mathématique de France}, 1985.

\bibitem{holographic_EMT}
S.~De~Haro, K.~Skenderis, and S.~N. Solodukhin.
\newblock {Holographic reconstruction of spacetime and renormalization in the AdS/CFT correspondence}.
\newblock {\em {Communications in Mathematical Physics}}, 217(3):595–622, March 2001.

\bibitem{vanBrunt}
B.~Van~Brunt.
\newblock {\em The calculus of variations}.
\newblock {Springer New York}, {New York}, January 2004.

\bibitem{Poincare}
M.~Bebendorf.
\newblock {A note on the Poincaré inequality for convex domains}.
\newblock {\em {Zeitschrift für Analysis und ihre Anwendungen}}, page 751–756, January 2003.

\bibitem{FlanaganWald}
E.~E. Flanagan and R.~M. Wald.
\newblock Does back reaction enforce the averaged null energy condition in semiclassical gravity?
\newblock {\em {Phys. Rev. D}}, 54:6233--6283, November 1996.

\bibitem{Bekenstein_1973}
J.~D. Bekenstein.
\newblock Black holes and entropy.
\newblock {\em {Phys. Rev. D}}, 7:2333--2346, April 1973.

\bibitem{Hawking_entropie}
S.~W. Hawking.
\newblock Particle creation by black holes.
\newblock {\em {Communications in Mathematical Physics}}, 43(3):199–220, August 1975.

\bibitem{dS_entropie_hawking}
G.~W. Gibbons and S.~W. Hawking.
\newblock Cosmological event horizons, thermodynamics, and particle creation.
\newblock {\em {Phys. Rev. D}}, 15:2738--2751, May 1977.

\bibitem{dS_entropie}
P.~C.~W. Davies.
\newblock Cosmological horizons and the generalised second law of thermodynamics.
\newblock {\em {Classical and Quantum Gravity}}, 4(6):L225, November 1987.

\bibitem{Friedmann_entropie}
P.~C.~W. Davies.
\newblock Cosmological horizons and entropy.
\newblock {\em {Classical and Quantum Gravity}}, 5(10):1349, October 1988.

\bibitem{SecondLaw_CausalHorizon}
T.~Jacobson and R.~Parentani.
\newblock {Horizon Entropy}.
\newblock {\em {Foundations of Physics}}, 33(2):323–348, February 2003.

\bibitem{Galloway_2000}
G.~J. Galloway.
\newblock Maximum principles for null hypersurfaces and null splitting theorems.
\newblock {\em {Annales Henri Poincaré}}, 1(3):543–567, July 2000.

\bibitem{hbar_expansion}
S.~J. Brodsky and P.~Hoyer.
\newblock The $\ensuremath{\hbar}$ expansion in quantum field theory.
\newblock {\em {Phys. Rev. D}}, 83:045026, February 2011.

\bibitem{Wall_entropy}
A.~C. Wall.
\newblock Ten proofs of the generalized second law.
\newblock {\em {Journal of High Energy Physics}}, 2009(06):021, June 2009.

\bibitem{Pippenger_2003}
N.~Pippenger.
\newblock The inequalities of quantum information theory.
\newblock {\em {IEEE Transactions on Information Theory}}, 49(4):773–789, April 2003.

\bibitem{shearfree}
G.~F.~R. Ellis.
\newblock Shear free solutions in general relativity theory.
\newblock {\em {General Relativity and Gravitation}}, 43(12):3253–3268, August 2011.

\bibitem{duff_anomaly}
M.~J. Duff.
\newblock Observations on conformal anomalies.
\newblock {\em {Nuclear Physics B}}, 125(2):334--348, 1977.

\bibitem{NonMinimalScalar}
D.~G. Figueroa, A.~Florio, T.~Opferkuch, and B.~A. Stefanek.
\newblock {Lattice simulations of non-minimally coupled scalar fields in the Jordan frame}.
\newblock {\em {SciPost Phys.}}, 15:077, 2023.

\bibitem{Hartle_2014}
J.~B. Hartle.
\newblock {\em {Gravity: An Introduction to Einstein’s General Relativity}}.
\newblock Pearson, 2014.

\bibitem{conforme_rand}
J.~Frauendiener.
\newblock {Conformal Infinity}.
\newblock {\em {Living Reviews in Relativity}}, 7(1), February 2004.

\bibitem{PeskinSchroeder}
M.~E. Peskin and D.~V. Schroeder.
\newblock {\em {An Introduction to Quantum Field Theory}}.
\newblock {CRC Press}, 2019.

\bibitem{Schwartz}
M.~D. Schwartz.
\newblock {\em {Quantum Field Theory and the Standard Model}}.
\newblock {Cambridge University Press}, 2014.

\bibitem{Wald_1994}
R.~M. Wald.
\newblock {\em Quantum field theory in curved spacetime and black hole thermodynamics}.
\newblock {The University of Chicago Press}, Chicago and London, January 1994.

\bibitem{CFT_Modave}
A.~Rovai.
\newblock {Introduction to Conformal Field Theory}.
\newblock {\em Proceedings of Science}, Modave VIII:001, 2013.

\bibitem{CFT_Rychkov}
S.~Rychkov.
\newblock {\em {EPFL lectures on Conformal Field Theory in $D \geq 3$ dimensions}}.
\newblock {Springer Cham}, 2017.

\bibitem{Wald_TraceAnomaly}
R.~M. Wald.
\newblock Trace anomaly of a conformally invariant quantum field in curved spacetime.
\newblock {\em {Phys. Rev. D}}, 17:1477--1484, March 1978.

\end{thebibliography}

\end{document}